\documentstyle[11pt,subfigure,graphicx,color,cancel,cite,ulem,framed,hyperref,url]{article}
\hypersetup{linktocpage}
\definecolor{shadecolor}{named}{Gray}

\newcommand{\newc}{\newcommand}

\def\Ord{\lower .7ex\hbox{$\;\stackrel{\textstyle <}{\sim}\;$}}
\def\OOrd{\lower .7ex\hbox{$\;\stackrel{\textstyle >}{\sim}\;$}}

\newc{\order}{{\cal O}}

\def\lum             {{\cal L}}

\newc{\be}{\begin{equation}}
\newc{\ee}{\end{equation}}
\newc{\br}{\begin{eqnarray}}
\newc{\er}{\end{eqnarray}}
\newc{\ba}{\begin{array}}
\newc{\ea}{\end{array}}
\newc{\bi}{\begin{itemize}}
\newc{\ei}{\end{itemize}}
\newc{\bn}{\begin{enumerate}}
\newc{\en}{\end{enumerate}}
\newc{\bc}{\begin{center}}
\newc{\ec}{\end{center}}
\newc{\ul}{\underline}
\newc{\ra}{\rightarrow}
\newc{\lra}{\longrightarrow}
\newc{\wt}{\widetilde}
\newc{\til}{\tilde}

\newc{\wh}{\widehat}
\newc{\ti}{\times}
\newc{\Dir}{\kern -6.4pt\Big{/}}
\newc{\Dirin}{\kern -10.4pt\Big{/}\kern 4.4pt}
\newc{\DDir}{\kern -10.6pt\Big{/}}
\newc{\DGir}{\kern -6.0pt\Big{/}}
\newc{\sig}{\sigma}
\newc{\sigmalstop}{\sig_{\lstoppair}}
\newc{\Sig}{\Sigma}  
\newc{\del}{\delta}
\newc{\Del}{\Delta}
\newc{\lam}{\lambda}
\newc{\Lam}{\Lambda}
\newc{\gam}{\gamma}
\newc{\Gam}{\Gamma}
\newc{\eps}{\epsilon}
\newc{\Eps}{\Epsilon}
\newc{\kap}{\kappa}
\newc{\Kap}{\Kappa}
\newc{\modulus}[1]{\left| #1 \right|}
\newc{\eq}[1]{(\ref{eq:#1})}
\newc{\eqs}[2]{(\ref{eq:#1},\ref{eq:#2})}
\newc{\etal}{{\it et al.}\ }
\newc{\ibid}{{\it ibid}.}
\newc{\ibidem}{{\it ibidem}.}
\newc{\eg}{{\it e.g.}\ }
\newc{\ie}{{\it i.e.}\ }

\newc{\nonum}{\nonumber}
\newc{\lab}[1]{\label{eq:#1}}
\newc{\dpr}[2]{({#1}\cdot{#2})}
\newc{\lt}{\stackrel{<}}
\newc{\gt}{\stackrel{>}}
\newc{\lsimeq}{\stackrel{<}{\sim}}
\newc{\gsimeq}{\stackrel{>}{\sim}}
\def\lsim{\buildrel{\scriptscriptstyle <}\over{\scriptscriptstyle\sim}}
\def\gsim{\buildrel{\scriptscriptstyle >}\over{\scriptscriptstyle\sim}}
\def\lapp{\mathrel{\rlap{\raise.5ex\hbox{$<$}}
                    {\lower.5ex\hbox{$\sim$}}}}
\def\gapp{\mathrel{\rlap{\raise.5ex\hbox{$>$}}
                    {\lower.5ex\hbox{$\sim$}}}}
\newc{\half}{\frac{1}{2}}

\newc{\bQ}{\ol{Q}}
\newc{\dota}{\dot{\alpha }}
\newc{\dotb}{\dot{\beta }}
\newc{\dotd}{\dot{\delta }}
\newc{\nindnt}{\noindent}

\newc{\matth}{\mathsurround=0pt}
\def\ML{\ifmmode{{\mathaccent"7E M}_L}
             \else{${\mathaccent"7E M}_L$}\fi}
\def\MR{\ifmmode{{\mathaccent"7E M}_R}
             \else{${\mathaccent"7E M}_R$}\fi}

\newc{\mr}{\mathrm}

\newc{\siminf}{\mbox{$_{\sim}$ {\small {\hspace{-1.em}{$<$}}}    }}
\newc{\simsup}{\mbox{$_{\sim}$ {\small {\hspace{-1.em}{$>$}}}    }}

\newc {\Zboson}{{\mathrm Z}^{0}}
\newc{\thetaw}{\theta_W}
\newc{\mbot}{{m_b}}
\newc{\mtop}{{m_t}}
\newc{\sm}{${\cal {SM}}$}
\newc{\as}{\alpha_s}
\newc{\aem}{\alpha_{em}}

\newc{\ppbar}{\mbox{$p\ol{p}$}}
\newc{\bbbar}{\mbox{$b\ol{b}$}}
\newc{\ccbar}{\mbox{$c\ol{c}$}}
\newc{\ttbar}{\mbox{$t\ol{t}$}}
\newc{\eebar}{\mbox{$e\ol{e}$}}
\newc{\zzero}{\mbox{$Z^0$}}

\newc{\wplus}{\mbox{$W^+$}}
\newc{\wminus}{\mbox{$W^-$}}
\newc{\ellp}{\ell^+}
\newc{\ellm}{\ell^-}
\newc{\elp}{\mbox{$e^+$}}
\newc{\elm}{\mbox{$e^-$}}
\newc{\elpm}{\mbox{$e^{\pm}$}}
\newc{\qbar}     {\mbox{$\ol{q}$}}

\newc{\Ebar}{{\bar E}}
\newc{\Dbar}{{\bar D}}
\newc{\Ubar}{{\bar U}}
\newc{\susy}{{{SUSY}}}
\newc{\msusy}{{{M_{SUSY}}}}

\def\photino{\ifmmode{\mathaccent"7E \gam}\else{$\mathaccent"7E \gam$}\fi}
\def\taugluino{\ifmmode{\tau_{\mathaccent"7E g}}
             \else{$\tau_{\mathaccent"7E g}$}\fi}
\def\mphotino{\ifmmode{m_{\mathaccent"7E \gam}}
             \else{$m_{\mathaccent"7E \gam}$}\fi}
\newc{\gl}   {\mbox{$\wt{g}$}}
\newc{\mgl}  {\mbox{$m_{\gl}$}}
\def \charginopm{{\wt\chi}^{\pm}}

\def \chonep {{\wt\chi_1^+}}

\def \ch2p {{\wt\chi_2^+}}
\def \chonem {{\wt\chi_1^-}}
\def \ch2m {{\wt\chi_2^-}}

\def \chonipm{{\wt\chi_i}^{\pm}}

\def \chonepm{{\wt\chi_1}^{\pm}}

\def \mchonepm{m_{\chonepm}}

\def \chtwopm{{\wt\chi_2}^{\pm}}

\newc{\dmchi}{\Delta m_{\wt\chi}}

\def \lspi{\wt\chi_i^0}

\def \lspone{\wt\chi_1^0}
\def \mlspone{m_{\lspone}}
\def \lsptwo{\wt\chi_2^0}
\def \mlsptwo{m_{\lsptwo}}

\newc{\sele}{\wt{\mathrm e}}
\newc{\sell}{\wt{\ell}}

\def \stauone{\wt\tau_1}
\def \stauonepm{{\wt\tau_1}^\pm}
\def \mstauone{m_{\stauone}}
\def \snu{\wt{\nu}}
\def \snutau{\wt{\nu}_{\tau}}
\def \nutau{{\nu}_{\tau}}                  

\newc{\snue}     {\mbox{$ \wt{\nu_e}$}}
\newc{\smu}{\wt{\mu}}
\newc{\stau}{\wt{\tau}}
\newc {\nuL} {\wt{\nu}_L}
\newc {\nuR} {\wt{\nu}_R}
\newc {\snub} {\bar{\wt{\nu}}}
\newc {\eL} {\wt{e}_L}
\newc {\eR} {\wt{e}_R}
\def \slepl{\wt{l}_L}

\def \slepr{\wt{l}_R}

\def \stau{\wt\tau}

\def \sq{\wt{q}}

\newc{\msqot}  {\mbox{$m_(\sq_{1,2} )$}}
\newc{\sqbar}    {\mbox{$\bar{\wt{q}}$}}
\newc{\ssb}      {\mbox{$\squark\ol{\squark}$}}
\newc {\qL} {\wt{q}_L}
\newc {\qR} {\wt{q}_R}
\newc {\uL} {\wt{u}_L}
\newc {\uR} {\wt{u}_R}
\def \ul{\wt{u}_L}

\newc {\dL} {\wt{d}_L}
\newc {\dR} {\wt{d}_R}
\newc {\cL} {\wt{c}_L}
\newc {\cR} {\wt{c}_R}
\newc {\sL} {\wt{s}_L}
\newc {\sR} {\wt{s}_R}
\newc {\tL} {\wt{t}_L}
\newc {\tR} {\wt{t}_R}
\newc {\stb} {\ol{\wt{t}}_1}
\newc {\sbot} {\wt{b}_1}
\newc {\msbot} {m_{\sbot}}
\newc {\sbotb} {\ol{\wt{b}}_1}
\newc {\bL} {\wt{b}_L}
\newc {\bR} {\wt{b}_R}

\newc{\csquark}  {\mbox{$\wt{c}$}}
\newc{\csquarkl} {\mbox{$\wt{c}_L$}}
\newc{\mcsl}     {\mbox{$m(\csquarkl)$}}

\newc {\stopl}         {\wt{\mathrm{t}}_{\mathrm L}}
\newc {\stopr}         {\wt{\mathrm{t}}_{\mathrm R}}
\newc {\stoppair}      {\wt{\mathrm{t}}_{1}
\bar{\wt{\mathrm{t}}}_{1}}
\def \lstop{\wt{t}_{1}}

\def \lstoppair{\lstop\lstop^*}

\newc{\tsquark}  {\mbox{$\wt{t}$}}
\newc{\ttb}      {\mbox{$\tsquark\ol{\tsquark}$}}
\newc{\ttbone}   {\mbox{$\tsquark_1\ol{\tsquark}_1$}}

\newc{\mix}{\theta_{\wt t}}
\newc{\cost}{\cos{\theta_{\wt t}}}
\newc{\sint}{\sin{\theta_{\wt t}}}
\newc{\costloop}{\cos{\theta_{\wt t_{loop}}}}

\newc{\mixsbot}{\theta_{\wt b}}

\newc{\tb}{\tan\beta}
\newc{\cb}{\cot\beta}
\newc{\vev}[1]{{\left\langle #1\right\rangle}}

\newc{\mhalf}{m_{1/2}}
\newc{\mzero} {\mbox{$m_0$}}
\newc{\azero} {\mbox{$A_0$}}

\newc{\lb}{\lam}
\newc{\lbp}{\lam^{\prime}}
\newc{\lbpp}{\lam^{\prime\prime}}
\newc{\rpv}{{\not \!\! R_p}}
\newc{\rpvm}{{\not  R_p}}
\newc{\rp}{R_{p}}
\newc{\rpmssm}{{RPC MSSM}}
\newc{\rpvmssm}{{RPV MSSM}}

\newc{\sbyb}{S/$\sqrt B$}
\newc{\pelp}{\mbox{$e^+$}}
\newc{\pelm}{\mbox{$e^-$}}
\newc{\pelpm}{\mbox{$e^{\pm}$}}
\newc{\epem}{\mbox{$e^+e^-$}}
\newc{\lplm}{\mbox{$\ell^+\ell^-$}}

\def\Ecm{\ifmmode{E_{\mathrm{cm}}}\else{$E_{\mathrm{cm}}$}\fi}
\newc{\rts}{\sqrt{s}}
\newc{\rtshat}{\sqrt{\hat s}}
\newc{\gev}{\,GeV}
\newc{\mev}{~{\rm MeV}}
\newc{\tev}  {\mbox{$\;{\rm TeV}$}}
\newc{\gevc} {\mbox{$\;{\rm GeV}/c$}}
\newc{\gevcc}{\mbox{$\;{\rm GeV}/c^2$}}
\newc{\intlum}{\mbox{${ \int {\cal L} \; dt}$}}
\newc{\call}{{\cal L}}
\def \met  {\mbox{${E\!\!\!\!/_T}$}}

\newc{\ptmiss}{/ \hskip-7pt p_T}

\def \etslash{\not \! E_T }

\newc{\PT}{\mbox{$p_T$}}
\newc{\ET}{\mbox{$E_T$}}
\newc{\dedx}{\mbox{${\rm d}E/{\rm d}x$}}
\newc{\ifb}{\mbox{${\rm fb}^{-1}$}}
\newc{\ipb}{\mbox{${\rm pb}^{-1}$}}
\newc{\pb}{~{\rm pb}}
\newc{\fb}{~{\rm fb}}
\newc{\ycut}{y_{\mathrm{cut}}}
\newc{\chis}{\mbox{$\chi^{2}$}}
\def\dzero{\emptyset}

\def \jet(s){\emph{jet(s) }}

\newc{\mpl}{M_{\rm Pl}}
\newc{\mgut}{M_{GUT}}
\newc{\mw}{M_{W}}
\newc{\mweak}{M_{weak}}
\newc{\mz}{M_{Z}}

\newc{\OPALColl}   {OPAL Collaboration}
\newc{\ALEPHColl}  {ALEPH Collaboration}
\newc{\DELPHIColl} {DELPHI Collaboration}
\newc{\XLColl}     {L3 Collaboration}
\newc{\JADEColl}   {JADE Collaboration}
\newc{\CDFColl}    {CDF Collaboration}
\newc{\DXColl}     {D0 Collaboration}
\newc{\HXColl}     {H1 Collaboration}
\newc{\ZEUSColl}   {ZEUS Collaboration}
\newc{\LEPColl}    {LEP Collaboration}
\newc{\ATLASColl}  {ATLAS Collaboration}
\newc{\CMSColl}    {CMS Collaboration}
\newc{\UAColl}    {UA Collaboration}
\newc{\KAMLANDColl}{KamLAND Collaboration}
\newc{\IMBColl}    {IMB Collaboration}
\newc{\KAMIOColl}  {Kamiokande Collaboration}
\newc{\SKAMIOColl} {Super-Kamiokande Collaboration}
\newc{\SUDANTColl} {Soudan-2 Collaboration}
\newc{\MACROColl}  {MACRO Collaboration}
\newc{\GALLEXColl} {GALLEX Collaboration}
\newc{\GNOColl}    {GNO Collaboration}
\newc{\SAGEColl}  {SAGE Collaboration}
\newc{\SNOColl}  {SNO Collaboration}
\newc{\CHOOZColl}  {CHOOZ Collaboration}
\newc{\PDGColl}  {Particle Data Group Collaboration}

\def\issue(#1,#2,#3){{\bf #1}, #2 (#3)}
\def\iss(#1,#2,#3){{\bf #1} (#3) #2}
\def\ASTR(#1,#2,#3){Astropart.\ Phys. \issue(#1,#2,#3)}
\def\AJ(#1,#2,#3){Astrophysical.\ Jour. \issue(#1,#2,#3)}
\def\AJS(#1,#2,#3){Astrophys.\ J.\ Suppl. \issue(#1,#2,#3)}
\def\APP(#1,#2,#3){Acta.\ Phys.\ Pol. \issue(#1,#2,#3)}
\def\JCAP(#1,#2,#3){Journal\ XX. \issue(#1,#2,#3)} 
\def\SC(#1,#2,#3){Science \issue(#1,#2,#3)}
\def\PRD(#1,#2,#3){Phys.\ Rev.\ D \issue(#1,#2,#3)}
\def\PR(#1,#2,#3){Phys.\ Rev.\ \issue(#1,#2,#3)} 
\def\PRC(#1,#2,#3){Phys.\ Rev.\ C \issue(#1,#2,#3)}
\def\NPB(#1,#2,#3){Nucl.\ Phys.\ B \issue(#1,#2,#3)}
\def\NPPS(#1,#2,#3){Nucl.\ Phys.\ Proc. \ Suppl \issue(#1,#2,#3)}
\def\NJP(#1,#2,#3){New.\ J.\ Phys. \issue(#1,#2,#3)}
\def\JP(#1,#2,#3){J.\ Phys.\issue(#1,#2,#3)}
\def\PL(#1,#2,#3){Phys.\ Lett. \issue(#1,#2,#3)}
\def\ZP(#1,#2,#3){Z.\ Phys. \issue(#1,#2,#3)}
\def\ZPC(#1,#2,#3){Z.\ Phys.\ C  \issue(#1,#2,#3)}
\def\PREP(#1,#2,#3){Phys.\ Rep. \issue(#1,#2,#3)}
\def\PRL(#1,#2,#3){Phys.\ Rev.\ Lett. \issue(#1,#2,#3)}
\def\MPL(#1,#2,#3){Mod.\ Phys.\ Lett. \issue(#1,#2,#3)}
\def\RMP(#1,#2,#3){Rev.\ Mod.\ Phys. \issue(#1,#2,#3)}
\def\SJNP(#1,#2,#3){Sov.\ J.\ Nucl.\ Phys. \issue(#1,#2,#3)}
\def\CPC(#1,#2,#3){Comp.\ Phys.\ Comm. \issue(#1,#2,#3)}
\def\IJMPA(#1,#2,#3){Int.\ J.\ Mod. \ Phys.\ A \issue(#1,#2,#3)}
\def\MPLA(#1,#2,#3){Mod.\ Phys.\ Lett.\ A \issue(#1,#2,#3)}
\def\PTP(#1,#2,#3){Prog.\ Theor.\ Phys. \issue(#1,#2,#3)}
\def\RMP(#1,#2,#3){Rev.\ Mod.\ Phys. \issue(#1,#2,#3)}
\def\NIMA(#1,#2,#3){Nucl.\ Instrum.\ Methods \ A \issue(#1,#2,#3)}
\def\EPJC(#1,#2,#3){Eur.\ Phys.\ J.\ C \issue(#1,#2,#3)}
\def\RPP (#1,#2,#3){Rept.\ Prog.\ Phys. \issue(#1,#2,#3)}
\def\PPNP(#1,#2,#3){ Prog.\ Part.\ Nucl.\ Phys. \issue(#1,#2,#3)}
\newc{\PRDR}[3]{{Phys. Rev. D} {\bf #1}, Rapid  Communications, #2 (#3)}

\def\PLB(#1,#2,#3){Phys.\ Lett.\ B  \iss(#1,#2,#3)}
\def\JHEP(#1,#2,#3){JHEP \iss(#1,#2,#3)}

\def\amususy{a_\mu^{\rm SUSY}}
\def\gmin2{(g-2)_\mu}

\catcode`\@=11 

\def\vev#1{\left\langle #1\right\rangle}
\def\lsim{\mathrel{\mathpalette\@versim<}}
\def\gsim{\mathrel{\mathpalette\@versim>}}
\def\@versim#1#2{\vcenter{\offinterlineskip
    \ialign{$\m@th#1\hfil##\hfil$\crcr#2\crcr\sim\crcr } }}
\def\etal{{\em et. al.}}

\def\r2{\sqrt 2}
\def\beq{\begin{equation}}
\def\eeq{\end{equation}}
\def\beqn{\begin{eqnarray}}
\def\eeqn{\end{eqnarray}}
\def\sinW2{\sin^2\theta_W}

\def\mz2{M_{z}^2}
\def\c2b{\cos 2\beta}

\def\m#1{{\tilde m}_#1}

\def\mw#1{{\tilde m}_{\omega #1}}

\def\mz{M_Z}
\def\m0{m_0}
\def\mhalf{m_{\frac{1}{2}}}

\def\cb{\cos\beta}
\def\sb{\sin\beta}

\def\sec2w{sec^2\theta_W}

\def\amususy{a_\mu^{\rm SUSY}}
\def\gmin2{(g-2)_\mu}

\catcode`\@=11 

\def\vev#1{\left\langle #1\right\rangle}
\def\lsim{\mathrel{\mathpalette\@versim<}}
\def\gsim{\mathrel{\mathpalette\@versim>}}
\def\@versim#1#2{\vcenter{\offinterlineskip
    \ialign{$\m@th#1\hfil##\hfil$\crcr#2\crcr\sim\crcr } }}
\def\etal{{\em et. al.}}

\def\tb{\tilde b}

\def\tL{\tilde L}

\def\ttau{\tilde \tau}

\def \charginopm{{\wt\chi}^{\pm}}

\def \chonep{{\wt\chi_1}^{+}}
\def \chonem{{\wt\chi_1^-}}
\def \chonep2{{\wt\chi_2^+}}
\def \chonem2{{\wt\chi_2^-}}

\def \chonipm{{\wt\chi_i}^{\pm}}

\def \chonepm{{\wt\chi_1}^{\pm}}

\def \mchonepm{m_{\chonepm}}

\def \chtwopm{{\wt\chi_2}^{\pm}}

\def \lstop{\wt{t}_{1}}

\def \lspi{\wt\chi_i^0}

\def \lspone{\wt\chi_1^0}
\def \mlspone{m_{\lspone}}
\def \lsptwo{\wt\chi_2^0}
\def \mlsptwo{m_{\lsptwo}}

\def\PL{Phys. Lett.}
\def\PRL{Phys. Rev. Lett.}

\def\PR{Phys. Rev.}

\def \lsptwo{\wt\chi_2^0}
\def \lspone{\wt\chi_1^0}
\def \chonem {{\wt\chi_1^\pm}}
\def \chargino1 {{\wt\chi_1^\pm}}
\def \chargino2 {{\wt\chi_2^\pm}}
\def \lstop{\wt{t}_{1}}
\def \ch2m {{\wt\chi_2^-}}
\def \lspi{\wt\chi_i^0}
\def \chonep {{\wt\chi_1^+}}

\def\mygraph#1#2{ \subfigure[]{
   \label{#1}
   \hspace*{-0.6in}
   \begin{minipage}[b]{0.5\textwidth}
   \centering
   \hspace*{4ex}
   \includegraphics[width=\textwidth]{#2}
   \vspace*{-4ex}
   \end{minipage}}
   \vspace*{-1ex}}
\begin{document}

\begin{center}
{\large \bf  The Electroweak Sector of the pMSSM in the Light of LHC - 8 TeV and Other Data}

\vskip 0.3cm
Manimala Chakraborti$^{a}$\footnote{tpmc@iacs.res.in},
Utpal Chattopadhyay$^{a}$\footnote{tpuc@iacs.res.in}, 
Arghya Choudhury$^{b}$\footnote{ arghyac@iiserkol.ac.in}, 
Amitava Datta$^{c}$\footnote{adatta@iiserkol.ac.in},
Sujoy Poddar$^{d}$\footnote{sujoy.phy@gmail.com}
\vskip 0.3cm
{$^a$
Department of Theoretical Physics, Indian Association
for the Cultivation of Science,\\
2A \& B Raja S.C. Mullick Road, Jadavpur,
Kolkata 700 032, India}\\
{$^b$
Department of Physical Sciences, Indian Institute of Science Education and \\ 
Research (IISER) - Kolkata, Mohanpur, Nadia, West Bengal - 741252, India
}\\
{$^c$
Department of Physics, University of Calcutta, 92 A.P.C. Road, 
Kolkata 700 009, India
}\\
{$^d$
Department of Physics, Netaji Nagar Day College, 170/436, N.S.C. Bose Road,\\ Kolkata - 700092, India}

\end{center}

\begin{abstract}
Using the chargino-neutralino and slepton search results from 
the LHC in conjunction with the WMAP/PLANCK and $(g-2)_{\mu}$ data,
we constrain several generic pMSSM models with decoupled strongly 
interacting sparticles, heavier Higgs bosons and 
characterized by different hierarchies among the EW sparticles.  
We  find that some of 
them are already  under pressure and this number increases if bounds 
from direct detection experiments like LUX are taken into 
account, keeping in mind the associated uncertainties.  
The XENON1T experiment is likely to scrutinize the remaining models 
closely. Analysing models with heavy squarks, a light 
gluino along with widely different EW sectors, we show that the limits on $\mgl$ 
are not likely to be below 1.1 TeV, 
if a multichannel analysis of the LHC data is performed.  Using this 
light gluino scenario we further illustrate that in future 
LHC experiments the models with different EW sectors can be 
distinguished from each other by the relative sizes of the 
$n$-leptons + $m$-jets + $\met$ signals for different choices of $n$.  
\end{abstract}

\newpage
\setcounter{footnote}{0}

\hrule
\tableofcontents
\vspace{0.2cm}
\hrule

\section{Introduction}
\label{section1}

The LHC experiments at $\sqrt{s}=$7/8 TeV have concluded recently. 
The painstaking searches for supersymmetry (SUSY) \cite{SUSYreviews1,SUSYreviews2,SUSYbooks}, 
the most popular and attractive extension of the standard model (SM) 
of particle physics have not observed any signal yet. Consequently stringent limits on the masses of the 
supersymmetric particles (sparticles) belonging to the strongly interacting sector, expected to be 
produced with large cross-sections, have been obtained by both the ATLAS and the CMS collaborations 
\cite{atlas0l, atlas1l, atlas2l, atlas-susy,cms-susy}
 \footnote{However, these stringent bounds are reduced significantly in compressed SUSY type scenarios \cite{compressed}.}. 
Whether these limits already put  question marks on the naturalness\cite{naturalness,naturalness_recent} of various SUSY models
may be debated in spite of the fact that it is 
hard to quantify the degree of naturalness.  Naturalness or the absence of it  should 
therefore be left at the stage of a healthy theoretical debate and not be regarded as the concluding remark on  SUSY.     

The minimal supersymmetric standard model (MSSM)\cite{SUSYreviews2,SUSYbooks} 
has another important component - the electroweak (EW) sector. 
The production cross-sections of the sparticles belonging to this sector at 
the LHC are rather modest. As a result 
there was no constraint on the properties of these sparticles until recently. 
Thus some weak mass limits from 
LEP\cite{lepsusy} and Tevatron \cite{cdf3l5.8fb, d03l2.3fb} were the only 
available information on this sector. 
The purpose of this paper is to focus on this sector in the light of the 
direct constraints from LHC 
\cite{atlas3lew,atlas2lew,cmsew} as 
well as indirect constraints like the observed value of the anomalous magnetic moment of the 
muon from the Brookhaven $\gmin2$ experiment\cite{g-2exp} and the 
relic density constraints for dark matter from WMAP\cite{wmap} or 
PLANCK\cite{planck} experiments. 
Using the combined constraints we then identify the allowed 
parameter space (APS).

We will also consider 
the constraints from direct \cite{xenon100,lux,xenon1t} and a few selected 
indirect searches \cite{fermi-lat-gamma} of dark matter which may involve considerable 
theoretical and astrophysical uncertainties 
(to be elaborated in a subsequent section).  
In view of this we present our results in such a way that the effect of 
each constraint may separately be seen. We also study the prospect of 
future LHC searches and the issue of 
distinguishing several EW scenarios having 
different dark matter (DM) annihilation/coannihilation mechanisms leading to 
correct relic density (we will often refer this as DM producing mechanisms).

Since the SUSY breaking mechanism leading to  a given pattern 
of sparticle masses is unknown,
in the most general MSSM the above two sectors are unrelated.  
Only in models with high scale physics inputs due to considering specific 
mechanisms of SUSY breaking like the minimal supergravity 
(mSUGRA)\cite{msugramodel}, the masses of the 
strong and the EW sparticles are correlated. 
As a result, the stringent bounds on the 
former sector translate into bounds on the masses of the latter some 
of which are 
apparently much stronger than the direct 
limits. However, since the mechanism of SUSY breaking is essentially 
unknown it is preferable to free ourselves from such model dependent 
restrictions. 

Apart from particle physics, the EW sparticles may play important roles in cosmology as well. 
An attractive feature of all models of SUSY with R-parity\cite{SUSYbooks} conservation is 
that the lightest supersymmetric particle (LSP) is stable. In many 
models the lightest neutralino $\lspone$ happens to be LSP. This weakly interacting massive particle is a popular 
candidate for the observed dark matter (DM) in the universe \cite{dm_rev1,dm_rev2,dmmany}. 
Moreover, the DM annihilation/coannihilation mechanisms leading to 
acceptable relic density for DM
may be driven entirely by the electroweak sparticles\cite{dm_rev1,dmmany,dmmssm}. Consequently the observed value 
of the DM relic density\cite{wmap,planck} may effectively be  
used to constrain the EW sector or a specific SUSY model in particular.

It was recently emphasized in Ref.\cite{arg_jhep1} that the physics of DM  and the stringent LHC bounds on the squark and 
gluino masses, obtained mainly from the jets + missing energy data, are controlled by two entirely different sectors 
of the phenomenological MSSM (pMSSM)\cite{pmssm}. While the DM producing 
mechanisms may broadly be insensitive to the strong sector\footnote{Except in situations like LSP-stop coannihilations.} of  
the pMSSM\cite{pmssm}, the response of the above LHC bounds 
to changes in the EW sector parameters is rather weak. It was demonstrated 
by simulations at the generator level that these bounds change modestly for a variety of EW sectors
with different characteristics all consistent 
with the DM relic density data\cite{arg_jhep1}. 
Thus the strong constraints on DM production in mSUGRA 
\cite{dmsugra, dmsugra_recent} due to squark-gluino mass bounds may be 
just an artifact of this model\footnote{For a recent review 
focussing on recent searches for dark-matter signatures at the LHC see 
Ref.\cite{vasiliki}.}. 

It was further noted that in the unconstrained MSSM, 
there are many possible DM producing mechanisms which are not 
viable in mSUGRA due to the constraints on the squark-gluino masses. Some  examples are LSP pair annihilation via  
Z or the lighter Higgs scalar (h) resonance, 
LSP-sneutrino coannihilation, coannihilation of a bino dominated LSP and a 
wino dominated chargino etc\cite{Baer:2005jq,arg_jhep1}. 
It may be emphasized that the  
discovery of the Higgs boson by the LHC collaborations\cite{higgs} 
has opened up the possibility of pinpointing the LSP pair annihilation via
h-resonance.

Subsequently both the CMS and the ATLAS collaborations published direct search limits on the masses of the electroweak 
sparticles in several models sensitive to the LHC experiments at 7 TeV \cite{atlas2l7, atlas3l7,cms2l3l7}. It 
was pointed out in Ref.\cite{arg_jhep2} the models constrained by the LHC experiments are important in the context of DM 
physics as well since many of these models contain light  sleptons either of L or R-type. It was demonstrated that even 
the preliminary mass bounds based on 13 $\ifb$ 8 TeV data\cite{atlas3l8tev13,cms3l8tev9} are able to put non-trivial 
 constraints on parameter space 
in regard to the neutralino relic density bounds. 
It was also pointed out that additionally if the gluinos are 
relatively light 
(just beyond the reach of the current 
LHC experiments) these models with the lightest neutralino as the LSP may 
lead to novel collider signatures. Especially in models with 
light sleptons the same sign dilepton (SSD) signal may indeed turn out to 
be stronger than the canonical jets + missing energy signal. 
Moreover, one is able to distinguish different relic density 
satisfying mechanisms by measuring the relative 
rates of the $n$-leptons + $m$-jets + missing energy 
events for different values of n.

More recently the LHC collaborations have published their analyses 
for EW sparticle searches based on 20 $\ifb$ data 
\cite{atlas3lew,atlas2lew,cmsew} which, as expected, yield stronger 
mass bounds. The results were interpreted in 
terms of several simplified models. In this approach only the masses of a limited number of sparticles
 relevant to a particular
signal are treated as free parameters, while the others are assumed to be decoupled. 
Moreover, in many cases the LSP 
is assumed to be bino dominated while the lighter chargino ($\chonepm$) to 
be wino dominated, but all the parameters that determine the masses and the mixings in the 
EW gaugino sectors are not precisely identified.
However, many of the above parameters which are moderately or 
marginally important for  
collider analyses, are quite important for computation of the 
indirect observables
such as the observed DM relic density bounds or $\gmin2$. In view of this 
we have computed the bounds by a PYTHIA \cite{pythia} based 
generator level analysis.  We use the full set of pMSSM parameters 
sufficient to determine all relevant observables. We also obtain bounds in 
related models not considered by the LHC collaborations in Refs.\cite{atlas3lew,atlas2lew,cmsew}.

We next consider a few indirect constraints in order of the level 
of stringency. 
We note that stringency of a constraint is increased if there is less 
model dependence while it is decreased if there is a large combined 
theoretical and experimental errors where some of the theoretical 
errors may not always even be precisely quantifiable.
With the details mentioned in 
Sec.~\ref{Section:DetailsOfConstraints}, 
the 
outline of the above constraints in the aforesaid order are given below:   
i) the precise dark matter relic density constraint from WMAP/PLANCK\cite{wmap,planck} 
within the ambit of standard model of cosmology\cite{kolb},   
ii) the $\gmin2$ data that deviates from the SM result by more than $3\sigma$\cite{g-2exp,g-2sm1,g-2sm2}, 
(which is becoming more and more potent 
with the gradual reduction of the disagreement between the $e^+$$e^-$ data based analyses 
and the ones that use hadronic $\tau$-decay data for evaluating the contributions for 
the hadronic vacuum polarisation part of the contributions to the theoretical estimation of $\gmin2$\cite{davierhadtaug-2}), 
iii) the bound on the spin-independent direct detection cross-section of 
DM ($\sigma_{\tilde \chi p}^{\rm SI}$) from 
XENON100\cite{xenon100} and LUX\cite{lux}. We also consider the reach 
of XENON1T\cite{xenon1t} and 
iv) the indirect detection constraint from photon signal as given by the FERMI data\cite{fermi-lat-gamma}. 
With a bino-dominated LSP 
the last constraint is hardly of any interest as we will see in Sec.~\ref{Section:DirectAndIndirectDetection}.

In the optimistic scenario of SUSY discovery in the LHC-13 TeV runs, 
it would still be difficult to pinpoint the underlying DM producing 
mechanism  by explicitly reconstructing the sparticle spectrum. This 
is  especially true for the early phase of the experiment. In this work  
 we address the possibility of distinguishing various 
pMSSM scenarios, with characteristic EW sectors constrained by the 
experiments discussed above. This may be  possible 
if at least one of the strongly interacting sparticles is within the reach of the LHC and its decays bear the
imprints of the underlying EW sector as we will show in a later section.

In our analysis we will particularly 
see the effects of variations of 
$\tan\beta$, the ratio of the vacuum expectation values of the two 
neutral Higgs bosons, $\mu$, the higgsino mass 
parameter, the slepton masses etc. This will be explored 
in a generic scenario with 
bino dominated LSP and wino dominated $\chonepm$ along with 
heavy squarks, gluino as well as large 
masses for the charged Higgs $H^{\pm}$, the heavier CP-even neutral Higgs $H$ 
and the pseudoscalar Higgs $A$ (${M_{H^{\pm}}, M_H, M_A}$ respectively). We will also 
consider a large top-trilinear parameter $A_t$ so that 
the lighter Higgs mass $m_h$ agrees with the observed value in the 
least possible mass reach of the super-partners.

The plan of this paper is as follows. In Sec.~\ref{Section:DetailsOfConstraints} 
we will review the effect of Higgs mass data 
as applied to pMSSM and indirect constraints like that from $\gmin2$, 
WMAP/PLANCK data for relic density of DM and the effect of XENON100, LUX 
and the future XENON1T on our analysis. 
In Sec.~\ref{actualanalysis} 
we will explore various electroweak sectors by having the left and right 
slepton masses 
(separately or together) in between the masses of the LSP and the 
lighter chargino. This will be analysed by considering sufficiently large values of 
$\mu$ such that one always obtains 
a bino-dominated LSP and a wino-dominated $\chonepm$. We will find 
the APS from collider bounds and constraints from the relic density as well as 
$\gmin2$. In Sec.~\ref{Section:DirectAndIndirectDetection} 
we will further impose the constraints for 
spin-independent direct detection cross-section limits from LUX and 
$\gamma$-ray constraints for indirect detection of DM from Fermi-LAT. 
In Sec.~\ref{Section:gluinomasslimit} 
we will analyse a few benchmark points chosen from the models of Sec.~\ref{actualanalysis}  
and discuss the prospects of distinguishing various models. 
We will conclude in Sec.~\ref{Section:Conclusion}. 

\section{The Constraints from $\gmin2$, DM Relic Density and Other Experiments}
\label{Section:DetailsOfConstraints}

We work in a  pMSSM framework where parameters are chosen such that the 
strongly interacting sector is beyond the reach of the
 LHC. We set all squark masses at 2 TeV. 
While probing the electroweak sector via the relevant constraints we 
remind ourselves that the mass eigenstates namely the charginos 
($\chonipm $, i = 1,2) and the neutralinos ($ \lspi $, i = 1-4) 
are composed of the  
$SU(2)$ gauginos (the winos), the $U(1)$ gaugino (the bino) and the higgsinos (the superpartners of the 
Higgs bosons) with appropriate charges. The degrees of mixing 
are essentially controlled by 4 free parameters - the 
gaugino mass parameters $M_1 $ and $M_2$, the higgsino mass parameter $\mu$ and tan$\beta$, the ratio of 
the vacuum expectation values of the two Higgs doublets. For  $|\mu|$ $>> |M_2| > |M_1|$, $\lspone$ 
is  bino ($\tilde B$) dominated and the lighter chargino $\chonepm$ (the second lightest neutralino 
$\lsptwo$) is mostly a charged (neutral) wino, but for $ |M_1| > |M_2|$, $\lspone$ ($\lsptwo$) is 
dominantly the neutral wino (bino). On the other hand, if $ |M_1| \simeq |M_2|$ the two lighter neutralinos are
admixtures of the neutral wino and bino. In the limit, $|\mu|$ $<< |M_1|, |M_2|$, $\lspone$ and $\lsptwo$
and the lighter chargino $\chonepm$ are all mostly higgsinos having approximately the mass $|\mu|$. 
A scenario with $|\mu|$ $\simeq |M_1| \simeq |M_2|$ would result into strong mixing for the concerned 
mass eigenstates. In this analysis we consider only bino-dominated LSP ($\lspone$) and 
wino-dominated $\chonepm$. The production cross-section of 
$\chonepm$, $\lsptwo$ would be drastically reduced for a consideration of a higgsino dominated $\chonepm$ 
which would in turn weaken the exclusion limits in 
the $\mlspone-\mchonepm$ plane.

We start our analysis by reviewing a few relevant constraints like the 
measured Higgs boson mass, gyromagnetic ratio of the muon and cold dark matter relic density. 
 
\subsection{Higgs at 125 GeV}
\label{section2.1}
We note that a study within MSSM should most importantly 
accommodate the lighter Higgs boson mass $m_h$ to be at 125 GeV \cite{higgs}. 
This has generally 
pushed up SUSY spectra to high masses in general for models like mSUGRA. However, the required 
large loop corrections to the Higgs boson mass primarily arise 
from loops involving top-squarks and these contributions 
can be controlled via considering large
trilinear coupling parameter $A_t$ ($\sim \rm -2~ {\rm to} -3$~TeV) 
leading to reduction of the average mass scale of the SUSY spectra \cite{djouadi}.     
We require the lighter Higgs scalar mass to be in the interval
$122<m_h<128$~GeV in MSSM. The spread is considered to accommodate 
a theoretical uncertainty of about  3~GeV in computing the Higgs mass. 
This indeed originates from uncertainties in the renormalisation scheme, scale 
dependence, the same in higher order loop corrections 
up to three loops or that due to the top-quark 
mass\cite{higgsuncertainty3GeV}.  The other Higgs bosons are 
assumed to be decoupled. 

Due to precise measurement of $m_h$ at LHC experiments \cite{higgs}, it
is now  possible to explore the specific regions of parameter space where
the LSP pair annihilation occurs via Higgs (h-resonance). 
We recall that this occurs for $\mlspone \approx m_h/2$. 
This enables us in examining critically the viability of this
mechanism in different models, as we
will show in the subsequent sections.

Limits on the masses of the charginos and the neutralinos from trilepton
data crucially depend on the
leptonic BR of these sparticle. When the decay mode $\lsptwo \ra h
\lspone$ is kinematically allowed,
the mass limits become reduced significantly\cite{higgsino}. The
information on the Higgs mass enables one in assessing the impact of this 
`spoiler mode'\footnote{A few recent analyses in this context may be seen in Refs.\cite{spoiler}.}
on the trilepton data in a more precise way. In a subsequent
section we shall take up the issue once more.                                     

\subsection{Anomalous Magnetic Moment of Muon}
\label{section2.2}
The Muon Anomalous Magnetic Moment ($a_\mu=\frac{1}{2}\gmin2$) 
is an important probe for 
the signatures of new physics\cite{muonrev}. 
A generic contribution to $a_\mu$ scales 
like $m_\mu^2/\Lambda^2$ where $\Lambda$ and $m_\mu$ refer to the scale of 
new physics and muon mass respectively. 
The experimental data of $a_\mu$ namely $a_\mu^{\rm exp}$\cite{g-2exp} 
 differs significantly from the Standard Model evaluation $a_\mu^{\rm SM}$\cite{g-2sm1,g-2sm2}.  
Thus $\Delta a_\mu=a_\mu^{\rm exp}-a_\mu^{\rm SM}$ can be an effective probe for a beyond the standard model (BSM) 
physics provided $\Lambda$ is not too large. 
$a_\mu^{\rm SM}$ may be broken into a part coming from pure 
quantum electrodynamics, a part coming 
from hadronic contributions and finally a part from Electroweak 
physics involving vector bosons and Higgs boson\cite{muonrev}. We note that 
the level of 
disagreement of $a_\mu^{\rm exp}$ from the 
SM result is of the same order as the 
contributions from electroweak corrections\cite{g-2sm1,g-2sm2}. 
$a_\mu^{\rm SM}$ itself has a significant amount of error primarily because 
of the uncertainties arising out of the hadronic vacuum polarization and 
the light-by-light scattering contributions\cite{g-2sm1,g-2sm2,muonrev}. 
We note that the hadronic 
vacuum polarization part has two different evaluations based on  
i) $e^+e^-$ and ii) hadronic $\tau$-decay data\cite{muonrev}.   
The difference of the two evaluations which has been diminishing over the years 
still affects 
$\Delta a_\mu$ to an appreciable degree\cite{davierhadtaug-2}.
The resulting discrepancy 
that amounts to more than 3$\sigma$ level of deviation is 
summarized as follows\cite{g-2sm2}. 
\begin{equation}
\Delta a_\mu=a_\mu^{\rm exp}- a_\mu^{\rm SM}=(29.3 \pm 9.0)\times 10^{-10}.
\label{gmin2equation}
\end{equation}
The contributions of different parts of $a_\mu^{\rm SM}$ may be seen in 
Ref.\cite{g-2sm2}\footnote{Considering all the uncertainties of $a_\mu^{\rm SM}$ including 
those arising from light-by-light scattering contributions there are 
analyses which estimate a much larger error going almost up to 5$\sigma$
(see the comments in Ref.\cite{g-2sm2}).}.

The supersymmetric 
contribution to $a_\mu$ namely $\amususy$  may be as large as the electroweak contribution for 
parts of parameter space associated with lighter electroweak sector 
super-partners like charginos, sneutrinos, neutralinos or smuons as well as 
for large $\tan\beta$\cite{oldSusyMuong}. 
It may, therefore, potentially explain
the discrepancy $\Delta a_\mu$ of Eq.\ref{gmin2equation}. 
Alternatively, SUSY parameter space can effectively be constrained with a given set of 
lower and upper bounds of $\Delta a_\mu$. Thus the limits of $\amususy$
at the level of 2$\sigma$ and 3$\sigma$ are as follows. 
\begin{equation}
11.3<\amususy \times 10^{10}<47.3 \quad (2\sigma) \quad {\rm and} \quad  2.3<\amususy \times 10^{10}<56.3 \quad (3\sigma).  
\label{amususylimits}
\end{equation}  
Details of $\amususy$ 
in the MSSM based scenarios 
including mSUGRA and various models with high scale physics input were 
studied several years ago for which a partial list may be seen in 
Refs.\cite{oldSusyMuong,susyg-2A,susyg-2B,endo}.  
At one-loop level, $a_\mu^{\rm SUSY}$ 
arises from 
loops containing chargino and sneutrino (${\widetilde \chi}_i^\pm-{\widetilde \nu}_\mu$)
and the same containing neutralino and smuon (${\widetilde \chi}_i^0-{\widetilde \mu}_j$). $a_\mu^{\rm SUSY}$ increases with  $\frac{1}{\cos\beta} \sim \tan\beta$ and in general 
for models like mSUGRA with universal boundary conditions 
the chargino loop containing the lighter chargino state 
is the most dominating one\cite{susyg-2A}. 
This dominance results into a 
correlation of 
the sign of $\mu M_2$ with that of $a_\mu^{\rm SUSY}$\cite{susyg-2A}, in
models like mSUGRA.
This is however not true 
in the general scenario of MSSM in spite of the fact that the lighter 
chargino loop (${\widetilde \chi}_1^\pm-{\widetilde \nu}_\mu$) still 
dominates over 
the other contributions for a large zone of parameter space\cite{susyg-2B}.   
The neutralino loop contributions can be significantly large 
for smaller smuon masses and for cases with 
large $|\mu M_1|$\cite{endo}. 
For the cases where neutralino loop contribution dominates  
the signs of $\amususy$ and $M_1 \mu$ become the 
same\footnote{We note that $\amususy$ can be large for a large left-right smuon mixing\cite{Endo:2013lva}.}.
In this work, the signs of $M_1$, $M_2$ and that of $\mu$ are considered positive.

In this analysis we will mostly focus on the pMSSM parameter space which is 
consistent with the $\Delta a_{\mu}$ constraint upto the level of 2$\sigma$ following
Eq.\ref{amususylimits}. Of course compared to a 2$\sigma$ level, requiring a consistency at 
the level 3$\sigma$ would be highly conservative but we have occasionally 
taken recourse to it. Henceforth 
we will require the APS to satisfy this level of consistency.

An important point to note is that 
a large range of $\amususy$ may put strong upper bounds 
on the super-partner masses in addition to indicating definite lower bounds 
for the same\cite{g-2sparticlebounds}. 
Particularly with the announcement of Higgs boson 
discovery, and/or with the latest LHC data of squark and gluino masses,   
models having limited number of high scale physics inputs such 
as mSUGRA can hardly accommodate the above constraint\cite{baer-prannath}. 
However, non-universal SUGRA models can still accommodate 
the above non-vanishing $\Delta a_\mu$   
apart from generic MSSM models with a larger set of inputs\cite{Nonunivg-2} .

\subsection{Dark Matter Relic Density and Results from Direct and Indirect Searches }
\label{section2.3}
   We will now come to the discussion of possible mechanisms of satisfying 
the observed relic density from WMAP and PLANCK data in our analysis. 
Similar to the limits used in 
Ref.\cite{Kozaczuk:2013spa} we consider a $2\sigma$ level of WMAP nine year data 
\cite{wmap}\footnote{We consider the eCMB+BAO+$H_O$ 
value of Table~4 of Ref.\cite{wmap}.} bound 
with a 10\% error in theoretical estimation as follows. This range also 
embraces the 3$\sigma$ limits from PLANCK\cite{planck}.
\begin{equation}
0.092 < \Omega_{\tilde \chi} h^2 <0.138.
\label{planckdata}
\end{equation}
Here, we will select only the lightest neutralino as the cold dark matter 
candidate. 
The LSP is sufficiently bino-dominated. Hence in general the 
possible annihilation mechanisms would be exchange of sleptons in the 
$t$-channel (bulk annihilation), LSP-annihilation via $s$-channel Higgs pole or even via Z-pole. 
The LSP 
can undergo coannihilation with a scalar particle like the stau or the
sneutrino, since top-squarks are assumed to be very heavy.
However, considering the present bounds of sparticle masses mSUGRA is not 
able to accommodate many of the above annihilation/coannihilation scenarios because of its 
associated correlations among sparticle masses as well as due to constraints like Higgs mass. 
For example, a neutralino with mass $ = M_Z/2$ is ruled 
out by LEP bound on chargino mass when the gaugino mass unification condition is applied.
We will identify the actual mechanisms in the parameter space of each model 
that would survive the combined analysis of LHC, CDM and precision 
data like $\gmin2$.

In addition to the constraint from dark matter relic density, we will also 
investigate the possibility of direct detection of dark matter via computing 
spin-independent LSP-proton scattering cross-section $\sigma_{\tilde \chi p}^{\rm SI}$ in relation to 
the XENON100 \cite{xenon100} and LUX\cite{lux} data. $\sigma_{\tilde \chi p}^{\rm SI}$ results from diagrams involving $t$-channel Higgs and $s$-channel 
squark exchanges. Unless the squark masses are close to the mass of the LSP 
which is certainly not our case after the LHC data, 
the Higgs exchange diagrams 
contribute dominantly to the above cross-section\cite{drees}. 
The effective couplings are dependent on 
the nature of composition of the LSP. Since the $h(H)-\lspone-\lspone$ 
couplings involves product of gaugino and higgsino components of the 
neutralino diagonalising matrix, only for the presence of a sufficient 
higgsino within $\lspone$ the direct detection cross-section 
$\sigma_{\tilde \chi p}^{\rm SI}$ may become appreciable\cite{Hisano:2009xv}.

We should however keep 
in mind various uncertainties in computing the cross-section $\sigma_{\tilde \chi p}^{\rm SI}$
arising from particle physics or astrophysics related issues\footnote{Apart from particle physics 
and astrophysics related uncertainties, see also Ref.\cite{gondolo} for the uncertainty arising 
out of poor knowledge of cosmic ray activation in detector materials in regard to direct detection backgrounds.}. 
There is a significant amount hadronic uncertainty in evaluating $\sigma_{\tilde \chi p}^{\rm SI}$. 
The strangeness content of nucleon is quite important 
for evaluating the cross-section. This is because, for WIMP-nucleon scattering the 
WIMP couplings with valence quarks like $u$ and $d$-quarks are small due to small Yukawa 
couplings. Thus the contributions to scattering amplitude due to 
heavy sea quarks become important (light quarks as sea quarks again have 
small contribution to the amplitude).  
Over the last few years the strangeness 
contribution to proton mass is effectively reduced via lattice 
computations\cite{LatticeStrangenessAndDM}.   
This in turn may potentially reduce the uncertainties in the 
evaluation of effective 
couplings of LSP-nucleon interactions leading to more precise results. 
We compute all the dark matter related 
quantities using micrOMEGAs (version-3.2)\cite{micromega3}. 
Unlike the previous versions, micrOMEGAs (version-3.2) treated  
the above error by 
using a different prescription for evaluating  
the strange quark content of a nucleon. An weighted 
average of $\sigma_s=m_s<p|{\bar s} s |p>$, a measure of strangeness content 
was obtained out of various 
lattice quantum chromodynamics (QCD) results. We must note that although we have used 
the default values of $\sigma_s$ as obtained by the weighted 
average as mentioned above, 
the individual lattice results used in this averaging 
vary widely from each other\footnote{See Table~1 of 
Ref.\cite{micromega3}} leading to enough uncertainty in the direct 
detection cross-section.    
Additionally, we should also 
keep in mind the uncertainties of astrophysical origin 
in finding the rate of dark matter events in a given detector.
Among the above, uncertainties may arise from determination of the local dark matter 
density\cite{beskidt,Bovy:2012tw}.  
Consideration of the existence of 
non-Maxwellian velocity distributions for WIMP also shows an adequate amount of variation 
in the direct detection 
rates\cite{Fairbairn:2012zs,Bhattacharjee:2012xm} specially for 
low mass DM.
Apart from the current data we will also relate our 
result with the reach of the future experiment  
XENON1T\cite{xenon1t} that would be about two orders of magnitude below 
the current LUX\cite{lux} or XENON100\cite{xenon100}  limit
for the scalar cross-section and can probe various SUSY models
even if the above uncertainties continues to persist.

Besides the
direct detection limits we would also explore the reach of indirect detection 
data from Fermi-LAT\cite{fermi-lat-gamma} for continuous $\gamma$-ray signal from 
dense astrophysical regions such as galactic center, 
dwarf galaxies etc. With a highly bino-dominated LSP, expectedly, our 
scenarios produce too little cross-section ($<\sigma v>$).

In the next 
section we intend to describe various models that are based on different relative 
masses of the EW sparticles. We will analyse these models particularly 
for interesting collider signatures while also imposing the 
necessity to satisfy the Higgs mass, the $\gmin2$ and the cold dark matter 
constraints and of course the LEP limits on chargino and slepton 
masses \cite{lepsusy}.  
Only after filtering out the APS we will explore the degree of constraints from 
the XENON100 and the LUX data keeping in mind the 
extent of theoretical and astrophysical uncertainties in the 
direct detection of dark matter which could at least be an order of magnitude or even more.

\section{Electroweak Sector of pMSSM Models in the Light of LHC and Other Constraints}
\label{actualanalysis}
The non-observation of the charginos, neutralinos 
as well as the sleptons at the LHC severely constrains several pMSSM 
models sensitive to the LHC searches. They are particularly important in the era of a known mass of the Higgs 
boson. We will focus on bino-dominated $\lspone$ and wino-dominated 
$\chonepm$/$\lsptwo$ which are very sensitive to the LHC searches.  This scenario can be easily 
realized  by considering a large $\mu$ and adjusting the 
gaugino mass parameters of the electroweak sector. We will analyse various scenarios 
of left and right slepton mass parameters ($M_{\tilde l_L}$, $M_{\tilde l_R}$) 
placed differently with respect to the electroweak 
gaugino mass parameters $M_1$ and $M_2$. 
The specific choices are motivated  by 
the direct production limits on  electroweak sparticle masses by ATLAS and 
CMS\cite{atlas3lew, atlas2lew, cmsew} and the other observables under consideration. Each scenario may have important 
signatures in regard to collider physics, dark matter relic density and precision observables like 
$\gmin2$. Our task is to find the APS after imposing the combined 
constraints and assess the possibility of observing 
EW SUSY particles in future LHC experiment.

For  the detailed study 
we choose the following pMSSM parameters.  All squark mass parameters as well as 
$M_3$ and $M_A$, which hardly affect the observables under consideration, are set to a large value of 2 TeV. A choice of the trilinear coupling  
$-3~{\rm TeV}<A_t<-2$~TeV is made for consistency with the measured  mass of the lighter Higgs boson without 
the need of a very large sparticle mass scale. All other trilinear couplings are 
vanishing namely $A_b=A_\tau=A_u=A_d=A_e=0$.   
$M_1$, $M_2$, $\mu$, $M_{\tilde l_L}$ and $M_{\tilde l_R}$ are varied in this study where 
the relevant SM parameters considered are $m_t^{pole}=173.2$~GeV, 
$m_b^{\overline {MS}}=4.19$~GeV and 
$m_\tau=1.77$~GeV.

\subsection{Light Gaugino and Left Slepton (LGLS) Scenario}
\label{section3.1}
In this model it is assumed that only left sleptons are lighter than $\chonepm$ and $\lsptwo$ 
while right sleptons are heavy. 
The ATLAS collaboration have searched for chargino-neutralino ($\chonepm - \lsptwo$) pair
production leading to the  trilepton signal for 20 $\ifb$\cite{atlas3lew} of data.
The results were interpreted in this simplified model. Here the L-sleptons (${\tilde l}_L$)
of all the generations have masses 
midway between the masses of $\chonepm$ and $\lspone$ whereas 
the R-sleptons (${\tilde l}_R$) are chosen to be very heavy 
leading to very small mixing 
effects in the slepton mass matrices. The sneutrinos are assumed to be degenerate with ${\tilde l}_L$, i.e.,
$ M_{\tilde l_L} = M_{\tilde \nu} = (\mlspone + \mchonepm)/2 $.  
It was further assumed
that the lightest neutralino is  
highly bino dominated and $\chonepm$ or $\lsptwo$  are  
wino dominated. As a result the branching ratio (BR) of chargino decay into  
slepton-neutrino and sneutrino-lepton modes of each flavour 
is the same. 
Similarly $\lsptwo$ would decay into neutrino-sneutrino and 
lepton-slepton pairs of each flavour with equal probability.  
The non-observation of signal yielded the exclusion contour 
in Fig.8a of Ref.\cite{atlas3lew} which is reproduced in Fig.\ref{LL_0.5_0.5_A} 
(see the black contour) for ready reference.

\begin{figure}[!htb]
\vspace*{-0.05in}
\mygraph{LL_0.5_0.5_A}{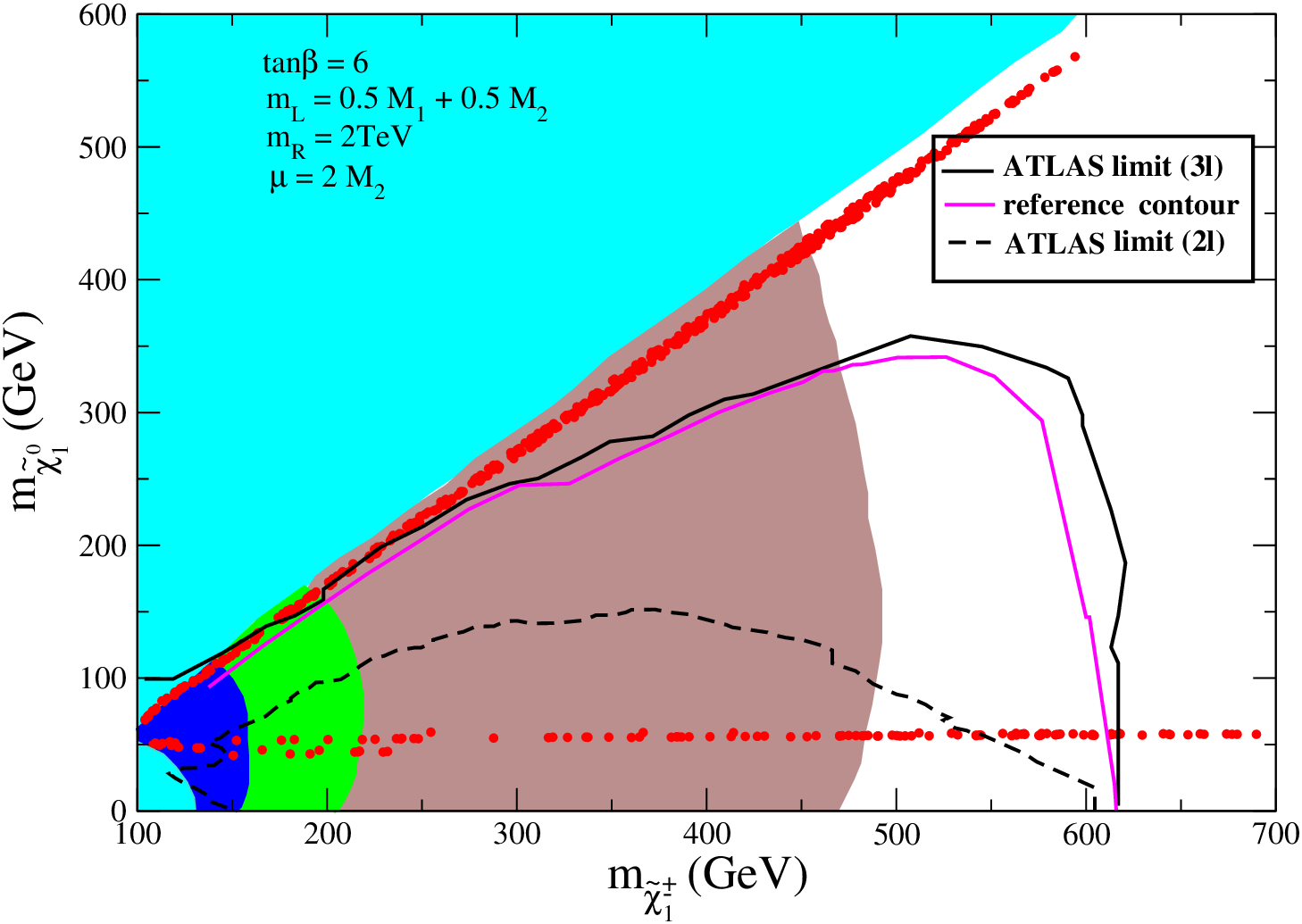}
\hspace*{0.5in}
\mygraph{LL_0.5_0.5_B}{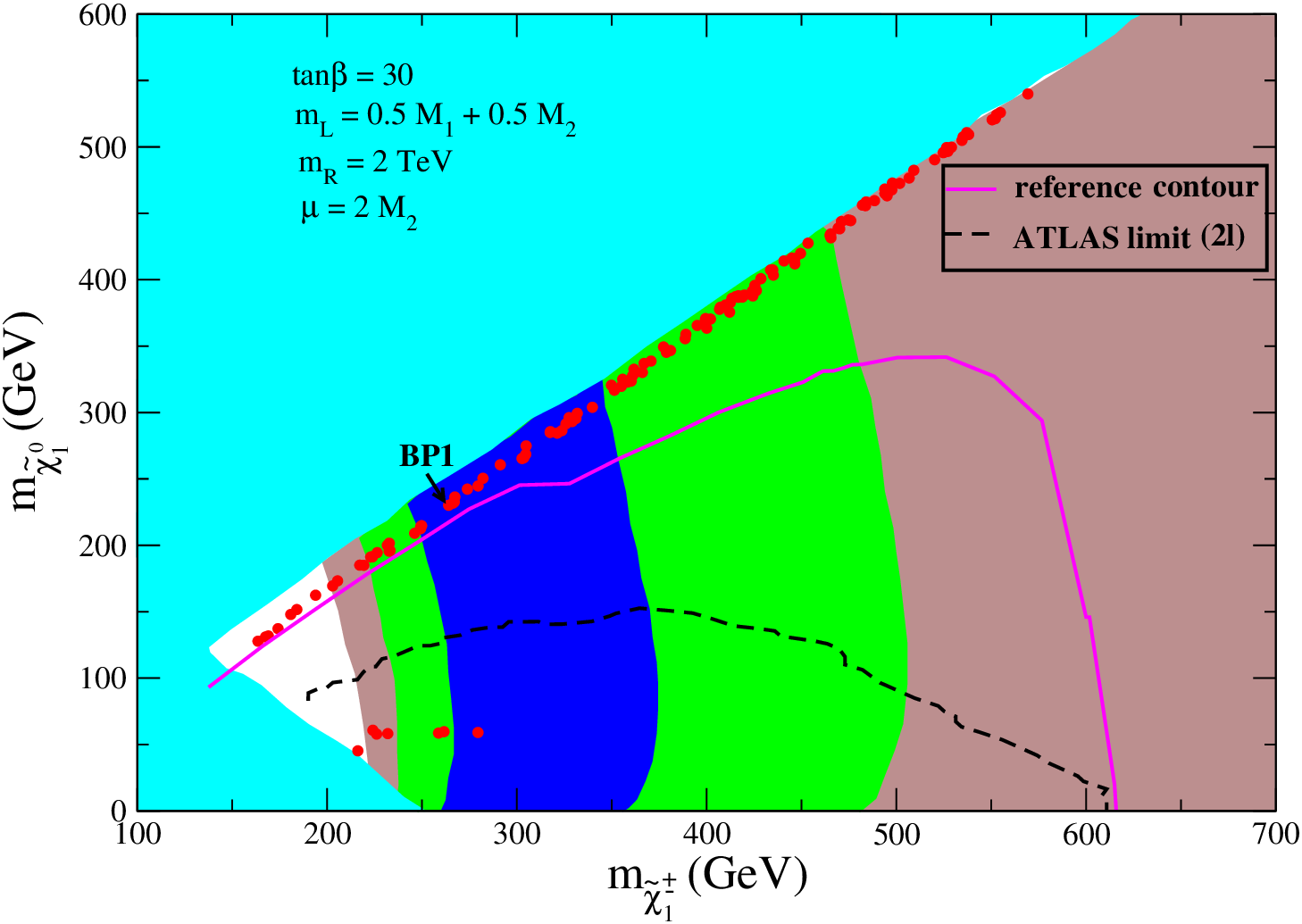}
\caption{{\it Plot in the $\mchonepm-\mlspone$ plane for the 
LGLS scenario with the slepton mass parameter satisfying 
$M_{{\tilde l}_L} = 0.5 M_1 +0.5 M_2$ for $\tan\beta=$~6 (a)~and~30 (b). 
 $M_{{\tilde l}_R}$ is chosen to be at 2 TeV.
Here, $m_{L/R} \equiv M_{\tilde l_{L/R}}$.
The blue, green and brown 
regions represent 
the parameter space where $\amususy$ is consistent with $\Delta a_\mu$ upto 
the level of $1\sigma$, $2\sigma$ and $3\sigma$ respectively.
The red points in the plot satisfy 
the relic density constraint from WMAP/PLANCK data.  The parameters used
for computing these and other observables are shown on the upper left corner
of each figure. The cyan region  
corresponds to the parameter space which is discarded by theoretical constraints and the  
LEP limits on the slepton mass\cite{lepsusy}.
The black line in the left plot (a) represents the exclusion contour at 95$\%$ CL 
obtained by the ATLAS collaboration at 8 TeV LHC
from trilepton searches\cite{atlas3lew}.
The magenta line (the reference contour) shows the exclusion limit obtained by our simulation. 
The dashed line refers to the boundary of the 
disallowed region corresponding to the 
slepton search limits from  8 TeV ATLAS data\cite{atlas2lew} (see Fig.\ref{30_LL_lsp_slep}).
For the case of tan$\beta$ = 30 (b), only the reference contour resulting from 
our simulation is shown. Throughout this paper we shall follow the same
colour coding and conventions as used in this figure.
}}
\label{LL_0.5_0.5}
\end{figure}

In order to validate our simulation we compute the above exclusion
contour using PYTHIA (v6.428)\cite{pythia}. 
The next to leading order (NLO) cross-section for the $\chonepm \lsptwo$ pair 
production have been computed by PROSPINO 2.1 \cite {prospino} with CTEQ6.6M PDF \cite {cteq6.6}.
Our simulation is based on those selection criteria of the ATLAS collaboration 
which may be implemented at the generator level. These are divided into several signal 
regions (SRs) : SRnoZa, SRnoZb and SRnoZc (see  Table~1 of Ref.\cite{atlas3lew} ). 
Each SR is characterized by a set of kinematical cuts and an 
upper bound on the effective cross-section
($\sigma_e$) $\equiv$ production cross-section $\times$ efficiency $\times$ 
acceptance or equivalently on $N_{BSM}$ (number of events from 
BSM physics) obtained from the observed number of events and the SM background.
  These constraints are also expressed in terms of $N_{BSM}$, 
the maximum allowed number of beyond standard model events.  
Any model point is excluded if its associated $\sigma_e$ 
exceeds the above upper bound for at least one of the above SRs.
Although we have not included the detector effects directly, we have 
introduced an approximate prescription for the combined trigger and electron (muon) 
identification efficiencies for different values of the transverse momentum (${\rm P_T}$)
following an analysis of ATLAS collaboration\cite{atlastrig}. 
We confirm that the above prescription reproduces the efficiencies 
mentioned in the Table~5 of Ref.\cite{atlas3lew}. 
The above efficiency is chosen to be 75\%  (90\%) for
electrons with $10 < {\rm P_T}< 30 ~({\rm P_T}> 30)$. 
The same is chosen to be 85\% for muons with ${\rm P_T}> 10$. 
We have implemented electron/muon - jet isolation according
to the ATLAS prescription\cite{atlas3lew}. 

Our exclusion contour, namely 
the magenta curve in Fig.\ref{LL_0.5_0.5_A}, for $\tan\beta=6$ 
validates the simulation. Henceforth this will be called the {\it reference contour}.
Our representative choice of a few other SUSY parameters 
essential for computing the observables discussed in Sec.~\ref{Section:DetailsOfConstraints} are 
given in the upper left corner. 
We emphasize that the LHC exclusion contours are 
in general fairly insensitive to such choices.
Additionally, we note that there is a less than 
10 percent disagreement 
between the two results for $\mchonepm >$ 500~GeV. 
We will come back to this issue soon. We note that $\mchonepm >$ 500~GeV
is disfavoured, in any case, either by the $\gmin2$ or LHC data or by both.
Henceforth, we will paste this reference contour in all 
the figures up to Fig.\ref{LLR_0.25_0.75} for comparison with  other models.

The following minor differences with the ATLAS paper may be noted.
For simplicity of computation we have scanned $M_1$ and $M_2$ while keeping 
L-slepton mass parameter midway, i.e., $M_{{\tilde l}_L}=\frac{1}{2}{(M_1+M_2)}$, 
instead of equating the physical slepton mass with $\frac{1}{2}(\mlspone+\mchonepm)$.
With a highly 
bino-dominated $\lspone$ and wino-dominated $\chonepm$, 
the above approximation would 
be good upto a few percent level.
Additionally, unlike what was used by ATLAS we 
do not assume any sneutrino-slepton mass degeneracy and 
entirely rely on the MSSM specified mass relations involving the D-term throughout our analysis. 
This increases the branching ratio of the decay 
$\lsptwo \ra \tilde {\nu} \bar{\nu}$ by a small but
non-negligible amount and reduces the trilepton signal 
resulting in a weaker
limit. Had we carried out our simulation following exactly 
the same assumptions as ATLAS
our limits on $\mlspone$ for $\mchonepm >$ 500 GeV 
would have been even closer to that obtained by ATLAS. 
Furthermore, we have shown the effect of the 
direct slepton search 
limit from 
the 8 TeV ATLAS data\cite{atlas2lew}\footnote{see Fig.\ref{30_LL_lsp_slep}} by the  black 
dashed line. The region within this contour is disfavoured. 
We denote the
physical masses of left and right sleptons of first two generations by 
$M^D_{\tilde l_{L/R}}$ taking into account the D-term contributions.
Similarly, for the sneutrinos we use the notation, $M^D_{\tilde \nu}$.
We clearly see that no additional parameter 
space is discarded by the slepton search 
limit in the LGLS scenario other than what is already excluded by the 
trilepton data.

We now incorporate the theoretical and indirect constraints like $\gmin2$ 
and the WMAP /PLANCK limits on  dark matter relic density. 
In Fig.\ref{LL_0.5_0.5_A} 
the upper cyan region corresponds to the parameter space 
which is discarded by the requirement of the 
LSP to be the lightest neutralino. The similarly coloured 
lower region is excluded via LEP limits on the slepton masses\cite{lepsusy}.
In the  dark blue, green and light brown 
regions $\amususy$  can explain the $\Delta a_\mu$ anomaly (Eq.\ref{gmin2equation})   upto the level of 
$1\sigma$, $2\sigma$ and $3\sigma$ respectively.  
Both lower and upper limits on $\amususy$ have been considered only for parameter 
regions satisfying theoretical/LEP constraints.
With almost a proportional dependence of 
$a_\mu^{\rm SUSY}$ on $\tan\beta$ 
the contribution of $a_\mu^{\rm SUSY}$ in 
Fig.\ref{LL_0.5_0.5_A} is small because of small value of tan$\beta$. We note that 
the right handed sleptons being heavy in all the LGLS scenarios, 
$a_\mu^{\rm SUSY}$ is dominantly contributed by the lighter 
chargino-sneutrino loop diagrams.

The WMAP/PLANCK 
allowed regions satisfying Eq.\ref{planckdata} for the dark matter 
relic density are shown as red circles\footnote{In all the figures in this paper
we shall follow the same colour convention.}. 
We note that the regions
satisfying the dark matter relic density limits are separated into top 
and bottom limbs. The parameter points denoted by
red circles in the lower limb 
satisfy the relic density limits 
by LSP annihilations via a $s$-channel light Higgs boson 
resonance of mass $\approx 125$ GeV. Additionally, 
there are some points that are associated with 
LSP pair annihilating via a $s$-channel Z resonance.
The upper red points satisfy the dark matter limits 
via coannihilation of LSP with a sneutrino or a slepton almost equally.  
Besides the above there can be coannihilations between 
sleptons and sneutrinos or even a lighter 
chargino and a sneutrino in this region. Furthermore, for low mass zones of 
the figure one finds some degree of bulk annihilations both for the 
upper and the lower limbs.

From the LHC data  at 8 TeV all parameter space which agrees with 
$\Delta a_\mu$ up to the $2\sigma$ level is almost
excluded leaving a tiny  region
consistent with the combined constraint.
 Moreover, LHC data exclude the Higgs resonance region for $\mchonepm < 620$ GeV. The part of
the parameter space with larger $\mchonepm$, however, is consistent with 
the $\Delta a_\mu$ constraint only at the level of $3 \sigma$.

Fig.\ref{LL_0.5_0.5_B} shows the analysis for a larger value of  
$\tan\beta$ $(=30)$ while keeping the same combination of 
other mass parameters. The colour codings are 
the same as in Fig.\ref{LL_0.5_0.5_A}.
The cyan shaded 
lower region is excluded via LEP limits on the slepton masses or 
sneutrinos becoming tachyonic due to its negatively contributing 
D-term part, where the latter increases with $\tan\beta$ in magnitude.
In the white region $\amususy$ differs from $\Delta a_\mu$ by 
more than $3\sigma$ because in this region of smaller  
$m_{{\tilde \chi}_1^\pm}$, $\amususy$ attains a very large value.

The prospect of finding a larger APS improves since $\amususy$ increases  
for large $\tan\beta$. On the other hand, 
an increased $\tan\beta$ hardly has any effect on the LHC constraints.
This is expected since the mixing effects in the stau mass matrix is not 
significant even for larger $\tan\beta$, a result of considering very heavy 
R-sleptons (2~TeV).  Thus with
lighter stau having similar mass with that of 
selectron the BRs of $\chonepm$ and 
$\lsptwo$ for leptonic decays remain unaltered while going from 
Fig.\ref{LL_0.5_0.5_A} to Fig.\ref{LL_0.5_0.5_B}. The same can be said about the upper limb of the WMAP/PLANCK allowed region.

Focusing on Fig.\ref{LL_0.5_0.5_B} we find that 
for relatively small $M_2$ or $\mchonepm$ 
LSP-pair annihilation via light Higgs boson resonance is possible 
for producing the right relic abundance but the 
parameter space is forbidden by the LHC data.
On the other hand, for larger $\mchonepm$, the above resonance annihilation 
is not sufficient to give rise to an acceptable relic abundance 
in Fig.\ref{LL_0.5_0.5_B}. Indeed, it disappears completely
outside the LHC forbidden region. There are two reasons that 
are important to note in this context. First, $h - \lspone -\lspone$ 
coupling decreases with increasing $\tan\beta$. Second, 
our choice of $\mu = 2 M_2$ that ensures 
$\chonepm$ to be wino-dominated, causes reduction  of the higgsino content 
of the LSP with increase of $M_2$, which in turn results into reduced 
LSP pair annihilation via $h$-resonance leading to over-abundance of dark 
matter.
For the rest of the analysis we will see that for a wino
dominated $\chonepm$ and bino dominated LSP, LSP-pair annihilation 
via the $h$-resonance  is disfavoured in 
general for large values of $\tan\beta$ for the above reasons.

\noindent
\subsubsection{Tilted LGLS Scenario}
\label{section3.1.1}
We now explore the situation where the L-slepton mass is shifted from the 
mean of the lighter chargino and the lightest neutralino masses.
We conveniently introduce the shift as follows\footnote{
The physical slepton mass is obtained by adding the D-term to
the RHS of Eq.\ref{slmass}.} 
\begin{equation}
M_{{\wt l}_L} = x M_1 + (1-x) M_2.
\label{slmass}
\end{equation}
where 
the tilting parameter $x$ (with $0 < x < 1$) determines the 
degree of closeness of $M_{{\wt l}_L}$ and 
$\mlspone$. The LGLS scenario analysed by ATLAS corresponds to $x=\frac{1}{2}$.    

We will consider two cases i) LGLS-$\lspone$: here $x=0.75$, indicating
L-slepton masses to be  closer to the mass of the LSP than that of $\chonepm$ and 
ii) LGLS-$\chonepm$: here $x=0.25$, making L-slepton mass parameters to be closer to the mass 
of $\chonepm$.

We will see soon that 
such variants of LGLS scenarios would hardly affect 
$a_\mu^{\rm SUSY}$, mildly change
the relic density satisfying properties for dark matter, 
but significantly 
change the size of the trilepton signal. The latter 
leads to changed exclusion contours compared to the 
LGLS scenario considered by ATLAS. This in turn may change  
the APS consistent with all the constraints.

\noindent
{\bf i) LGLS-$\lspone$:} \\
In the analysis leading to Fig.\ref{LL_0.75_0.25_A}, we consider
$x=0.75$, while  
all other relevant parameters are kept same as in 
Fig.\ref{LL_0.5_0.5_A}. 
The  
lower cyan region is excluded due to tachyonic sneutrinos, sneutrino becoming the
LSP and the LEP limits on $\chonepm$ masses. 
In regard to $\gmin2$ the dominant SUSY diagrams contributing to 
$a_\mu^{\rm SUSY}$ are not different
from those of Fig.\ref{LL_0.5_0.5_A}.  
As a result the $\gmin2$ constraint is almost insensitive to the 
modest variation of $M_{{\tilde l}_L}$. Hence, the $1\sigma$, $2\sigma$,
and $3\sigma$ allowed regions do not change appreciably with respect to 
Fig.\ref{LL_0.5_0.5_A}. \\

\begin{figure}[!htb]
\vspace*{-0.05in}
\mygraph{LL_0.75_0.25_A}{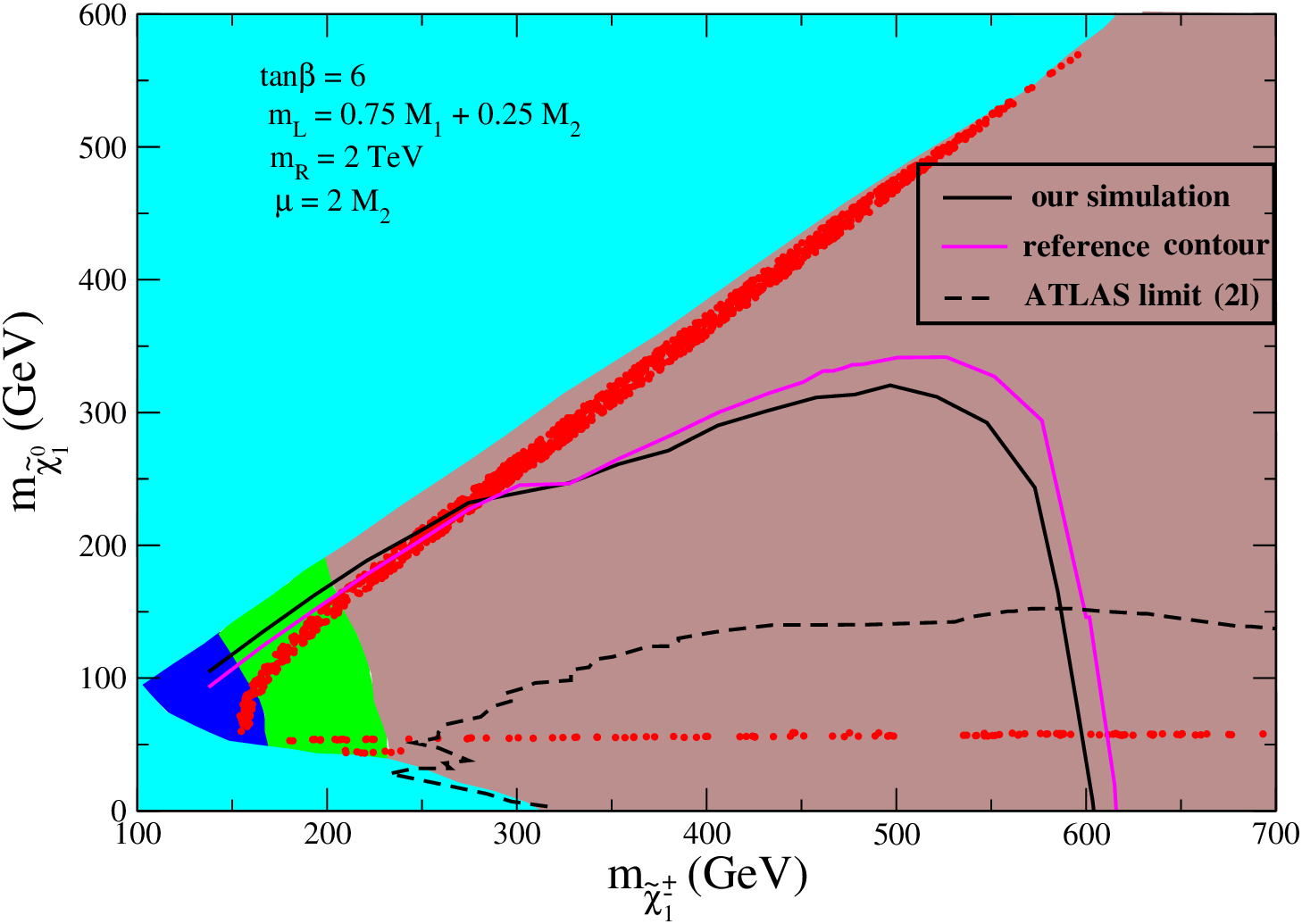}
\hspace*{0.5in}
\mygraph{LL_0.75_0.25_B}{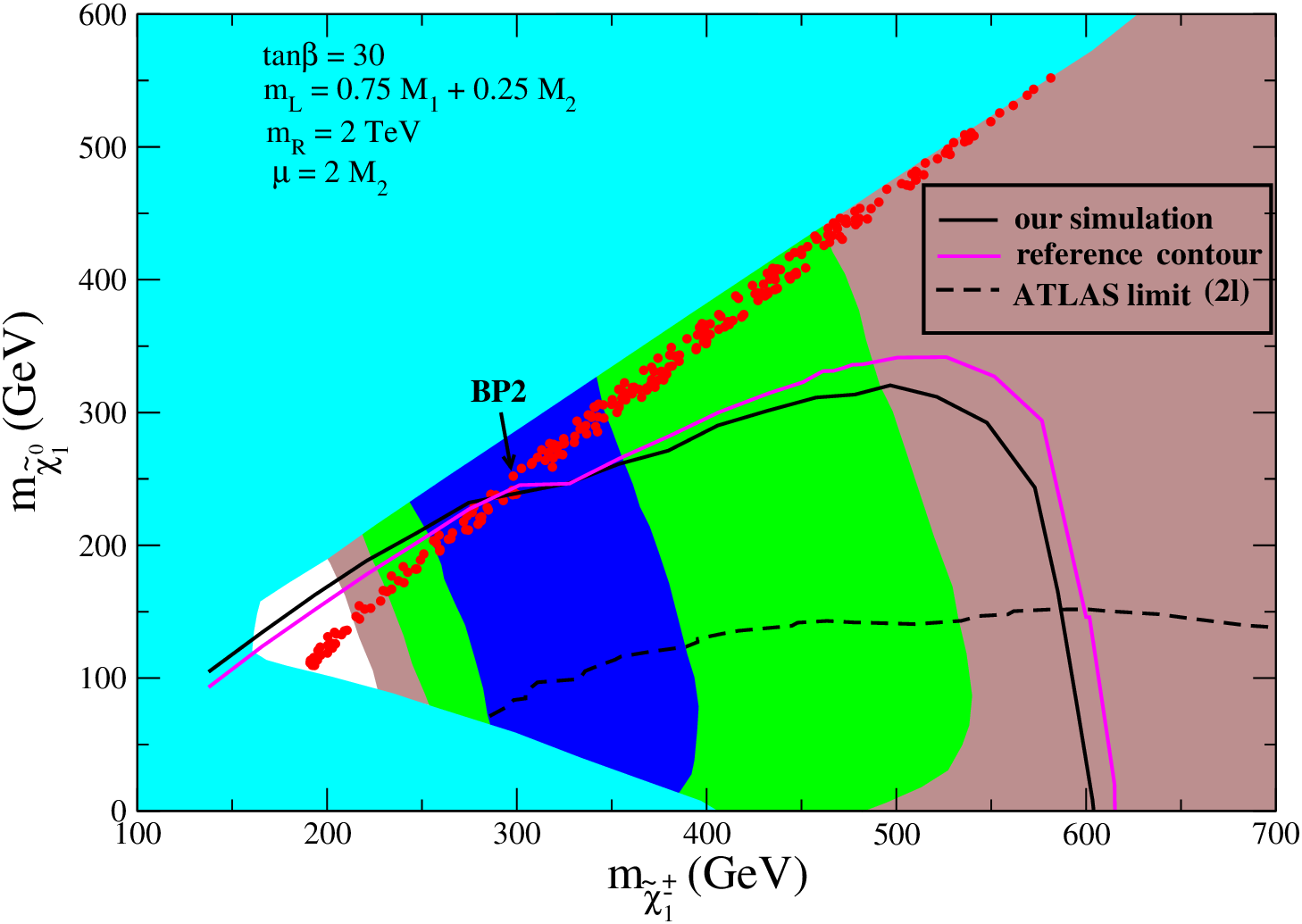}
\caption{{\it Plot in the $\mchonepm-\mlspone$ with the slepton mass parameter satisfying 
$M_{{\tilde l}_L} = 0.75 M_1 +0.25 M_2$ for $\tan\beta=$~6 (a)~and~30 (b).  
Colours and conventions are the same as in Fig.\ref{LL_0.5_0.5}. 
The exclusion contour for this scenario obtained by us is
represented by the black line.  The lower discarded 
region depending on the parameter point may be sensitive to the choice of 
the scale in the REWSB conditions.
}}
\label{LL_0.75_0.25}
\end{figure}

Since the sleptons are closer in mass to that of $\lspone$,  
the leptons arising from decays  $\wt l^{\pm} \ra l^{\pm} \lspone$  
would be softer. This in turn would reduce the trilepton 
detection efficiency. Consequently, 
the limit on $\mlspone$ for a fixed 
$m_{\charginopm_1}$ may decrease by 10-25~GeV compared to  
Fig.\ref{LL_0.5_0.5_A}.
In regard to the cold dark matter results in Fig.\ref{LL_0.75_0.25_A}, 
the annihilation/coannihilation properties of LSP are 
almost unchanged from the LGLS scenario. 
However, this scenario is in tension with the $\Delta a _\mu$ 
constraint at 2$\sigma$ level.

The direct 
slepton search limits also disallow a large part of the parameter space which is
allowed by the trilepton searches. In fact the 
bottom limb of the relic density satisfied region  corresponding to LSP pair annihilation into the h-resonance is disfavoured 
even if the $\gmin2$ constraint is relaxed to 3$\sigma$.

In Fig.\ref{LL_0.75_0.25_B}, we consider tan$\beta = 30$.
The results in regard to DM production via LSP - sneutrino coannihilation 
and $\gmin2$ studies are similar to what has been described for 
Fig.\ref{LL_0.5_0.5_B} for the reasons 
discussed above.  
On the other hand, with large $\tan\beta$ and for small values of 
$M_2$  the parameter region in the ($\mchonepm-\mlspone$) 
plane where DM pair-annihilation into the $h$-resonance could possibly 
occur as in Fig.\ref{LL_0.5_0.5_B} is 
already excluded here because sneutrinos turn out to be the LSP or even 
tachyonic.  In regard to muon anomaly, 
Fig.\ref{LL_0.75_0.25_B} shows an agreement even up to 
$1\sigma$ level.
The nature of the two discarded cyan
regions is similar to those of Fig.\ref{LL_0.75_0.25_A}, but the shape of the lower discarded 
region depends on the choice of 
the scale in the radiative electroweak symmetry breaking (REWSB) conditions\cite{SUSYbooks}. 
We have employed the canonical choice of the scale as the geometric mean of the two top-squark 
scalar mass parameters.

\noindent
{\bf ii) LGLS-$\chonepm$:}\\
In the analysis leading to
Fig.\ref{LL_0.25_0.75_A} we use $x=0.25$.  Thus, here L-sleptons 
are closer in mass with that of $\charginopm_1 / \lsptwo$. 
As a result the leptons
arising from decays via $\charginopm_1 \ra l^{\pm} \wt \nu$ or
$\lsptwo \ra \wt l^{\pm} l^{\mp} $ would be softer. This would reduce the 
trilepton efficiency and relax the LHC constraints. Compared to Fig.\ref{LL_0.5_0.5_A} we find that 
the limit on $\mlspone$ relaxes by 20-40~GeV which allows the 
parameter space to become available at $1\sigma$ limit of the $(g-2)_{\mu}$ 
constraint. Consequently, parameter points corresponding to low mass 
sparticles with masses as low as 
$m_{{\tilde \chi}_1^\pm} \simeq 135$~GeV and $m_{{\tilde \chi}_1^0} \simeq 100$~GeV 
in Fig.\ref{LL_0.25_0.75_A} become allowed.  DM relic density production 
is driven by sneutrino-LSP coannihilation in the parameter space 
consistent with LHC and $\gmin2$ constraints.
Fig.\ref{LL_0.25_0.75_B} shows the result for $\tan\beta=30$. 
Here satisfying DM constraint by the Higgs resonance is 
disfavoured for reasons similar to 
what was described for Fig.\ref{LL_0.5_0.5_B}.

We also note that depending on  $M_1$ and 
$M_2$, situations may arise when the masses of the sleptons with positive D-term 
contributions may become larger than $\mchonepm$ or $\mlsptwo$, but 
the sneutrinos which have negative D-term contributions for their masses, may become lighter than the above gauginos. 
Then, $\lsptwo$ decays into 
neutrino-sneutrino pairs with large BRs (100\%).  The latter in turn would undergo invisible decay into 
neutrino and the LSP. In each LGLS-$\chonepm$ scenario there is a value of $x$ 
which will deplete the trilepton signal due to such blind spots. 
Because of the above there are several blind spots in Figs.\ref{LL_0.25_0.75_A} and \ref{LL_0.25_0.75_B}.  
This scenario with three invisible sparticles (the LSP, $\lsptwo$ and the sneutrino) 
have interesting collider phenomenology\cite{ADBMMG, ADDPADMD}. 
In particular at a high energy $e^+$$e^-$ collider \cite{ilc} it would lead to a 
significantly enhanced signal in the single photon + missing energy channel \cite{ADSRAD} 
compared to a pMSSM scenario with LSP as the lone carrier of missing energy \cite{pandita}. \\

\begin{figure}[!htb]
\mygraph{LL_0.25_0.75_A}{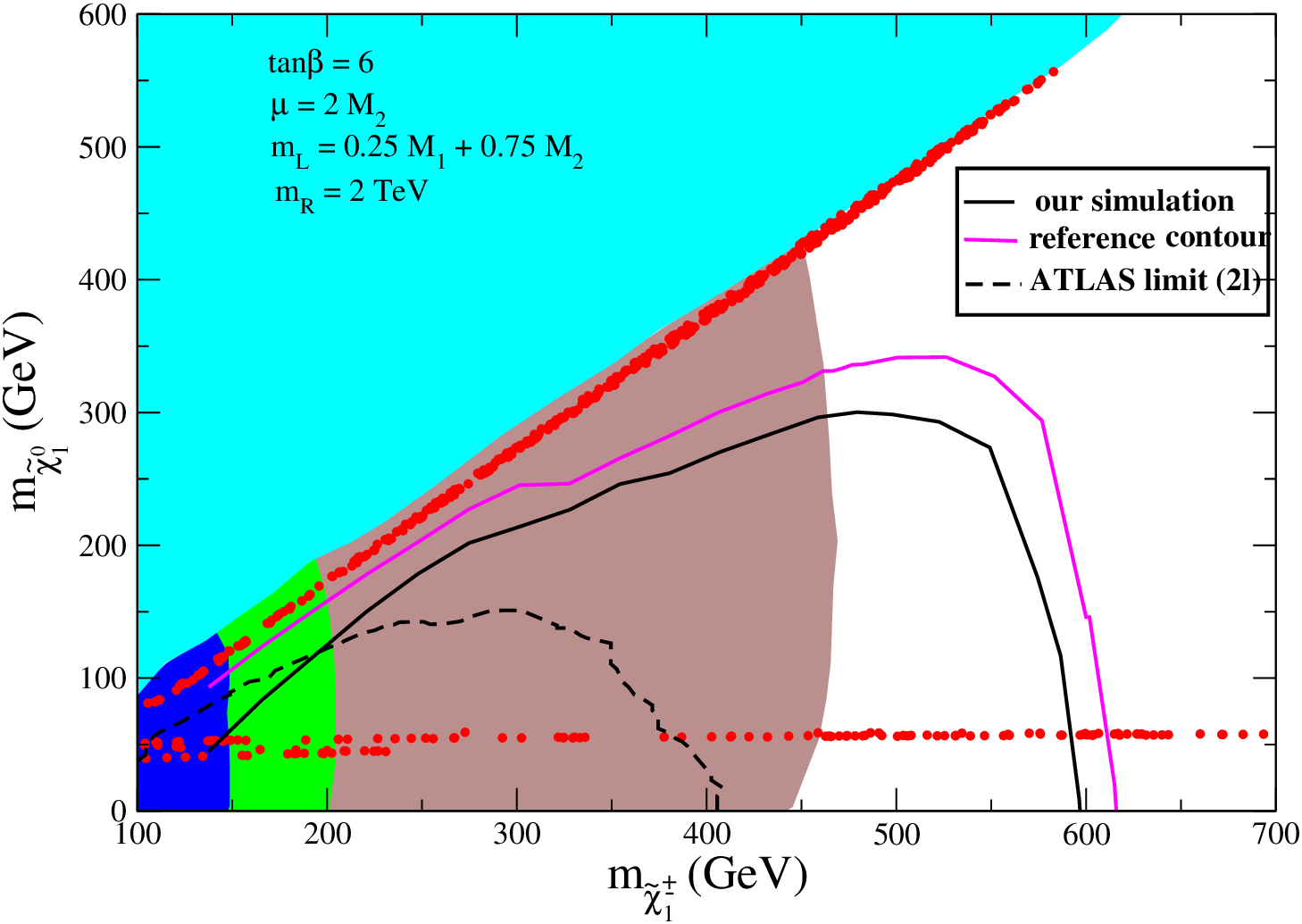}
\hspace*{0.5in}
\mygraph{LL_0.25_0.75_B}{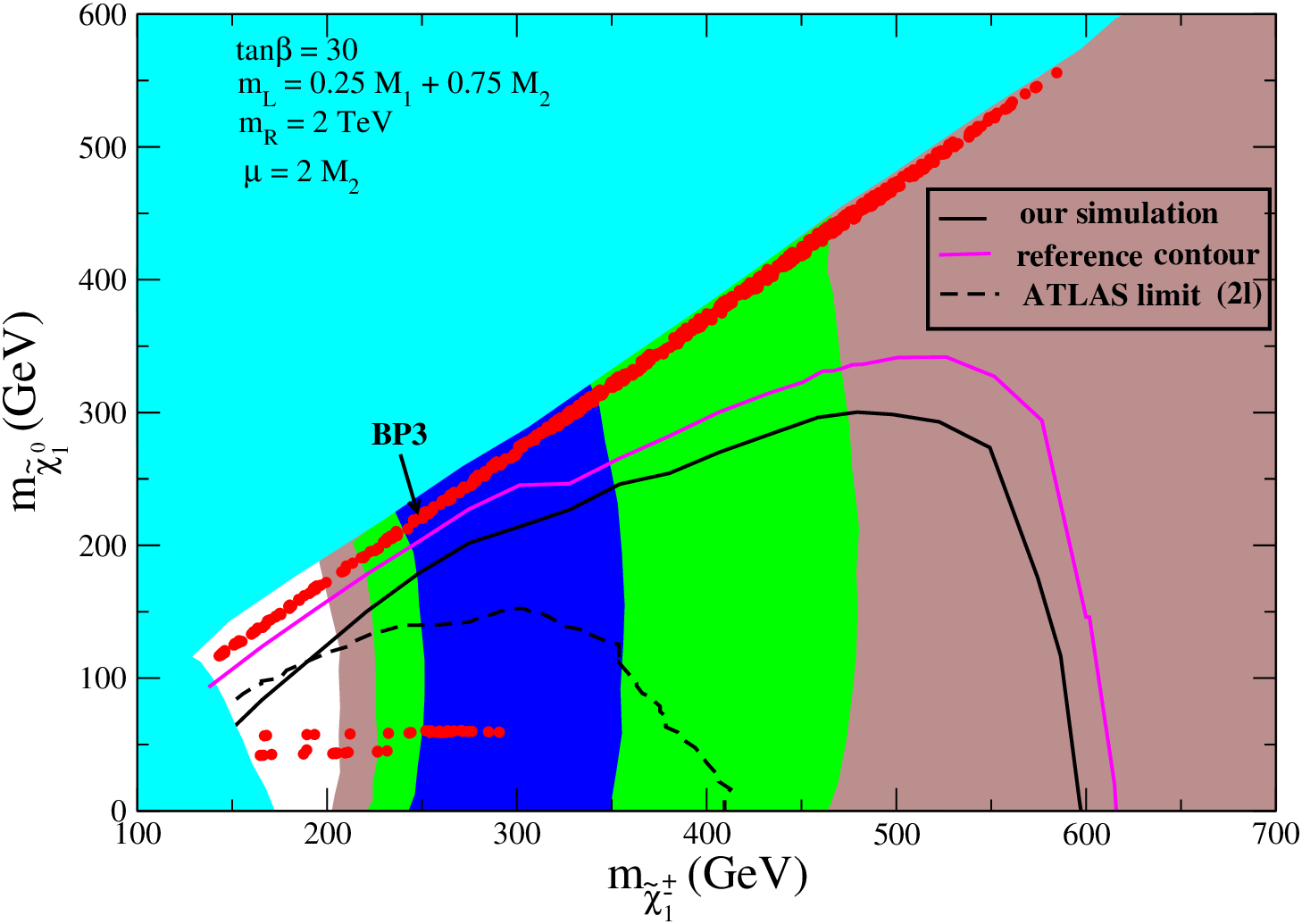}
\caption{{\it (a) Plot in the $\mchonepm-\mlspone$ plane for 
the tilted LGLS scenario choosing
$M_{{\tilde l}_L} = 0.25 M_1 +0.75 M_2$ and tan$\beta =$ 6 (a)
and 30 (b).
Colours and conventions are the same as in Fig.\ref{LL_0.5_0.5}.  
The lightly shaded (cyan) upper region 
is discarded by the requirement of the 
LSP to be the lightest neutralino.
The exclusion contour for this scenario is
represented by black line. 
}}
\label{LL_0.25_0.75}
\end{figure}


\subsection{Light Gaugino and Light Left and Right Slepton (LGLRS) Scenario}
\label{section3.2}
We now come to the analyses of the LGLRS  
scenario.  This was not considered by the ATLAS collaboration\cite{atlas3lew}.
We assume the R-slepton mass parameters ($M_{\tilde l_R}$) to be same as that 
of the L-sleptons ($M_{\tilde l_L}$).  The principal difference of this scenario with 
LGLS is that the L-R mixing effect becomes prominent 
in the third generation slepton sector. As a result the  
${\tilde \tau}_1$ instead of the sneutrino often 
becomes a charged NLSP or even the LSP leading to a forbidden region.  
For a given value of 
$m_{\charginopm_1}$ this results into 
elimination of larger values of $\mlspone$,   
causing a shrinkage of parameter space  
for the upper $\mlspone$ region in comparison to a corresponding 
LGLS case. There is a significant region in the smaller   
$m_{{\tilde \chi}_1^0}-m_{{\tilde \chi}_1^\pm}$ zone that is 
discarded due to the appearance of
tachyonic stau or stau becoming the LSP.    

We start with the case of slepton mass 
parameters (L and R) at the average of $M_1$ and $M_2$ as in 
Fig.\ref{LLR_0.5_0.5_A}.  In regard to the DM relic density
the upper branch arises via LSP-stau coannihilation and some bulk 
annihilations for low mass regions. The lower 
branch as usual occurs due to the $h$-resonance and some Z-resonance as well 
as some bulk-annihilations for the low mass regions. \\

\begin{figure}[!htb]
\mygraph{LLR_0.5_0.5_A}{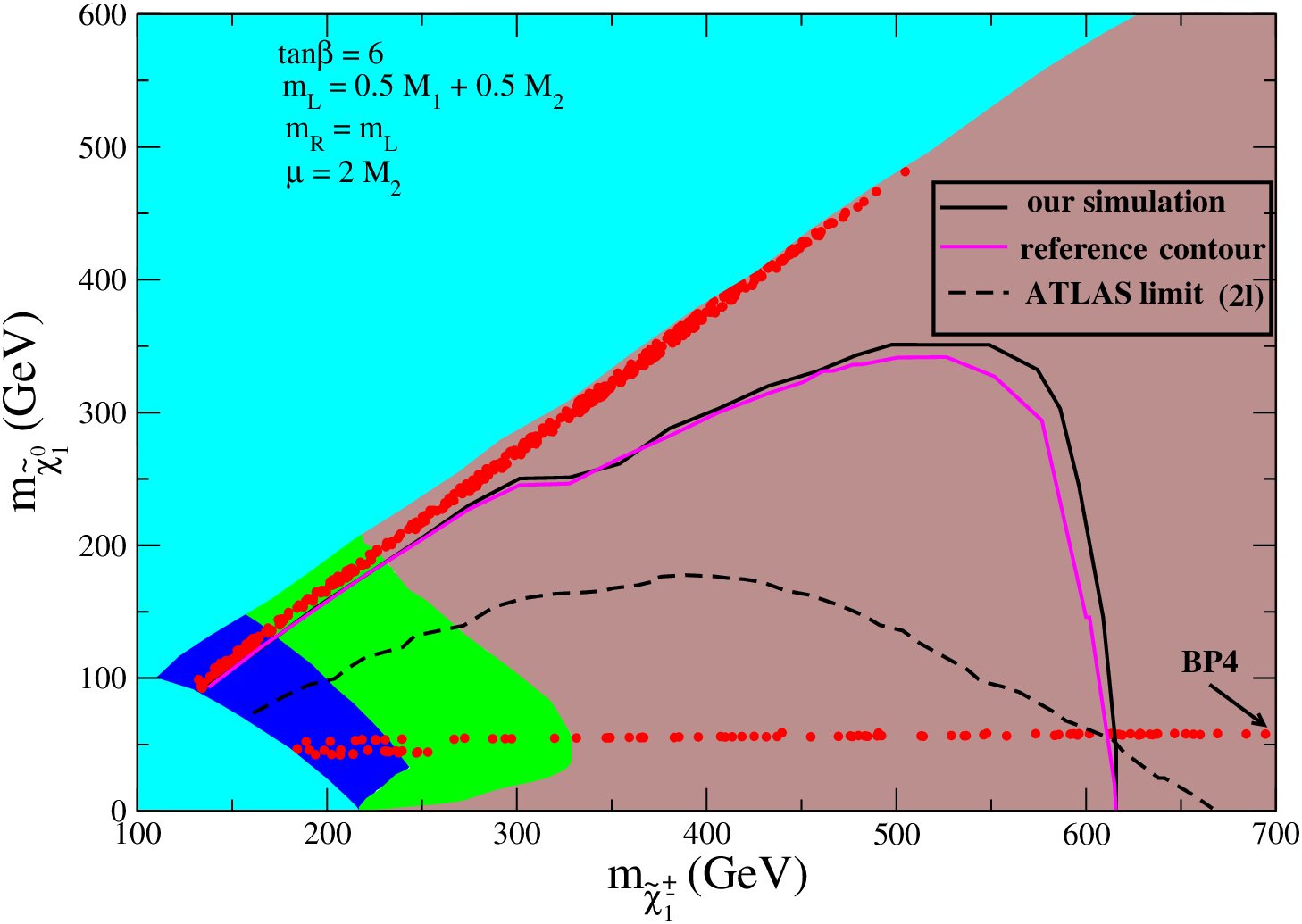}
\hspace*{0.5in}
\mygraph{LLR_0.5_0.5_B}{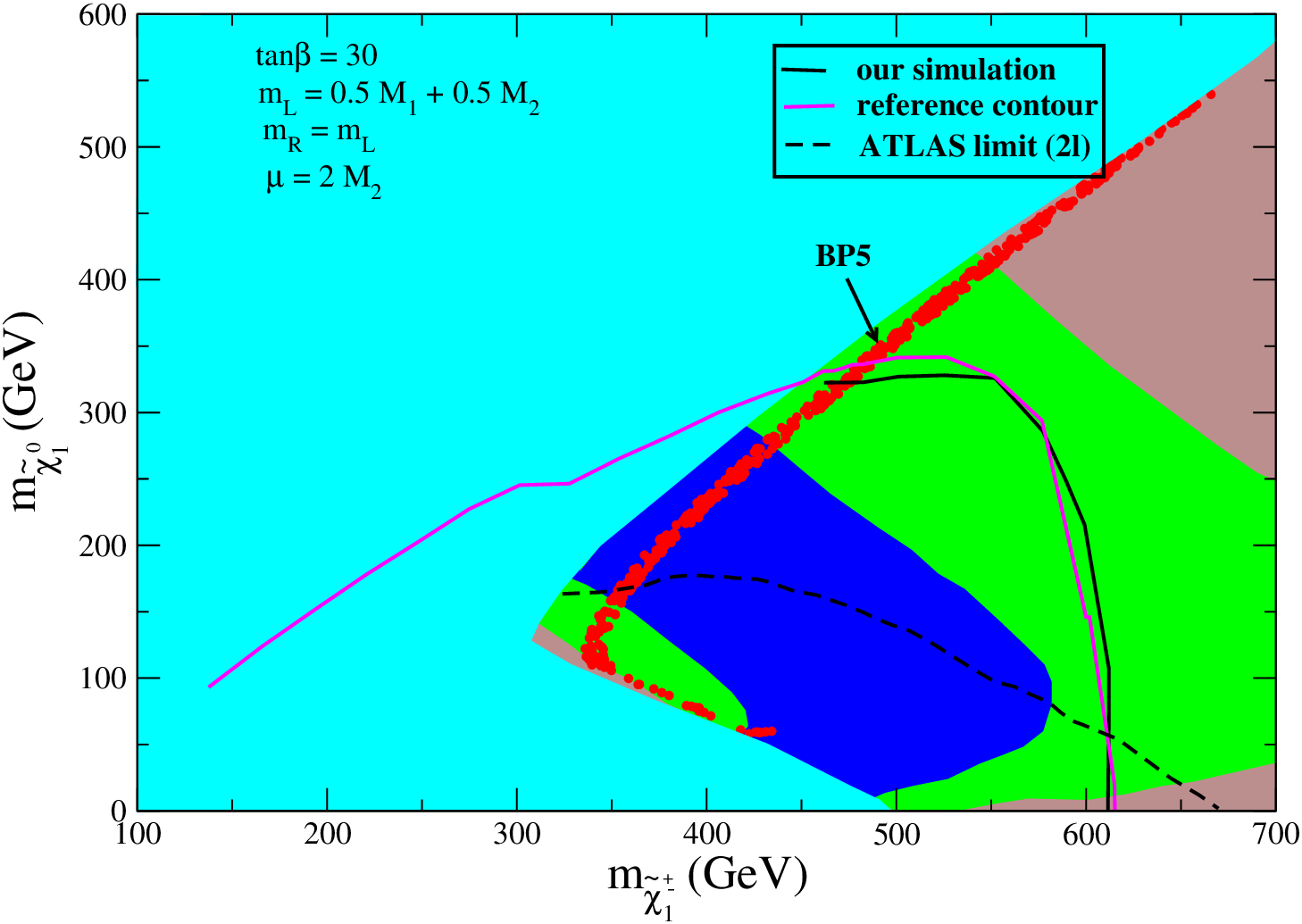}
\caption{{\it (a)  Plot in the $\mchonepm-\mlspone$ plane for the LGLRS scenario with
$M_{{\tilde l}_L} = M_{{\tilde l}_R} = 0.5 M_1 +0.5 M_2$ and tan$\beta =$6 (a) and 30 (b).  Colours and
conventions are the same as in Fig.\ref{LL_0.5_0.5}.
The exclusion contour for this scenario is
represented by the black line.  
}}
\label{LLR_0.5_0.5}
\end{figure}

Since both $\chonepm$ and $\lsptwo$ are wino dominated, 
they primarily decay into left sleptons.  Thus the inclusion of 
right sleptons does not alter BR of $\chonepm$ and $\lsptwo$  decaying into left sleptons.
But as the trilepton efficiency increases, 
the collider limit on $\mlspone$ becomes stronger by 20-30~GeV for 
$\mchonepm > 450~{\rm GeV}$ 
compared to the reference contour of Fig.\ref{LL_0.5_0.5_A}.
On the other hand,  
since a part of neutralino-smuon loop contribution 
scales as $\frac {m_{\mu}^2 M_1 \mu}{M^D_{{\tilde \mu_L}^2}
M^D_{{\tilde \mu_R}^2}} tan\beta$ \cite{g-2sm1}, $a_\mu^{\rm SUSY}$ is significantly 
boosted because both the left and the right slepton mass parameters 
are the same (unlike the LGLS scenario). A larger $\amususy$ does 
not however make more and more smaller mass region in 
the $\mchonepm-\mlspone$ plane to be finally available. Much of such 
low mass regions become unavailable because ${\tilde \tau}_1$ turns out to be 
lighter than the LSP due to L-R mixing 
or even it can become tachyonic. The unavailable regions fall 
in the cyan shaded zone.  
We must however keep in mind that 
an effort to nullify the L-R mixing by considering an appropriate  
non-vanishing $A_\tau$ parameter would open up the low mass 
region that would also satisfy the constraints of collider and 
the WMAP/PLANCK data as well as $\gmin2$ in this LGLRS scenario.

An analysis for $\tan\beta=30$ is presented in 
Fig.\ref{LLR_0.5_0.5_B}. Here in comparison with 
Fig.\ref{LL_0.5_0.5_B} the effects of 
L-R mixing (leading to unacceptably light ${\tilde \tau}_1$)   
is significantly strong causing an appreciable 
shrinkage of the available parameter space. $\amususy$  
is enhanced due to a large value of $\tan\beta$.  
As before $\lspone-{\tilde \tau}_1$ coannihilation is the dominant DM producing
mechanism. The mechanism via h-resonance occurs in a region forbidden
by unacceptable ${\tilde \tau}_1$ mass.
The lowest mass combination within the valid parameter space
is about $m_{{\tilde \chi}_1^\pm} \simeq 470$~GeV and
$m_{{\tilde \chi}_1^0} \simeq 330$~GeV that falls in the
$2\sigma$ zone of $\gmin2$.

\noindent
\subsubsection{Tilted LGLRS Scenario}
\label{section3.2.1}
{\bf i) LGLRS-$\lspone$:} 

\begin{figure}[!htb]
\vspace*{-0.05in}
\mygraph{LLR_0.75_0.25_A}{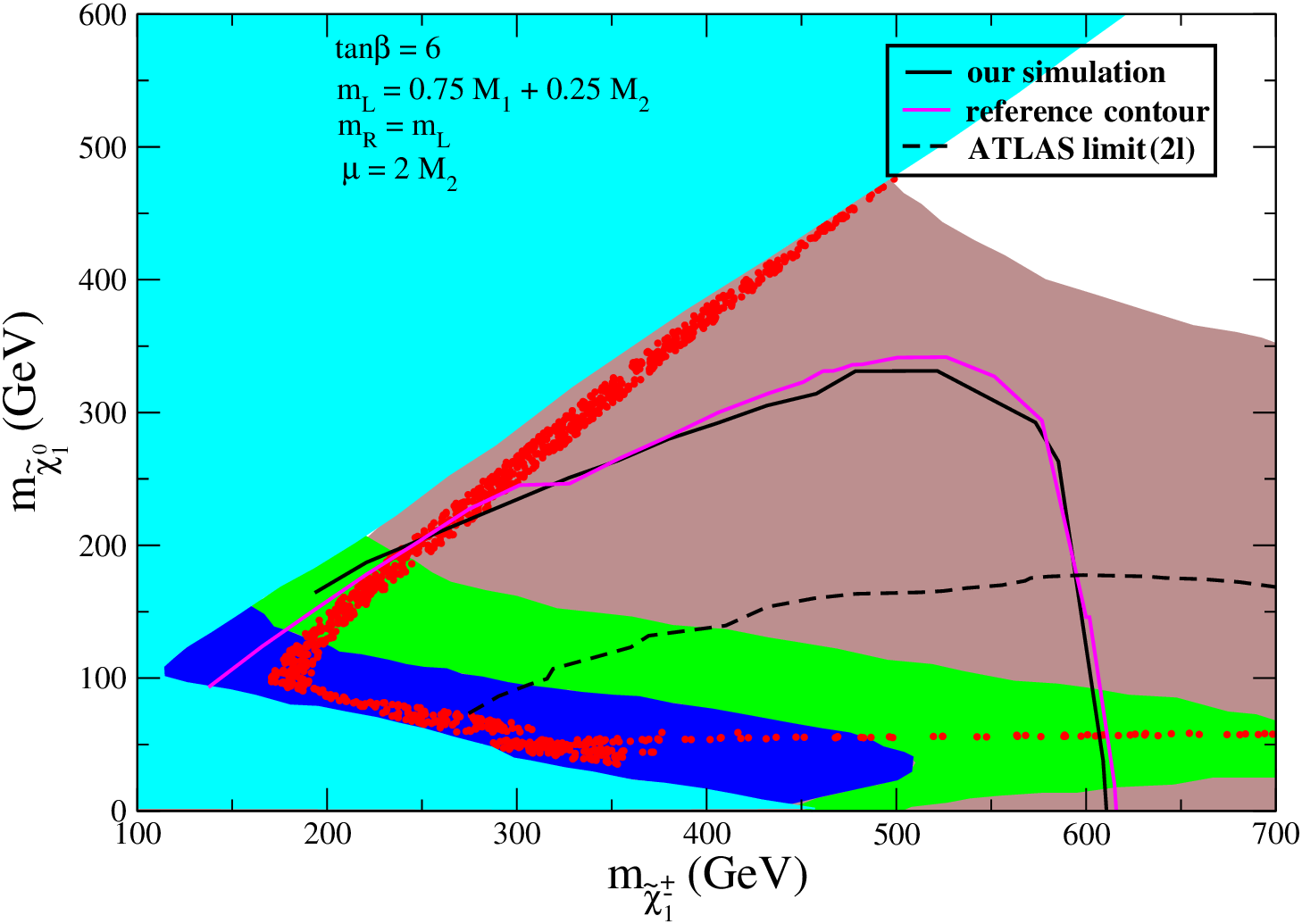}
\hspace*{0.5in}
\mygraph{LLR_0.75_0.25_B}{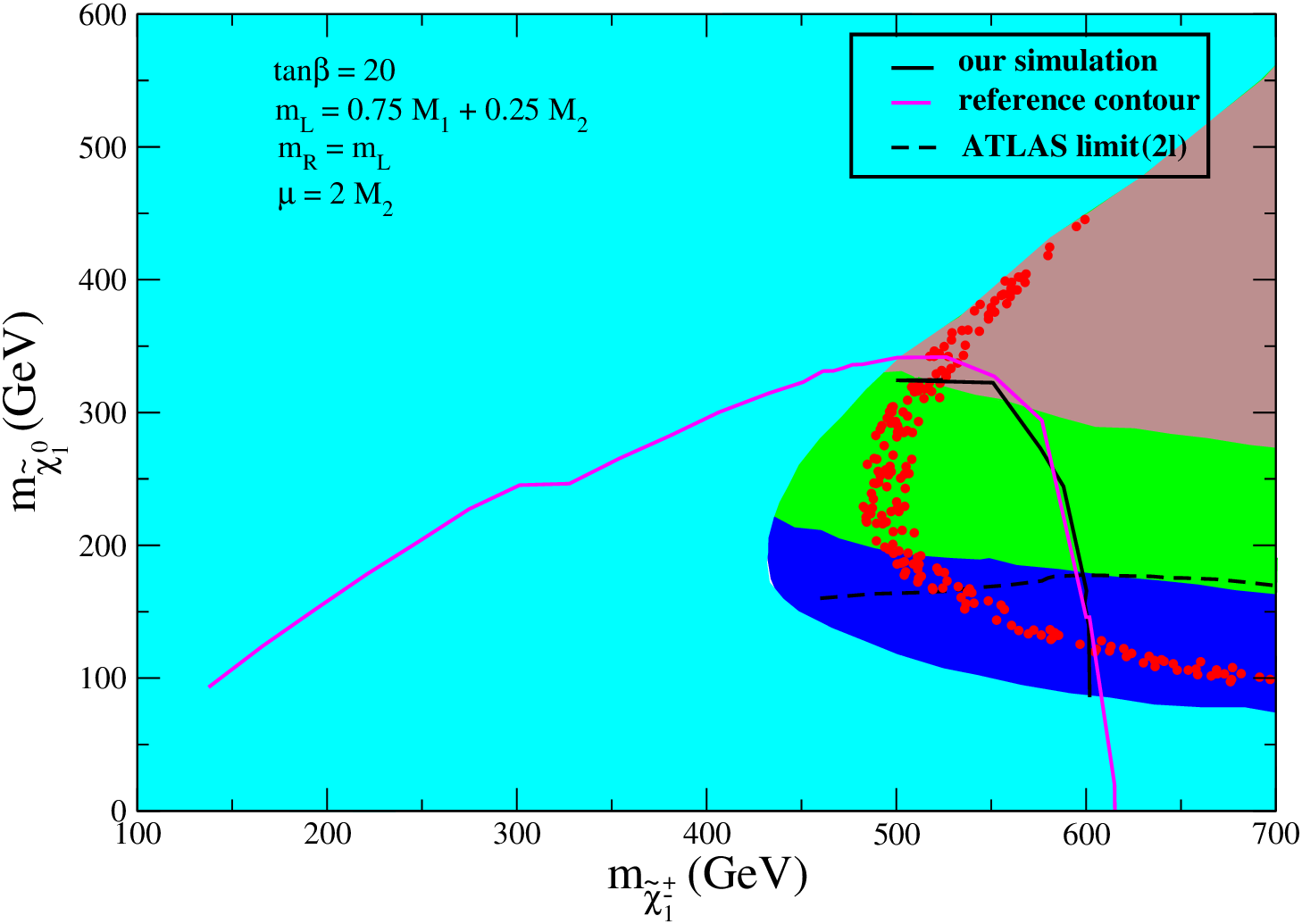}
\caption{{\it (a) Plot in the $\mchonepm-\mlspone$ plane for the LGLRS scenario with
$M_{{\tilde l}_L} = M_{{\tilde l}_R} = 0.75 M_1 +0.25 M_2$ and tan$\beta =$6 (a) and 20 (b).
Colours and conventions are same as Fig.\ref{LL_0.5_0.5}. The exclusion contour for this scenario is
represented by the black line.  
}}
\label{LLR_0.75_0.25}
\end{figure}

In Fig.\ref{LLR_0.75_0.25_A} we explore the case where
both L and R-sleptons are 
closer to the mass of the LSP
via $M_{{\tilde l}_L} =M_{{\tilde l}_R}= 0.75 M_1 +0.25 M_2$. 
While sleptons become light, similar to what happens for 
Fig.\ref{LLR_0.5_0.5_A} 
the dominant contribution to $a_\mu^{\rm SUSY}$ comes from 
the one-loop neutralino-smuon loop diagram as discussed before.
As a result $m_{\charginopm_1}$ becomes unconstrained 
leading to increase of the upper limit of the same for 
a given error corridor of $\gmin2$ compared to what appears in 
Fig.\ref{LLR_0.5_0.5_A}. 
In this case, as discussed before, the trilepton efficiency 
would decrease due to the fact that the sleptons are shifted more 
towards the LSP.
Here it almost overlaps with the limit corresponding to 
Fig.\ref{LL_0.5_0.5_A}.  
Additionally, there is a large discarded region where   
$\tilde \tau_1$ becomes the LSP or tachyonic because of 
mixing between the components of the third generation of slepton fields. 
The allowed region satisfying the relic density constraint and the 
collider limits mostly occurs in the $3\sigma$ region of $\gmin2$. 
We note that the direct  
slepton mass bounds from ATLAS disallow the entire 
bottom limb of the relic density satisfied region that is associated 
with the $h$-pole annihilation unless $\mchonepm$ is very large.
Thus we do not find any APS in this scenario if the $\gmin2$ constraint 
is imposed at the level of 2$\sigma$.

In Fig.\ref{LLR_0.75_0.25_B}, we are compelled to use a relatively smaller 
value of $\tan\beta$~$(= 20)$ unlike previous results, where we could 
comfortably analyse a larger value of $\tan\beta~(=30)$. 
This is simply because, in this case the slepton masses are closer to the 
LSP mass and the masses of the left and right slepton 
partners are almost similar in magnitude (apart from D-term contributions).  
The effect of mixing is dominant in the stau sector and this 
leads to $\tilde \tau_1$ to become the LSP or even tachyonic for a larger 
value of 
$\tan\beta$. Even for $\tan\beta=20$, as may be seen in Fig.\ref{LLR_0.75_0.25_B} there is a considerable region that becomes discarded because of the 
above reason.  The collider limits on the other hand
remain almost unchanged 
with respect to that of Fig.\ref{LLR_0.75_0.25_A}.

The dominant diagrams contributing to $a_\mu^{\rm SUSY}$ 
are the neutralino-smuon loop diagrams similar to the other LGLRS models.   
Here, the regions allowed via $\gmin2$  that also satisfy the 
collider limits and the DM relic density 
occur i) in the $3\sigma$ zone for which 
the mass of LSP is higher and ii) in the $1\sigma$ zone for which 
the mass of $\chonepm$ is higher ($> 600$~GeV). 
The DM relic density satisfied points result 
mainly from LSP-${\tilde \tau}_1$ and ${\tilde \tau}_1-{\tilde \tau}_1$ 
coannihilations in the upper zone. 
In the lower region there are some points for which the LSP
undergoes self-annihilations via
t-channel slepton exchange mechanism thus producing the right amount of 
abundance. The importance of the direct slepton search is showcased by
this scenario. It rules out the LGLRS-$\lspone$ model for high tan$\beta$ 
discussed above, which is consistent with $\gmin2$, WMAP/PLANCK data and 
trilepton searches at the LHC. 

\noindent
ii) {\bf LGLRS-$\chonepm$:}\\
Fig.\ref{LLR_0.25_0.75_A} describes the constraints in a scenario with 
the common slepton mass parameter closer to $\mchonepm$ ($M_{{\tilde l}_L}=M_{{\tilde l}_R} 
= 0.25 M_1 +0.75 M_2$) for $\tan\beta=6$. 
The dominant corrections contributing to $a_\mu^{\rm SUSY}$ 
come from the neutralino-smuon loop diagrams similar to other 
cases of small left and right slepton masses.
Since the slepton masses are closer to $\mchonepm$ than $\mlspone$, the 
trilepton efficiency decreases. This weakens  
the collider limit of $\mlspone$ by 15-45~GeV 
compared to the reference contour.  
As seen from the figure this shrinkage of limit in 
turn opens up a
parameter space to the $\gmin2$ constraint at 1$\sigma$ level. 
The DM relic density satisfying mechanisms are annihilations via 
s-channel Higgs resonance and some t-channel slepton exchange for a 
small $\mchonepm$ for the lower horizontal branch of red 
points only.
This branch is, however, strongly disfavoured by the LHC data.
For the upper branch, the relic density is satisfied via a multitude of processes like 
LSP annihilations via 
chargino mediation and various coannihilations such as those 
between LSP-stau, LSP-sneutrino, stau-stau, 
stau-sneutrino, sneutrino-sneutrino, and chargino-sneutrino.\\

\begin{figure}[!htb]
\mygraph{LLR_0.25_0.75_A}{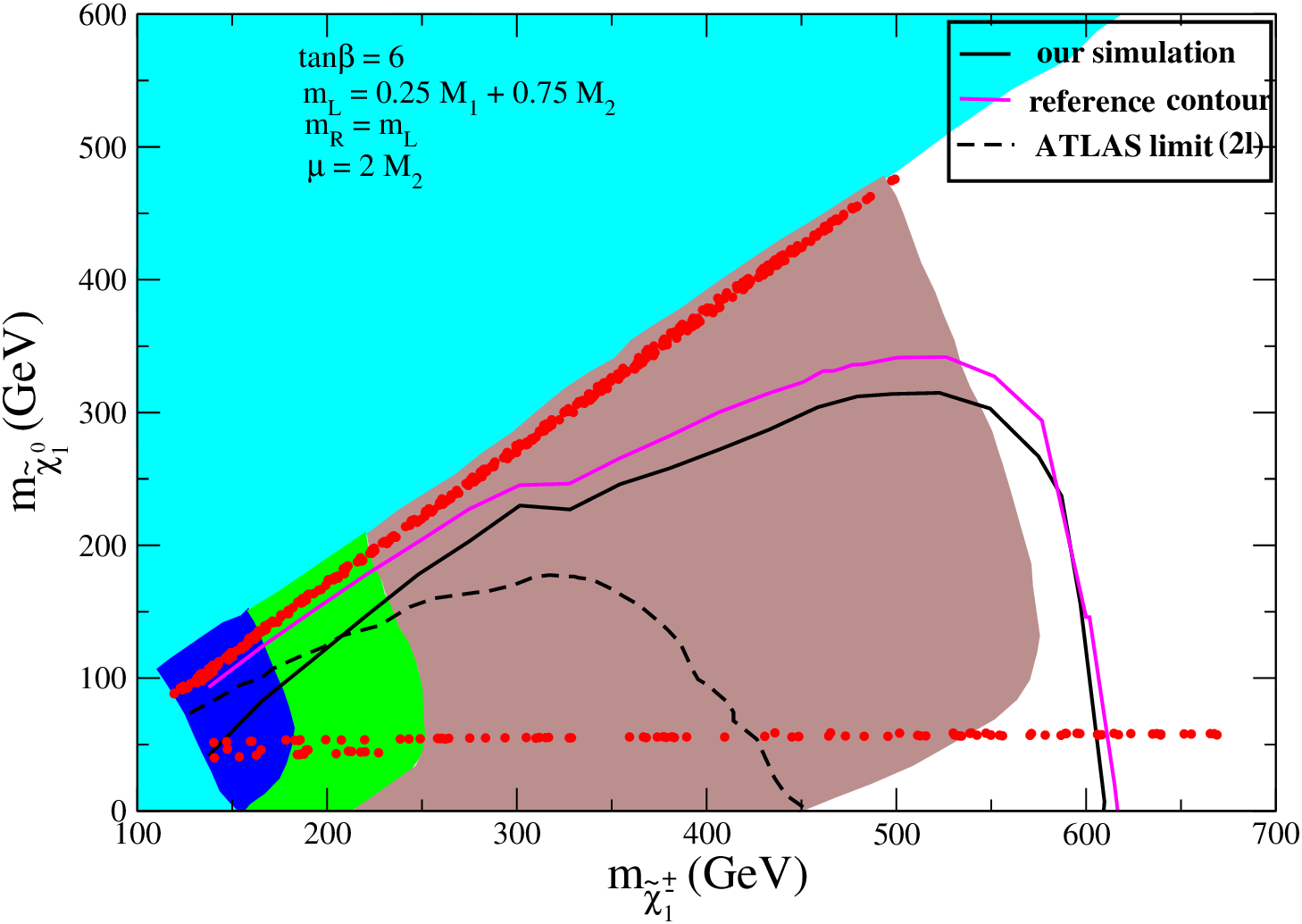}
\hspace*{0.5in}
\mygraph{LLR_0.25_0.75_B}{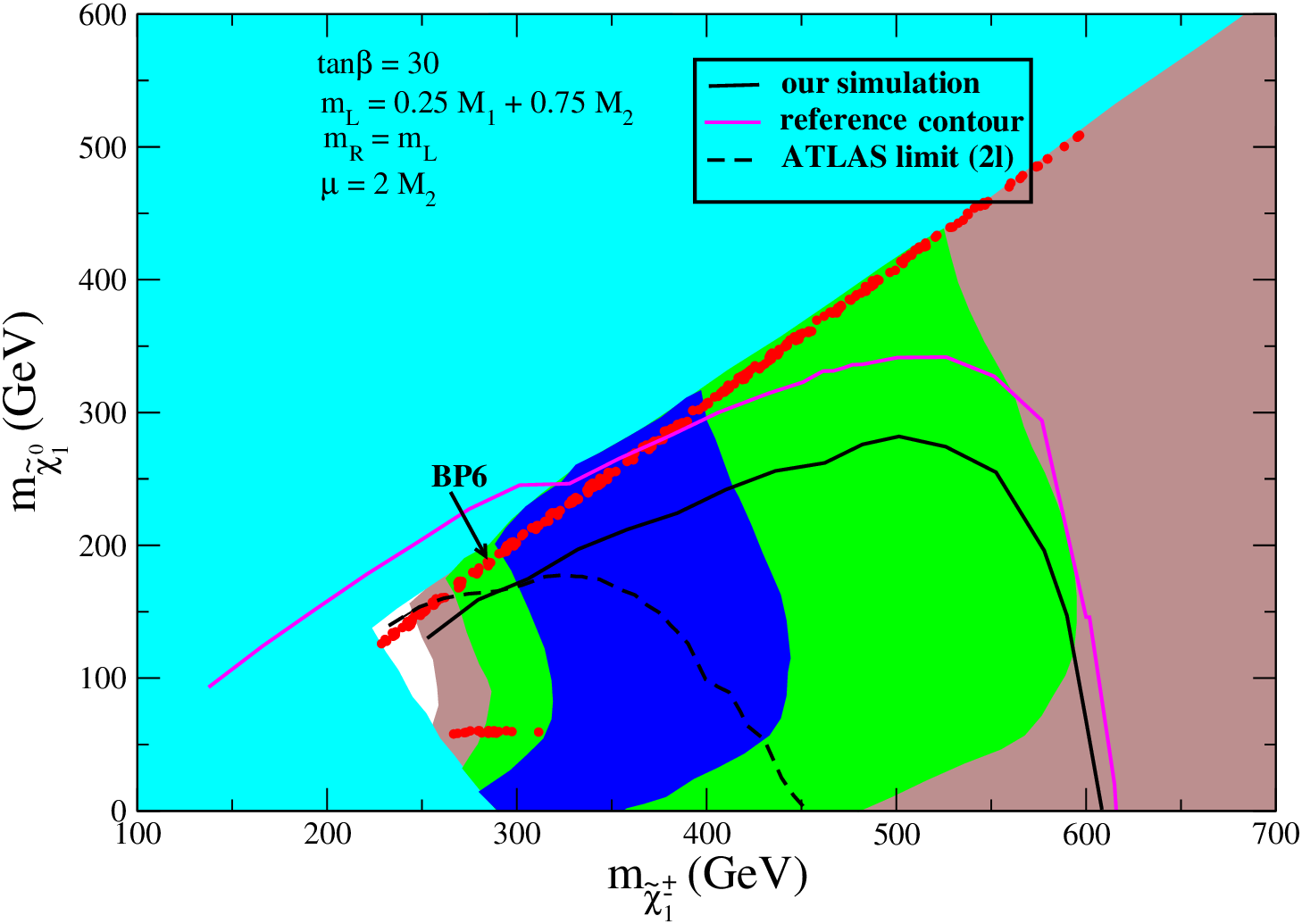}
\caption{{\it (a) Plot in the $\mchonepm-\mlspone$ plane for the LGLRS scenario with
$M_{{\tilde l}_L} = M_{{\tilde l}_R} = 0.25 M_1 +0.75 M_2$ and tan$\beta =$6 (a) 
and 30 (b). Colours and conventions are same as Fig.\ref{LL_0.5_0.5}.
The exclusion contour for this scenario is
represented by the black line.  
}}
\label{LLR_0.25_0.75}
\end{figure}

Fig.\ref{LLR_0.25_0.75_B} refers to $\tan\beta=30$.
The $(g-2)_\mu$ allowed regions are extended to larger values of
$M_2$.  The trilepton efficiency is smaller 
here  even in comparison with Fig.\ref{LLR_0.25_0.75_A}.  
This is due to the fact that there is a large mixing in the stau sector
leading to an increase in the branching ratio of 
$\chonepm / \lsptwo$ decaying into $\tilde \tau_1$, which in 
turn decreases the number of trilepton events.  
The combined effect weakens the collider 
limit upto 65-75~GeV for most of the  parameter space.  
The DM relic density satisfying mechanisms for the upper branch 
are mainly LSP-$\ttau_1$
and $\ttau_1$-$\ttau_1$
coannihilations.  For the tiny lower branch there is not much difference 
with the situation encountered earlier for large tan $\beta$.

\subsection{Light Gaugino and Right Slepton (LGRS) Scenario}
\label{section3.3}

In this case, we consider the R-slepton mass for  all the three generations to lie between 
$\mlspone$ and $\mchonepm$ so that  $M_{{\tilde l}_R} = \frac{1}{2}(M_1 + M_2)$. The 
corresponding L-slepton mass parameter is taken to be greater than the lighter
chargino mass: $M_{{\tilde l}_L} = M_2 + 200$ GeV.
\vspace*{+0.2cm}
\begin{figure}[!htb]
\begin{center}
\includegraphics[scale=0.35]{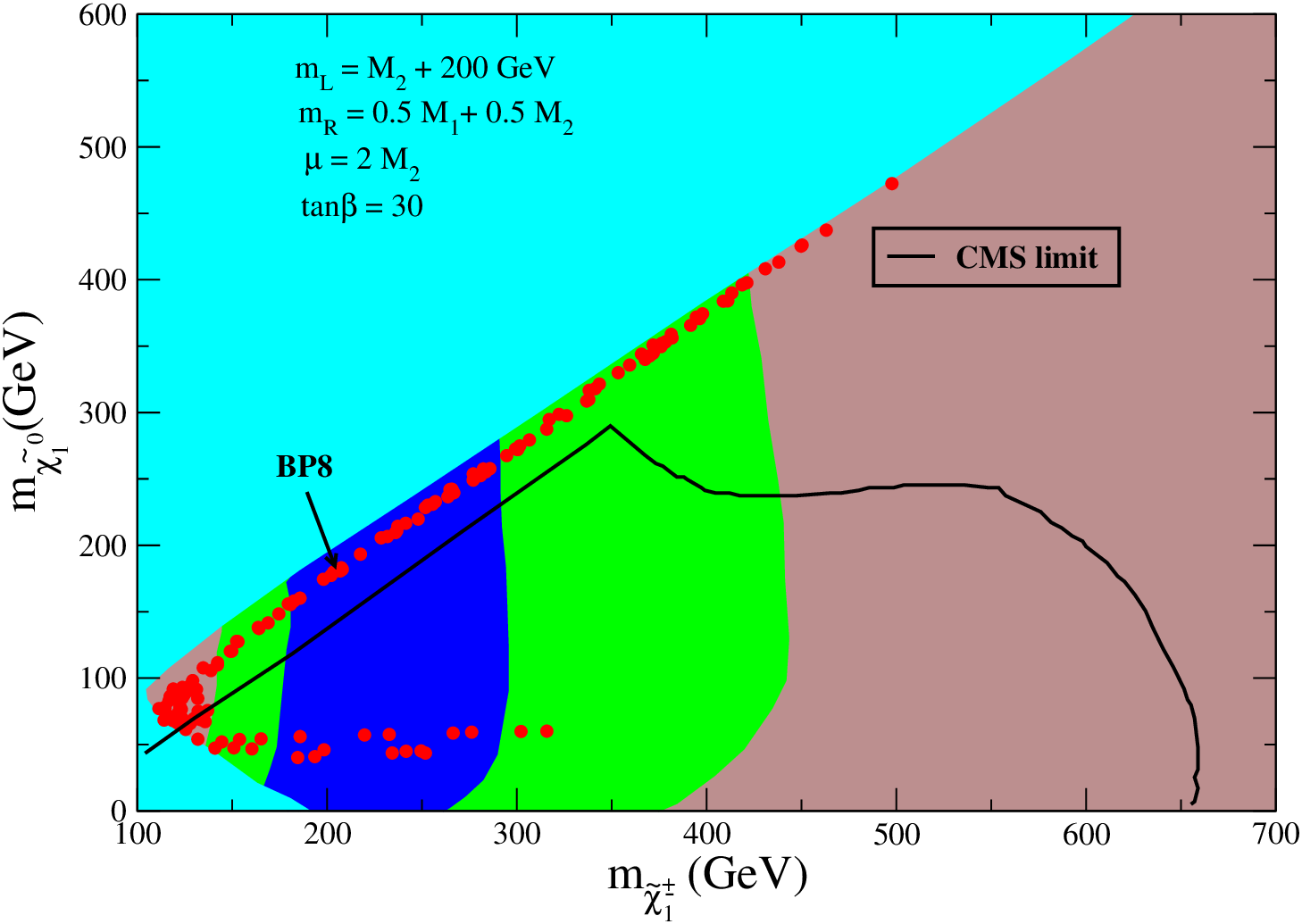}
\end{center}
\caption{{\it Result of scanning the $\mlspone$-$\mchonepm$ plane for the LGRS scenario with
tan$\beta = $30.  Here, $M_{{\tilde l}_L} = M_2 +$ 200 GeV, and 
$M_{{\tilde l}_R} = 0.5 M_1 + 0.5 M_2$.
Colours and conventions are same as those of 
Fig.\ref{LL_0.5_0.5}.  The CMS exclusion contour is shown as a black line 
(Fig.21 of Ref.\cite{cmsew}).}}
\label{30_R_slep_gt_m2}
\end{figure}
Fig.\ref{30_R_slep_gt_m2} shows the results for the  LGRS scenario with tan$\beta=$30.  The CMS exclusion 
contour (Fig.21 of Ref.\cite{cmsew}) is shown as a black line.  The main contribution 
to $\amususy$ comes from the neutralino-smuon
loop.  For moderate values of $M_2$, the contribution coming from the
chargino-sneutrino (bino-higgsino-$\tilde {\mu}_R$) loop is also significant.

 The PLANCK/WMAP allowed points for the upper branch undergo LSP-stau, 
as well as stau-stau coannihilations. However, the region at the lower end of this branch corresponding to bulk annihilation is disfavoured by the
$\gmin2$ data. There also exists 
a small amount of coannihilation of LSP/stau with
the right handed slepton of the first two generations 
and annihilations via chargino exchange.
For the lower branch disfavoured by the LHC data, there are resonant Higgs/Z exchange annihilation processes and also bulk annihilation.
As can be seen from the figure, there is a significant area of parameter space  which satisfies WMAP/PLANCK data,
 $\gmin2$ at the level of 1$\sigma$ along with collider constraints.

\subsection{Light Gaugino and Heavy Slepton (LGHS) Scenario}
\label{section3.4}

The ATLAS group has also searched for the trilepton signal in the light 
gaugino heavy slepton (LGHS) model. All sleptons 
with equal masses for the left and the right components are assumed to be 
heavier than $\chonepm$ or $\lsptwo$. 
\vspace*{+0.2cm}
\begin{figure}[!htb]
\begin{center}
\includegraphics[scale=0.35]{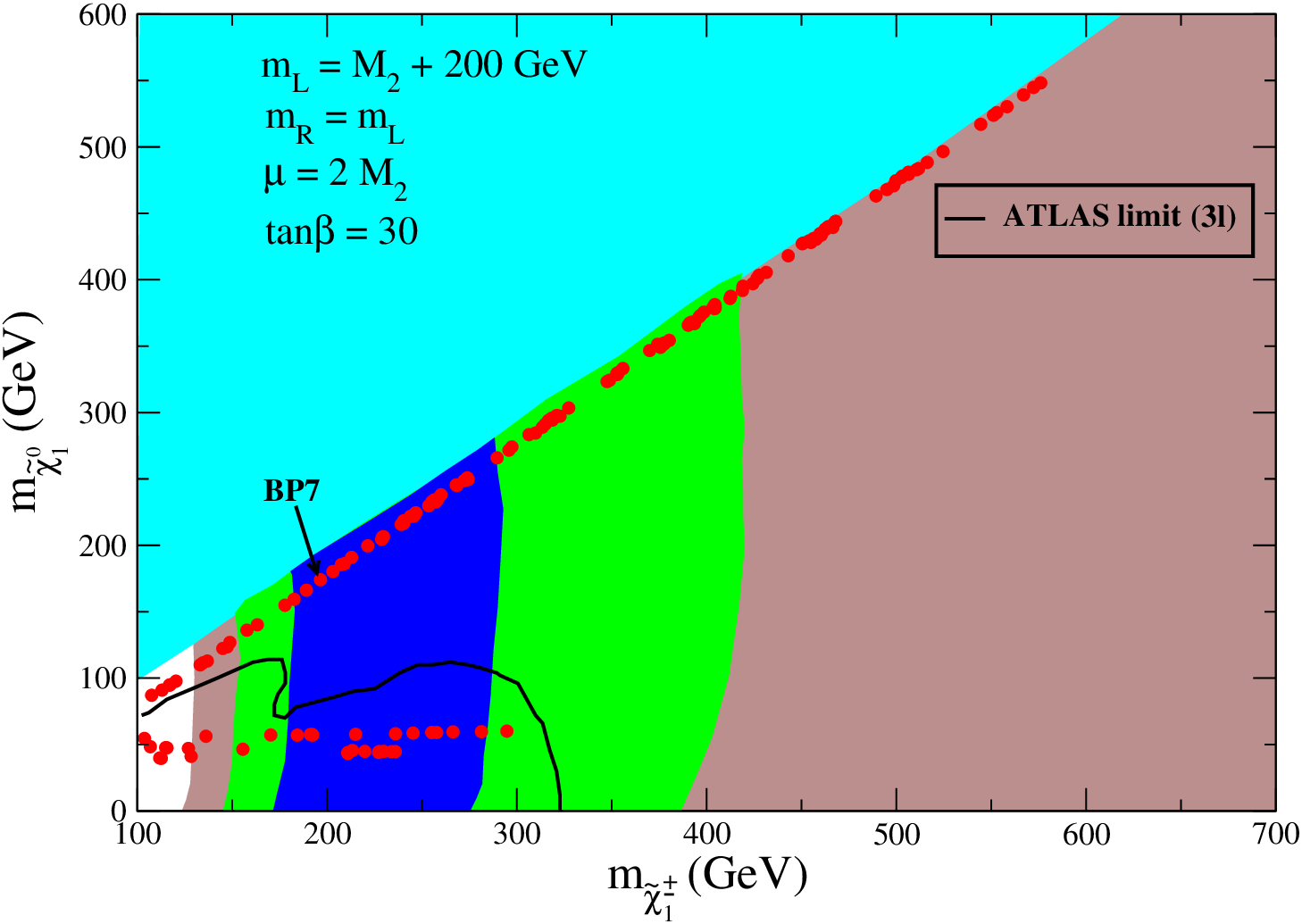}
\end{center}
\caption{{\it Plot in the $\mchonepm-\mlspone$ plane for the 
LGHS scenario with $M_{{\tilde l}_R} = M_{{\tilde l}_L} = M_2 +200$~GeV and
tan$\beta = $30.  Colours and conventions are same as Fig.\ref{LL_0.5_0.5}.
The black line represents the
exclusion contour at 8 TeV by the ATLAS collaboration \cite{atlas3lew}.}}
\label{30_slep_gt_m2}
\end{figure}
The 
bounds mainly depend on the chargino and the LSP mass 
(see the exclusion contour in Fig.8b of ATLAS\cite{atlas2lew} reproduced in
Fig.\ref{30_slep_gt_m2}  for ready reference). The sensitivity 
to the other MSSM parameters is rather mild.

We consider the representative choice $M_{\tilde l_R} = M_{\tilde l_L} = M_2 +200$~GeV as in 
Fig.\ref{30_slep_gt_m2}. It may be noted that with this choice the sleptons 
contribute neither to
the LHC signal nor do they affect LSP annihilation/coannihilation.
Here, since 
$\chonepm$ or $\lsptwo$ are unable to
decay into sleptons, 
they decay via gauge bosons with a $100 \%$
branching ratio. As a result, each collider limit becomes
independent of the SUSY input parameters like tan$\beta$.  
Here the choice  tan$\beta = 30,  $ simply yields a large 
$\amususy$ leading to widening of the 1$\sigma$ allowed region for 
$\gmin2$ (the dominant contributions to $a_\mu^{\rm SUSY}$ come from the neutralino-smuon 
loops)\footnote{For $\tan\beta=6$ LSP pair annihilation via Higgs 
resonance would be quite efficient but consistency of 
$\amususy$ with the measured value is only at the 3$\sigma$ level.}. 
The relic density producing mechanisms for the lower red points are 
annihilations via s-channel Higgs and Z resonances which are disfavoured by the LHC data. 
Points in the upper branch primarily undergo $\chonepm$/$\lsptwo$ 
coannihilations. 

In obtaining the LHC exclusion contour in Fig.\ref{30_slep_gt_m2} it is assumed that the
decay $\lsptwo \rightarrow Z \lspone$ occurs with 100\% BR. However, in
parts of the excluded parameter space, the spoiler mode $\lsptwo
\rightarrow h \lspone$ may occur with significant BR and weaken the
limits\cite{higgsino}. It is particularly interesting to note that in the Higgs resonance
region the BR of this mode is appreciable for $\mchonepm \approx \mlsptwo
> m_h + \mlspone \approx 1.5 m_h$. As a result this region, particularly
the points close to the exclusion contour, cannot be excluded beyond
doubt. On the other hand the exclusion obtained by assuming that $\lsptwo
\rightarrow h \lspone$ occurs with 100\% BR is too weak to affect the
Higgs resonance region \cite{higgsino}.

\subsection{Light Left Slepton (LLS) Scenario}
\label{section3.5}
In the Light Left Slepton model, the left sleptons are light but 
the right sleptons and all the charginos and the 
neutralinos except the LSP are heavy.  
The ATLAS collaboration has reported the results of 
slepton search in the LLS model\cite{atlas2lew}. Their
exclusion contour is reproduced in Fig.\ref{30_LL_lsp_slep}. 

\begin{figure}[!htb]
\begin{center}
\includegraphics[scale=0.35]{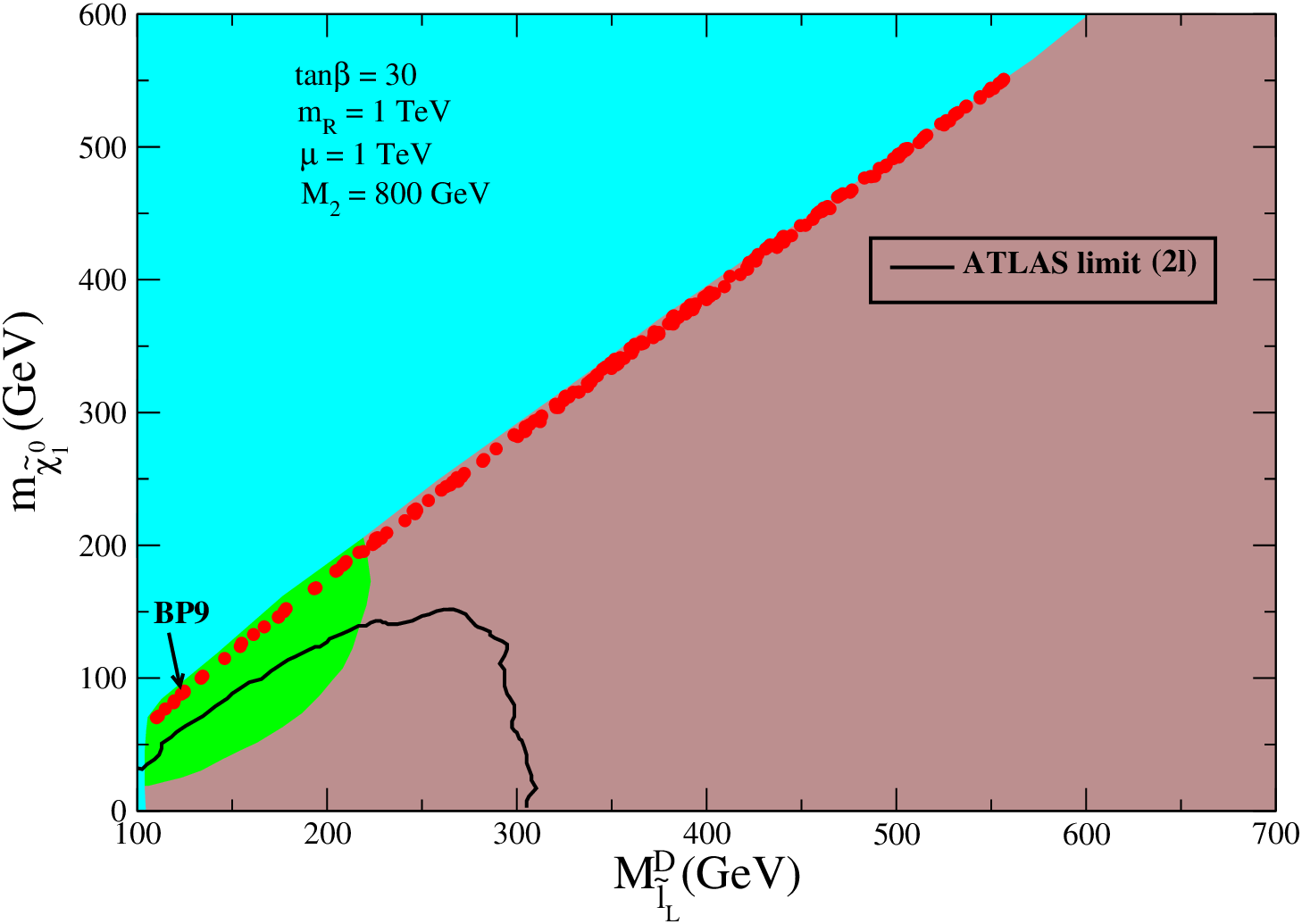}
\end{center}
\caption{{\it Plot in the $M^D_{\tilde l_L}-\mlspone$ plane  for the LLS scenario with $M_2 =800$~GeV, $\mu=1$~TeV and
$\tan\beta=30$. Here $M^D_{\tilde l_L}$ represents physical left slepton masses.
Colours and conventions are same as those of Fig.\ref{LL_0.5_0.5}.The black line
represents the exclusion contour at 8 TeV by the ATLAS collaboration\cite{atlas2lew}.}}
\label{30_LL_lsp_slep}
\end{figure}

With the choice of a heavy right slepton ($M_{{\tilde l}_R}=1$~TeV), we 
scan $M_1$ and $M_{{\tilde l}_L}$ and show the results in the 
$\mlspone-M^D_{{\tilde l}}$ plane of Fig.\ref{30_LL_lsp_slep}. 
We fix a wino dominated lighter chargino with the choice of $M_2=$800~GeV  
and $\mu=1$~TeV for $\tan\beta=30$. This choice of $M_2$ is motivated
by the chargino mass bounds in the LGLS models considered in Sec.~\ref{section3.1}.  $M_1$ 
is varied upto 600 GeV for the given choice of $M_2$ and $\mu$ so as to have a  bino-like LSP.    
With the right slepton being heavy, the contribution from 
neutralino-smuon loop to $a_\mu^{\rm SUSY}$ is suppressed.  
Again, since $\mu$ and $ M_2$ are sufficiently large in magnitude, 
the chargino-sneutrino loop is also suppressed.  
Nevertheless, we have acceptable $a_\mu^{\rm SUSY}$, though
at the 2$\sigma$ level, consistent with all other constraints.
The red points satisfy DM relic density constraint by primarily 
LSP-sneutrino coannihilations.  There are also 
sneutrino-sneutrino, sneutrino-stau coannihilations.

\subsection{Light Left and Right Slepton (LLRS) Scenario}
\label{section3.6}
Here the right and the left sleptons are assumed to be degenerate in mass and  are 
lighter than the lighter chargino (Fig.\ref{LLR_lsp_slep}).  
The ATLAS collaboration has also reported slepton pair production in the 
LLRS model in addition to LLS\cite{atlas2lew}.

Since the $\chonepm$ and $\lsptwo$ are taken to be heavier than the sleptons, 
the sleptons decay into leptons and $\lspone$ with
100$\%$ branching ratio.  Thus, the exclusion limits would be independent 
of the input parameters like $M_2$, $\mu$, tan$\beta$ etc. 
Here we use the ATLAS exclusion contour\cite{atlas2lew} as shown in Fig.\ref{LLR_lsp_slep}.

\begin{figure}[!htb]
\mygraph{LLR_lsp_slep_A}{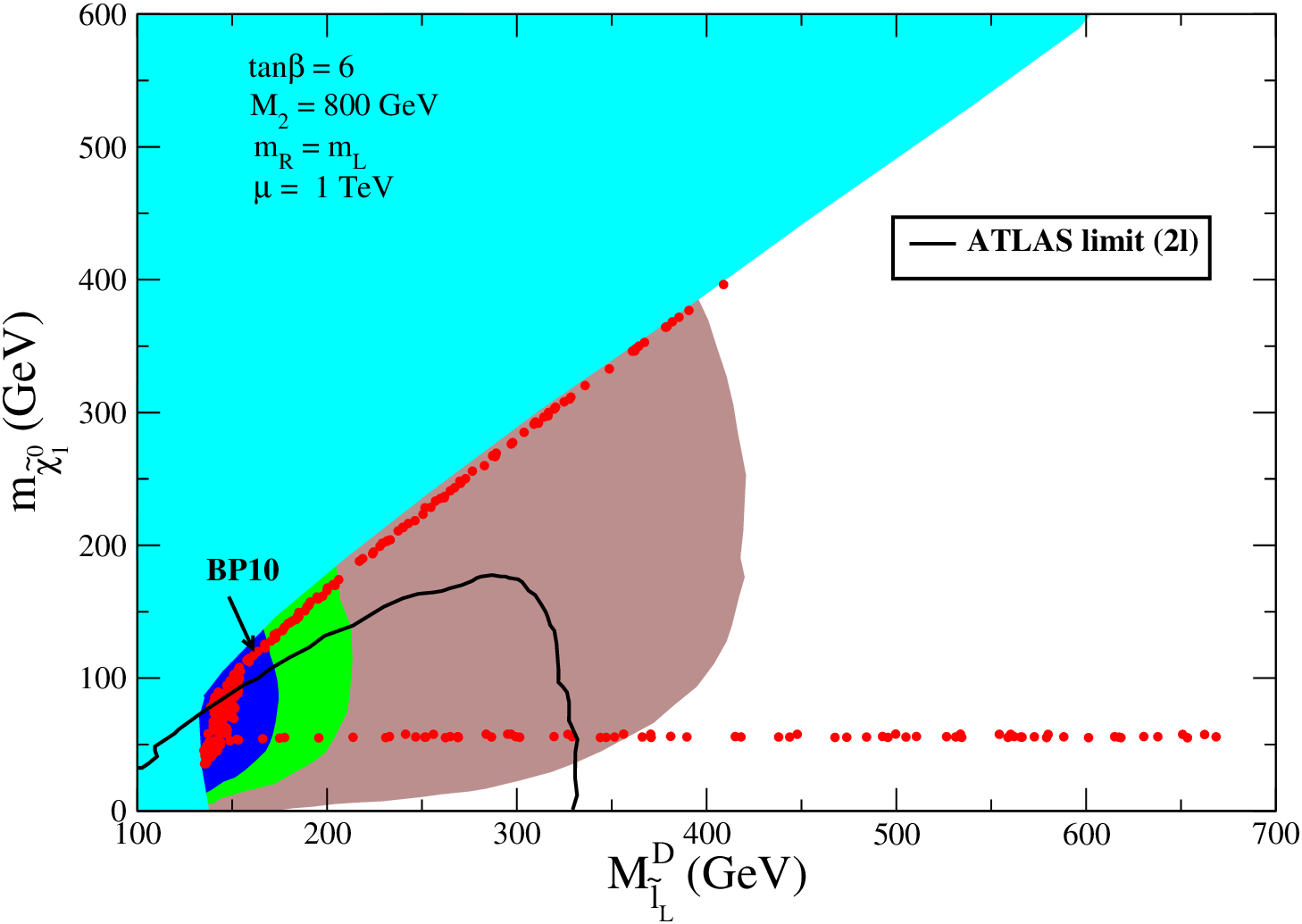}
\hspace*{0.5in}
\mygraph{LLR_lsp_slep_B}{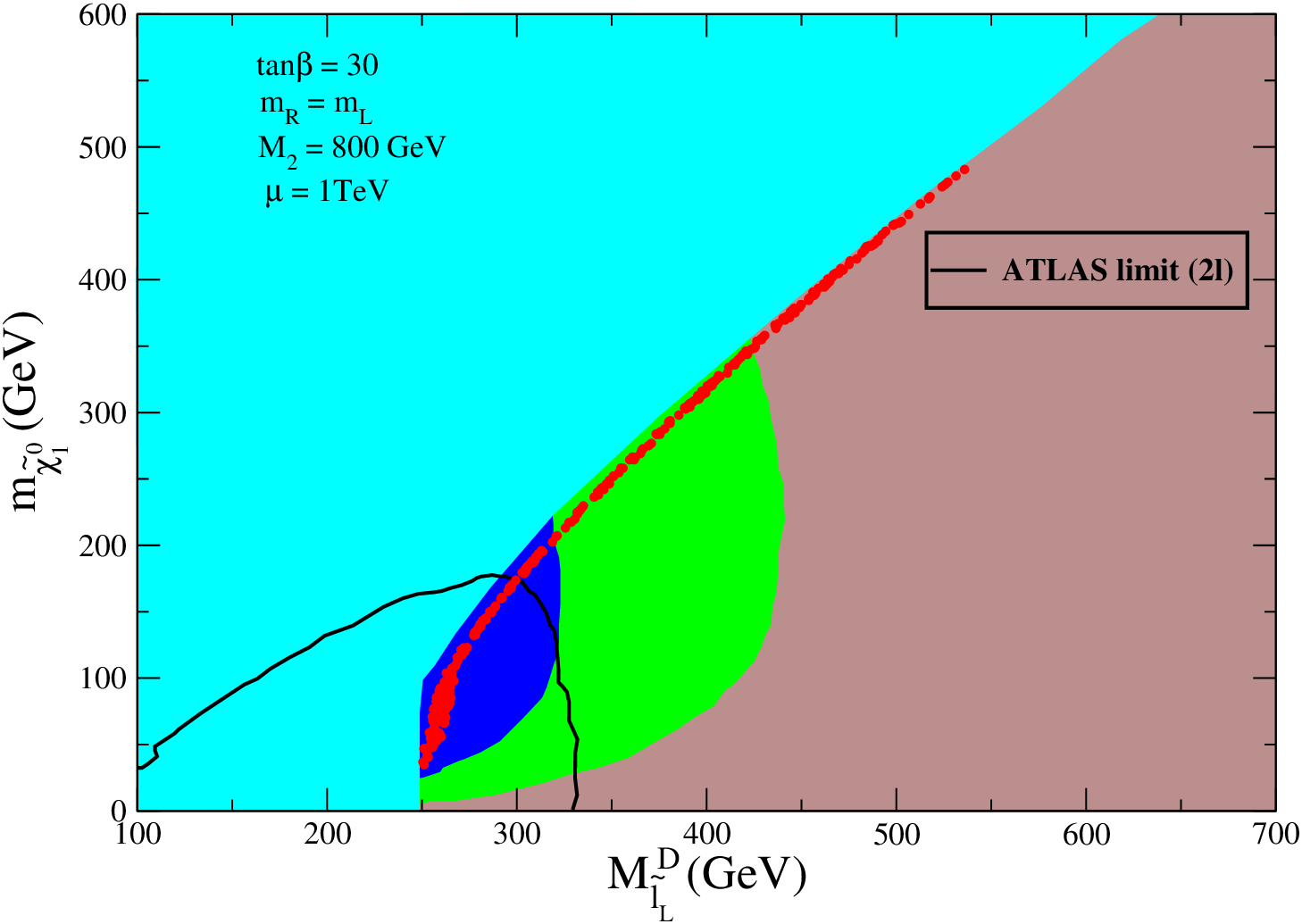}
\caption{{\it (a) Plot in the $M^D_{\tilde l_{L/R}}-\mlspone$  
plane for the LLRS scenario with $M_2 =$ 800 GeV and
tan$\beta = $6 (a) and 30 (b).  The common masses of sleptons are varied so 
that these are always smaller than $\mchonepm$.  
Colours and conventions are same as those of Fig.\ref{LL_0.5_0.5}.  The black line represents the
exclusion contour at 8 TeV by the ATLAS collaboration\cite{atlas2lew}. 
}}
\label{LLR_lsp_slep}
\end{figure}

\noindent
Fig.\ref{LLR_lsp_slep_A} shows the results for the case of light and 
degenerate left and right sleptons ($M_{{\tilde l}_L} = M_{{\tilde l}_R}$) for 
tan$\beta=6$.  
There is a significant 
amount of parameter space which is allowed by the 
collider data and $\gmin2$ constraint at the level of 1$\sigma$.
The principal contributions to $a_\mu^{\rm SUSY}$ come 
form the neutralino-smuon diagrams.  
The DM relic density satisfying mechanisms for the upper branch are LSP-stau coannihilations.  The 
s-channel light Higgs 
resonance process is viable only 
if $M_{{\tilde l}_L} = M_{{\tilde l}_R} > 360$ GeV. However, for this 
region $\amususy$ is satisfied only at the level of 3$\sigma$ for a narrow range of slepton masses. 
At the lower left corner of the parameter space, 
there is a nearly vertical strip of DM relic density satisfied points
with low values of input slepton mass.  Only a small part of this region 
corresponding to bulk annihilation is allowed by the LHC data.

In Fig.\ref{LLR_lsp_slep_B} we show a similar study with tan$\beta=30$ that 
shows the effect of enhanced $a_\mu^{\rm SUSY}$ leading to opening of 
1$\sigma$ region for larger values of the slepton masses. 
The region with $M_{{\tilde l}_L} \leq 250$ GeV is discarded because here stau 
becomes the LSP.
Similar to the case of tan$\beta=6$, 
there is a region with low 
values of slepton mass that 
arises because of bulk annihilation which is disfavoured by the LHC data.  
The upper red points satisfy relic density constraint
through LSP-stau coannihilation. 

\section{Direct and Indirect Detections of Dark Matter} 
\label{Section:DirectAndIndirectDetection}
\subsection{Direct Detection}
We probe the direct search prospects of dark matter for 
the scenarios discussed above keeping in mind the uncertainties stated in Sec.~\ref{section2.3}.
The spin independent scattering of the LSP with a proton may 
occur via t-channel Higgs 
exchange or s-channel squark exchange processes.  
Since the squarks are very heavy in view of the LHC bounds,
the Higgs exchange processes would dominantly contribute to $\sigma_{\tilde \chi p}^{\rm SI}$. 
However, since we consider only 
a bino-like LSP, we do not expect the scattering cross-section to be too large\cite{Hisano:2009xv}.
In the following figures we only show the points which satisfy PLANCK/WMAP
constraint, $\gmin2$ data upto the level of 2$\sigma$ and collider limits.

\vspace*{+0.4cm}
\begin{figure}[!htb]
\mygraph{dd_LL}{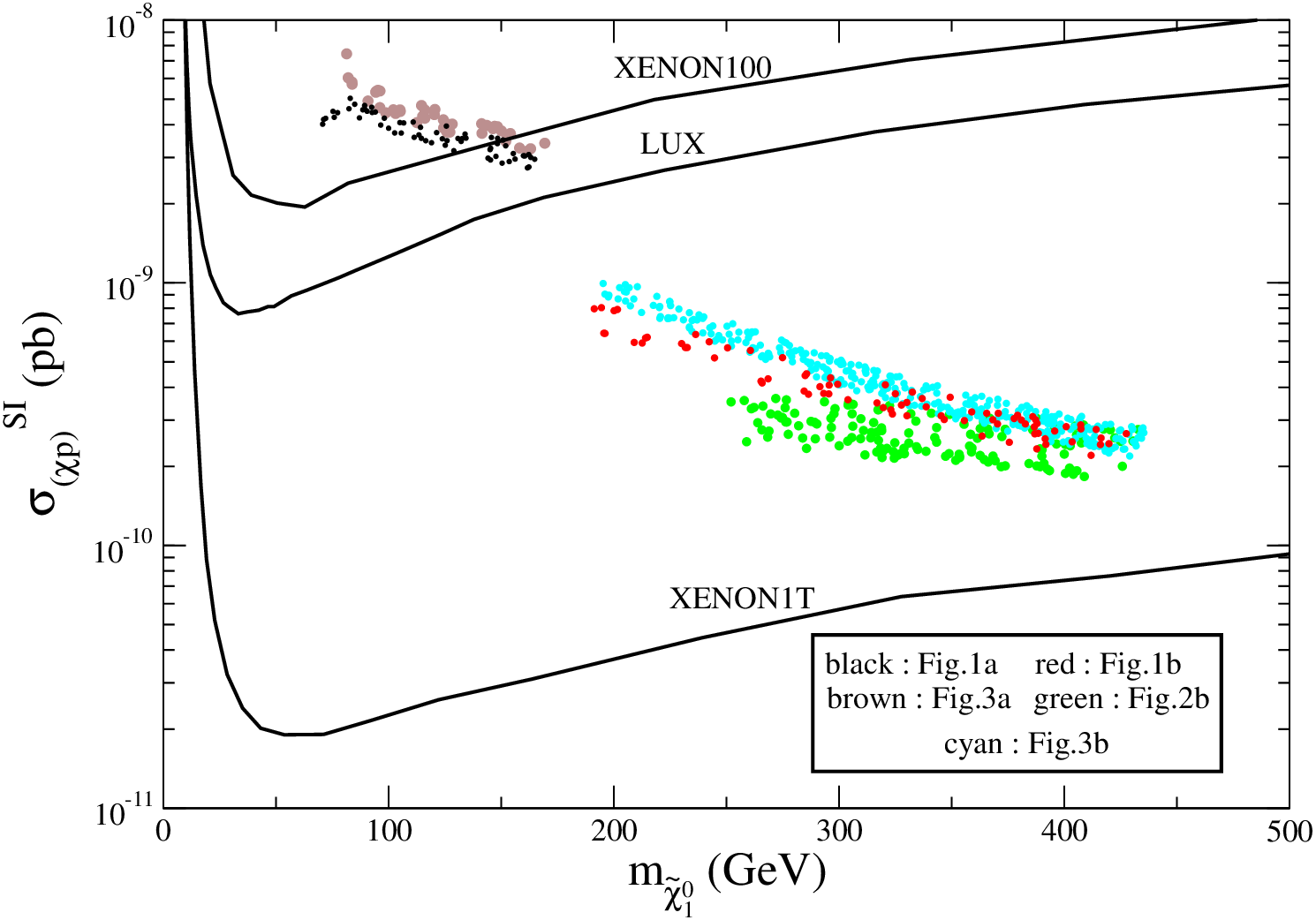}
\hspace*{0.5in}
\mygraph{dd_LLR}{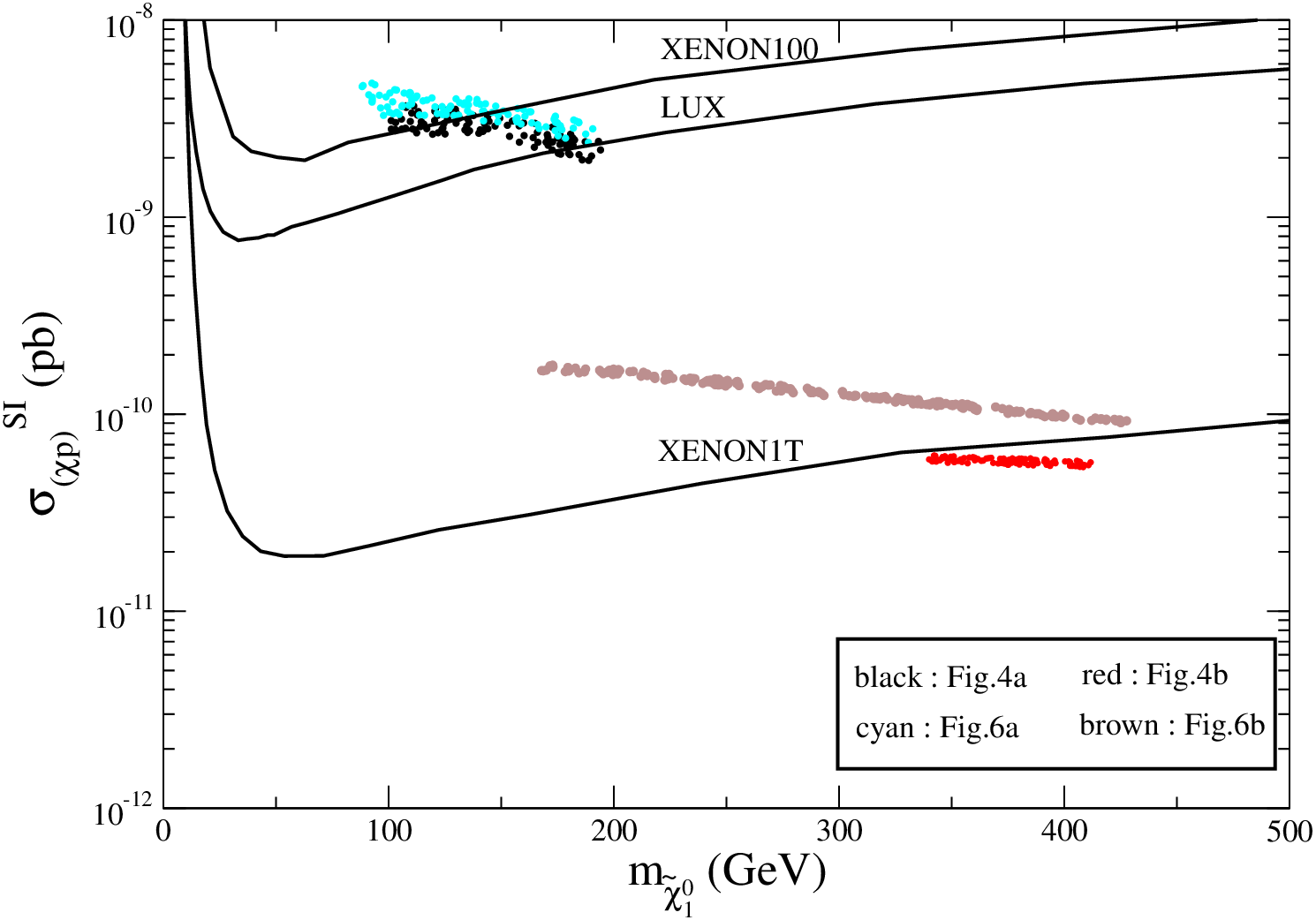}
\caption{{\it (a) Plot of spin independent scattering cross-section 
$\sigma_{p \tilde \chi}^{\rm SI}$ for scattering of proton with $\lspone$
 as a function of the mass of the LSP for the LGLS scenarios.  
Only the points which satisfy WMAP/PLANCK, 
$\gmin2$ upto the level of 2$\sigma$ and collider constraints 
are shown in the figure.
The exclusion contours for XENON100, LUX and XENON1T
experiments are shown as black lines.
Black and red points represent the case of Fig.\ref{LL_0.5_0.5_A} and 
Fig.\ref{LL_0.5_0.5_B} respectively.  
Green, brown and cyan points represent the case of Fig.\ref{LL_0.75_0.25_B},
Fig.\ref{LL_0.25_0.75_A} and Fig.\ref{LL_0.25_0.75_B} respectively.  
(b) Similar plot as (a) for the LGLRS scenarios. 
Black, Red, cyan and brown points 
represent the cases of Fig.\ref{LLR_0.5_0.5_A}, \ref{LLR_0.5_0.5_B}, 
\ref{LLR_0.25_0.75_A} and \ref{LLR_0.25_0.75_B} respectively.}}
\label{dd_LL_and_LLR}
\end{figure}

In Fig.\ref{dd_LL} we plot $\sigma_{\tilde \chi p}^{\rm SI}$ vs the mass of LSP 
for the LGLS scenarios (see Sec.~\ref{section3.1}) using 
micrOMEGAs (version 3.2)\cite{micromega3}. 
The exclusion limits specified 
by the present XENON100\cite{xenon100}, LUX\cite{lux} and future XENON1T\cite{xenon1t} 
experiments are shown as black lines.   
It follows from Sec.~\ref{actualanalysis} that the 
tilted LGLS-$\lspone$ model at low $\tan\beta$ (Fig.\ref{LL_0.75_0.25_A}) is excluded. Hence it 
does not appear in this figure. It also follows from Fig.\ref{dd_LL} that two 
other models at low $\tan\beta$ namely the LGLS model (Fig.\ref{LL_0.5_0.5_A}) and the tilted LGLS-$\chonepm$ 
model (Fig.\ref{LL_0.25_0.75_A}) of Sec.~\ref{section3.1} are disfavoured by the direct detection experiments. However,
as discussed in Sec.~\ref{Section:DetailsOfConstraints}, $\sigma_{\tilde \chi p}^{\rm SI}$
could have at least an order of magnitude of uncertainties. We therefore do 
not take the disfavoured points as finally excluded. 
We note that because of decreased coupling there
is a significant reduction in cross-section 
while moving from tan$\beta = 6$ to tan$\beta = 30$. 
We further note that the remainder of 
this class of models will be closely probed by XENON1T\cite{xenon1t} experiment.

\vspace*{+0.4cm}
\begin{figure}[!htb]
\begin{center}
\includegraphics[scale=0.35]{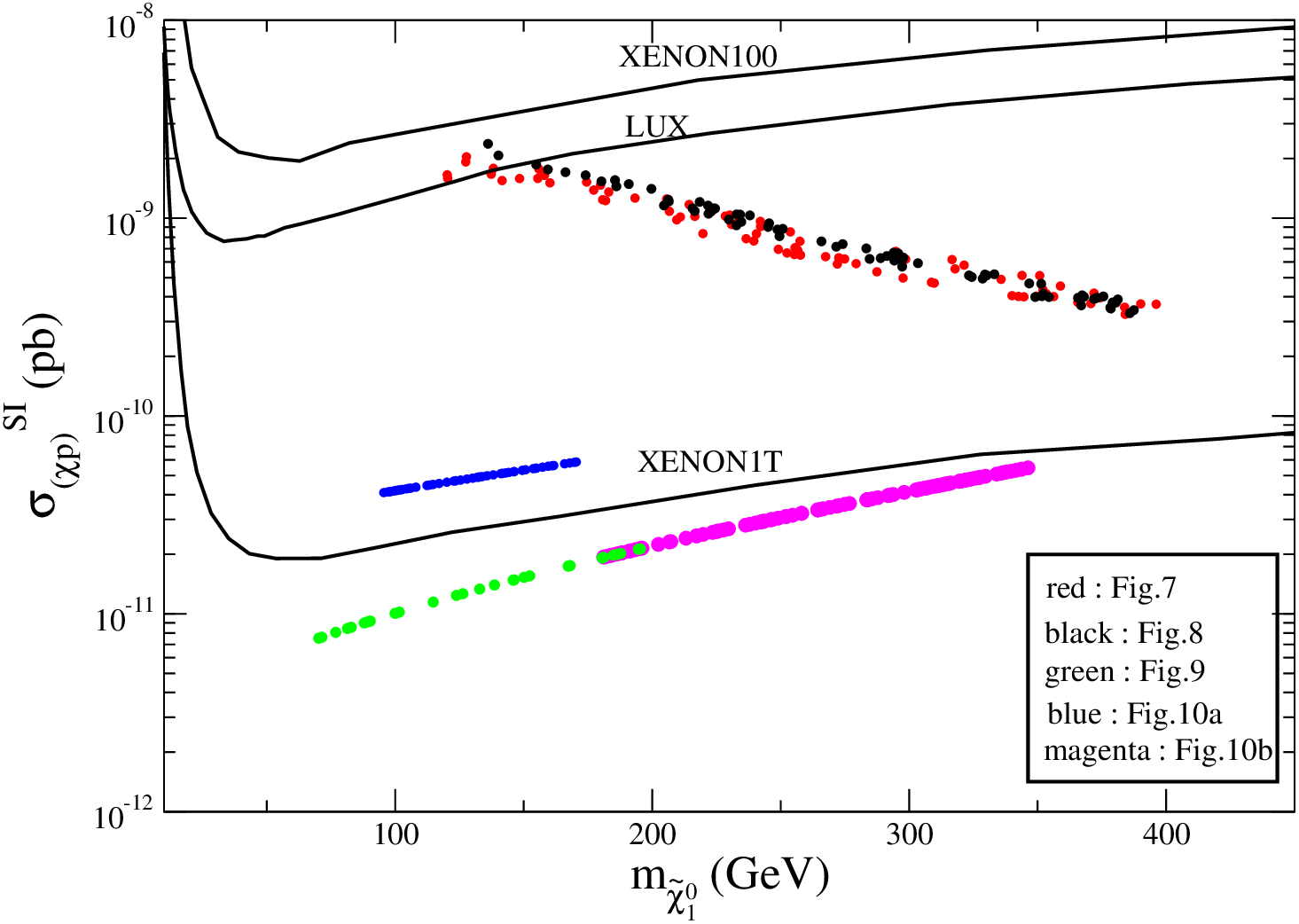}
\end{center}
\caption{{\it Similar plot as Fig.\ref{dd_LL_and_LLR} for the 
scenarios described in Figs. 
\ref{30_R_slep_gt_m2} to \ref{LLR_lsp_slep}. Red, black, green, blue and magenta points represent the cases
of Figs.\ref{30_R_slep_gt_m2}, \ref{30_slep_gt_m2}, \ref{30_LL_lsp_slep}, \ref{LLR_lsp_slep_A} and
\ref{LLR_lsp_slep_B} respectively.
}}
\label{dd_fig7_to_10}
\end{figure}

Our results 
for the LGLRS scenarios (see Sec.~\ref{section3.2}) are shown in Fig.\ref{dd_LLR}.  
We note that Fig.\ref{LLR_0.75_0.25_A} and Fig.\ref{LLR_0.75_0.25_B} corresponding to tilted LGLRS-$\lspone$ scenarios 
for low and high $\tan\beta$ have already been disfavoured by the analysis of 
Sec.~\ref{section3.2.1}. Modulo the aforesaid uncertainties, the available points 
corresponding 
to LGLRS (Fig.\ref{LLR_0.5_0.5_A}) and tilted LGLRS-$\chonepm$ (Fig.\ref{LLR_0.25_0.75_A}) 
scenarios at low $\tan\beta$ are 
disallowed via LUX\cite{lux} data. These models will be conclusively probed via 
the XENON1T. In addition, XENON1T will tightly scrutinize 
the remaining scenarios (LGLRS and tilted LGLRS-$\chonepm$)  
at high $\tan\beta$.

The direct detection cross-section for all the 
other cases namely LGRS, LGHS, LLS and LLRS (see Fig.\ref{30_R_slep_gt_m2} 
to Fig.\ref{LLR_lsp_slep} ) are plotted in Fig.\ref{dd_fig7_to_10}. 
These models are fairly insensitive to XENON100\cite{xenon100} and LUX\cite{lux} data. 
They can only probe the cases like LGHS and LGRS models for low mass range of LSP. 
The large $\mlspone$ region of these models and the remaining models 
will be probed by the XENON1T. Moreover, some of the models can even be excluded
if the theoretical uncertainties 
are brought under control in future. 

\subsection{Indirect Detection of DM through Photon Signal}

Indirect detection of DM via photon signals may be useful for probing 
certain types of DM candidates. In general, weakly interacting massive 
particles (WIMP) may undergo nuclear scattering that would reduce 
the velocity of the WIMP leading to gravitational capture 
within dense regions of astrophysical objects such as the galactic 
center, dwarf galaxies or even the Sun or the Earth\cite{dm_rev1}. 
\begin{figure}[!htb]
\vspace*{+0.4cm}
\mygraph{1a_sigmav}{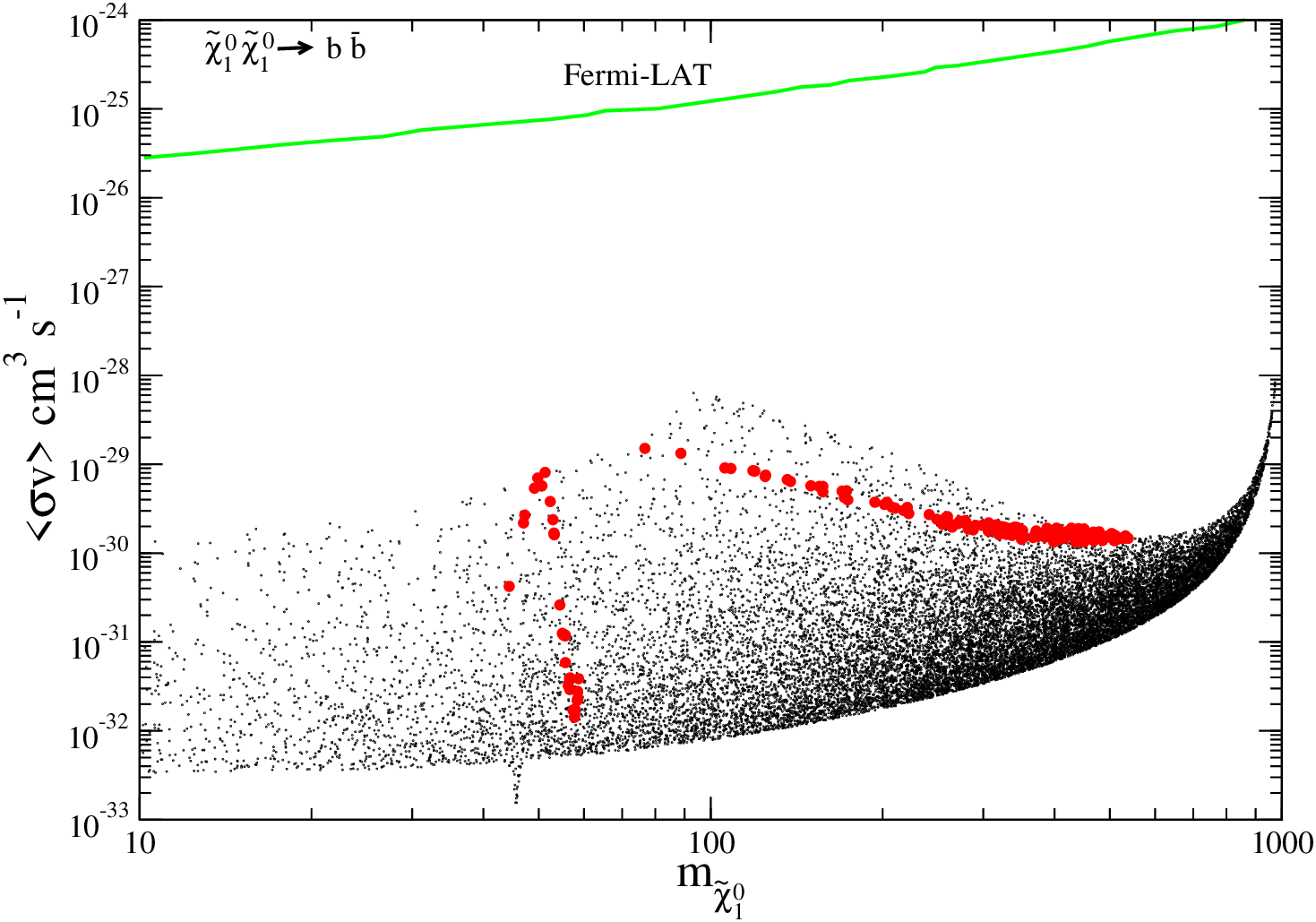}
\hspace*{0.5in}
\mygraph{1b_sigmav}{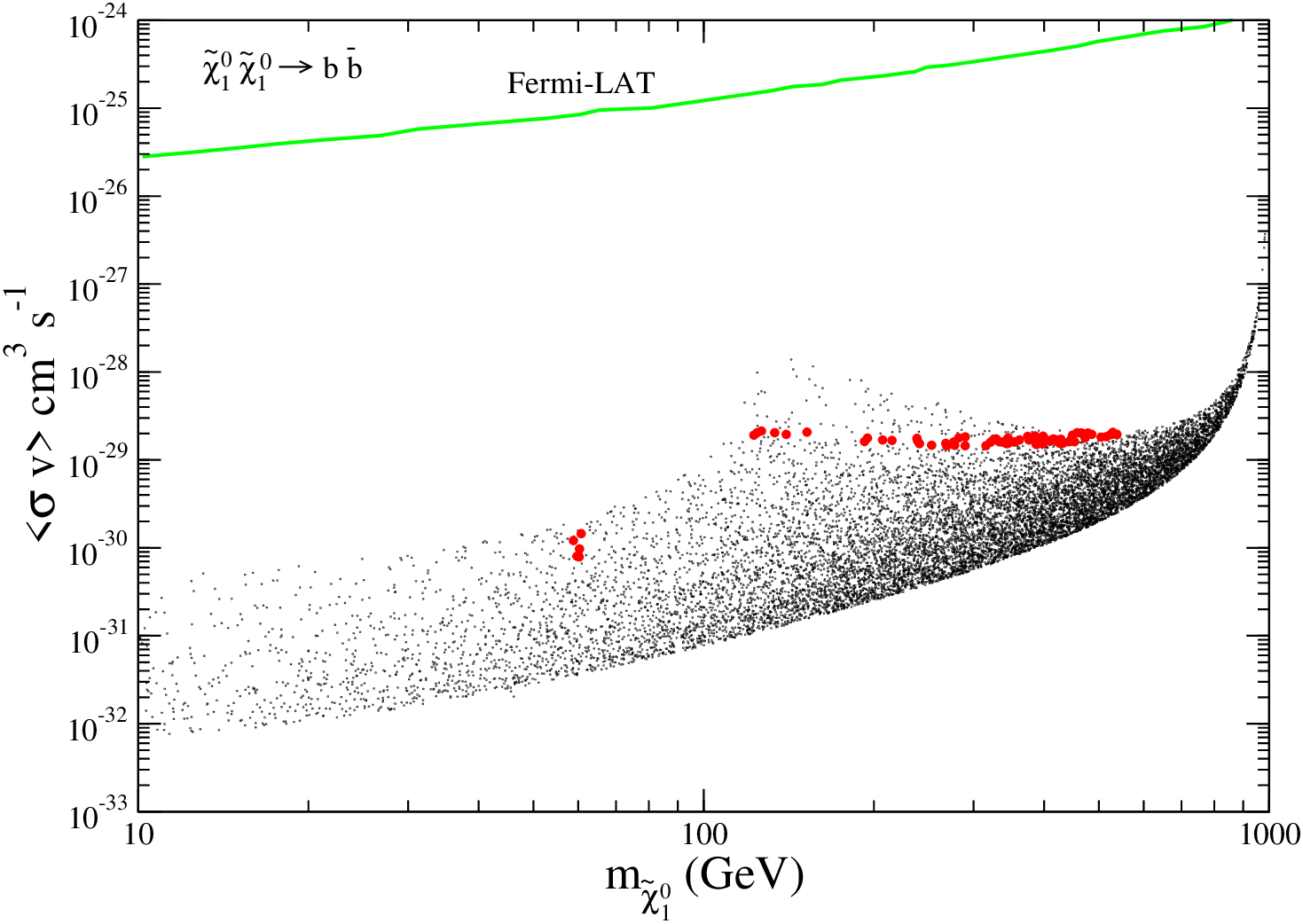}
\caption{(a) Scatter plot of DM self-annihilation cross-section against LSP mass for 
the scenario described in Fig.\ref{LL_0.5_0.5_A}.
The red points satisfy WMAP relic density constraint.  Fermi-LAT exclusion limit for
$\lspone \lspone \rightarrow b \bar b$ channel is shown as a green line. (b) Similar
plot as (a) for the case of Fig.\ref{LL_0.5_0.5_B}.}
\label{LL_0.5_sigmav}
\end{figure}
At tree level, WIMPs or LSPs may annihilate into fermion-antifermion 
pairs (quarks or leptons) or Electroweak bosons.
Hadronisation and decays of the product of primary annihilations 
may produce $\pi^0$ that would lead to photons. This is apart from the  
photons belonging to the final state radiation of primary particles.     
We note that unlike the annihilations that occurred at the freeze-out 
temperature when LSP would have a velocity that is an appreciable fraction 
of the speed of light $c$, in the present day 
environment of gravitational capture of LSPs the latter have a much smaller 
velocity $v \sim 300$~km/s or $v/c \sim 10^{-3}$ \cite{arindam}. Thus, there is a large 
$p$-wave suppression ($\sim {(v/c)}^2$) 
in the annihilation of the LSPs.

\begin{figure}[!htb]
\vspace*{+0.4cm}
\mygraph{4a_sigmav}{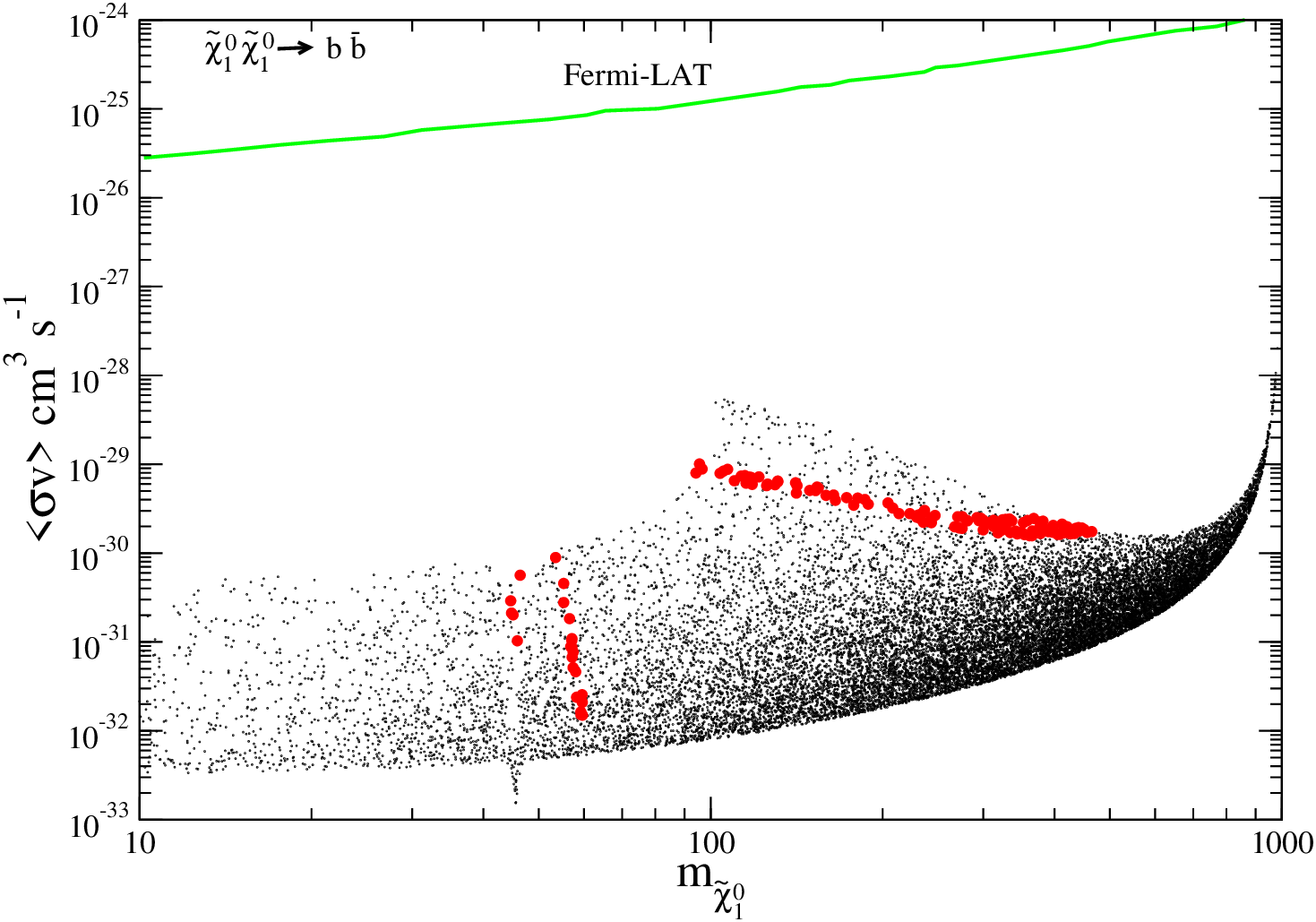}
\hspace*{0.5in}
\mygraph{4b_sigmav}{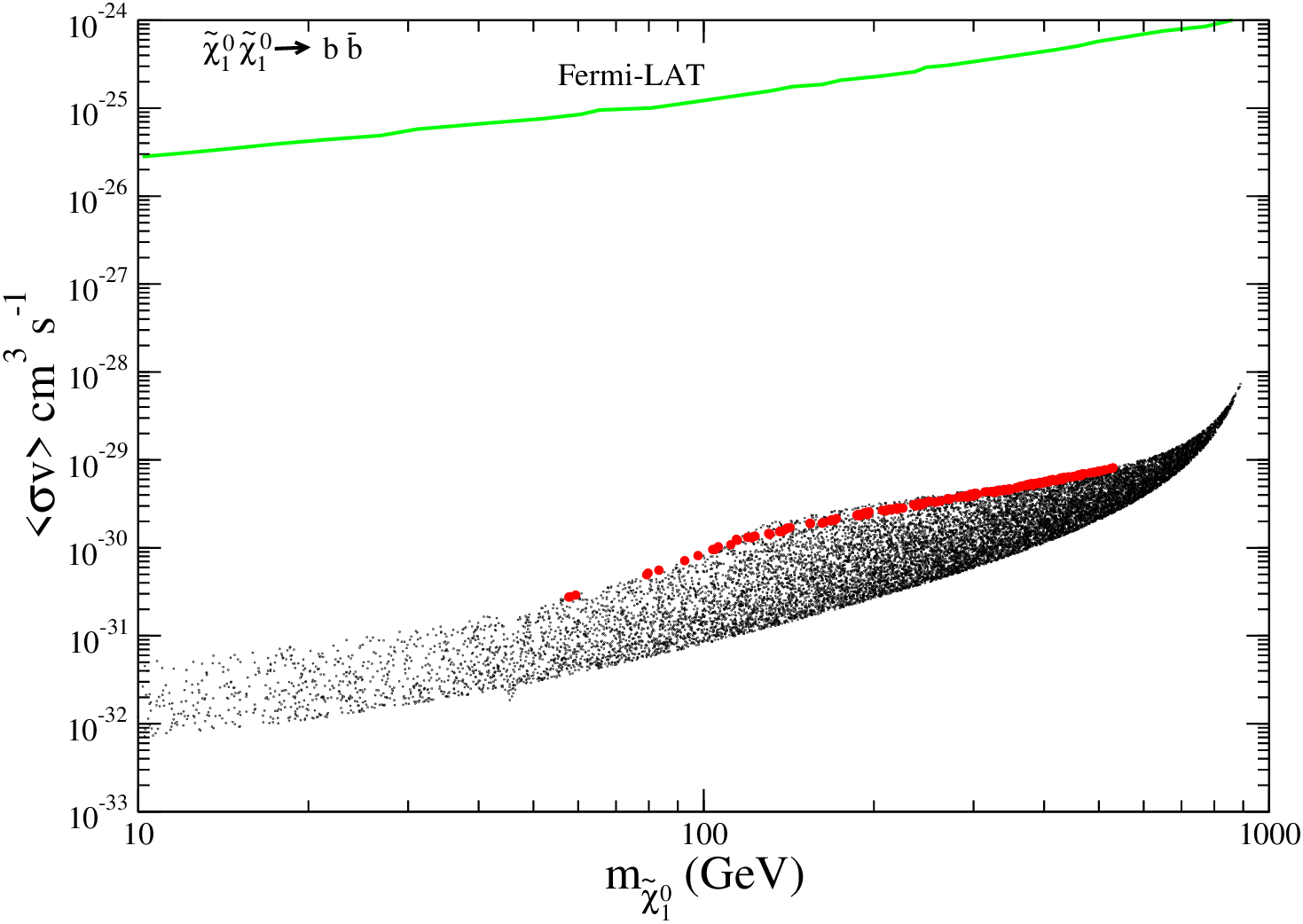}
\caption{(a) Scatter plot of DM self-annihilation cross-section against LSP mass for 
the scenario described in Fig.\ref{LLR_0.5_0.5_A}.
The red points satisfy WMAP relic density constraint.  Fermi-LAT exclusion limit for
$\lspone \lspone \rightarrow b \bar b$ channel is shown as a green line. (b) similar
plot as (a) for the case of Fig.\ref{LLR_0.5_0.5_B}.}
\label{LLR_0.5_sigmav}
\end{figure}

On the other hand, with $s$-wave annihilation becoming the dominating mode  
there is a strong helicity 
suppression that disfavours light leptons/quarks in the final state. 
We note that 
for the combined $s$-wave state, the parity of the LSP-pair is negative.  
Neutralino being a Majorana particle 
({\it i.e.} same as its anti-particle) the combined CP property 
of the LSP pair is same as the combined parity of the LSP-pair,  
which is negative. 
Hence one finds that the CP-odd Higgs boson resonance channel to contribute 
dominantly toward the photon signal. This can obviously increase if there 
is a sufficient higgsino component within the LSP.  
Thus with a principally bino type of LSP along with 
a large $M_A$ (2 TeV) we do not expect any large photon 
signal for our models.  Nevertheless, we compute 
the signal for two cases namely the LGLS scenarios (see Sec.~\ref{section3.1}, Fig.\ref{LL_0.5_0.5}) 
and LGLRS scenarios (see Sec.~\ref{section3.2}, Fig.\ref{LLR_0.5_0.5}).   
 We display the thermally averaged DM self-annihilation cross-section
in Fig.\ref{LL_0.5_sigmav} and Fig.\ref{LLR_0.5_sigmav}. The results obtained by using micrOMEGAs 
(version 3.2) corresponds to the 
NFW profile \cite{nfw} for the DM density distribution.  The Fermi-LAT 
exclusion bound for the above quantity\cite{fermi-lat-gamma} for the annihilation channel 
$\lspone \lspone \rightarrow b \bar b$ corresponding 
to the given profile is as shown.  
The figures show that the cross-sections for our case 
stay way below the Fermi exclusion limits and there is a rise
in cross-section when $\mlspone$ goes close to $M_A/2$, as 
expected, from the discussion made above. \\


\section{Gluino Mass Limits in Different Models and Their Characteristic Signatures}
\label{Section:gluinomasslimit}

We now study the feasibility of distinguishing different pMSSM scenarios introduced 
in Sec.~\ref{actualanalysis}. For this purpose we assume the gluino to be light while 
all squarks are heavy. We derive the gluino mass limit in each scenario discussed in Sec.~\ref{actualanalysis}, 
using the ATLAS data on $N_{BSM}$ (see below) in the generic $n$-leptons + $m$-jets + $\met$ 
channel \cite{atlas0l,atlas1l,atlas2l} for $n$ = 0, 1 and 2 (the same sign dilepton (SSD) signal). 
The variation of each mass limit  indicates the sensitivity of the corresponding scenario to 
the search channels. This motivates us to choose observables with different values of $n$ which can 
potentially  distinguish the models. In the process we also derive the most stringent limits on 
$\mgl$ in the above scenarios and compare them with the corresponding LHC limits on mSUGRA 
and other simplified models. 

We essentially follow the procedure of  Ref.\cite{arg_jhep2} and 
introduce appropriate ratios of the cross-sections of channels characterized by different 
values of $n$ and for $\mgl$ 
beyond the LHC limits. However, this analysis is based on recent data along with one more observable 
compared to Ref.\cite{arg_jhep2}. 
It is worth recalling that these ratios are almost free from theoretical uncertainties 
like the choice of the QCD scale, the parton density function etc.

For illustrating our main points, we chose several benchmark points (BPs) representing different 
EW sectors. All points except one (see below), are consistent with the LHC constraints from EW sparticle
 searches, the WMAP/PLANCK data, the 
${(g - 2)}_\mu$ (at the level of 2$\sigma$) and LUX data for direct detection of DM. 
Table~\ref{tab1} contains the sparticle 
spectra and the values of different observables 
corresponding to the BPs. The decay modes relevant for the gluino signals for $\mgl$ = 1.2 TeV and their 
branching ratios (BRs) are included in Table~\ref{tab2}. It may be noted that
in this table the BRs of the gluino do not add upto 100 \% in all cases. This is
due to the fact that in some scenarios the gluino also decays into the heavier chargino 
and the heavier neutralinos with small but non-negligible BRs. However, all modes are taken into account while simulating the gluino decay signal. Similarly $\lsptwo$ decays into $\wt{\tau}_2$, the heavier stau mass eigenstate,  with $\approx 7.0\%$ BR for  BP4 and BP10 which is not shown in Table~\ref{tab2}.

\begin{table}[!htb]
{\tiny
\begin{center}
\begin{tabular}{|c|c|c|c|c|c|c|c|c|c|c|c|c|}
\hline
Benchmark  & $M_1$ & $M_2$ & $\mlspone$ &$\mchonepm$ &$\mlsptwo$ &$M^D_{\tilde l_{L}}$ & $M^D_{\tilde l_{R}}$& $\mstauone$ &  $M^D_{\tilde \nu}$ & $\Omega_{\tilde \chi} h^2$ & $\sigma_{SI}$ & $a_\mu^{\rm SUSY}$\\
Points	    &	      &		   &	       &	&       &($l=e,\mu$)    &($l=e,\mu$)  &	  &              &              
&$\times 10^{-10}$	&$\times 10^{-9}$	\\
& (GeV)     &(GeV)	   &(GeV)      &(GeV)	&(GeV)    &(GeV)	      &(GeV)	   &(GeV) & (GeV) 	      &     
&(pb)			&			\\
\hline
BP1 (Fig.1b) & 240 & 262 & 232  & 266 	& 267   & 255	& 2000       &255	& 243	& 0.116	& 5.7 & 2.9\\
\hline
BP2 (Fig.2b)& 248 & 289 & 240  & 298    & 299   & 263   	& 2000      & 263 & 251   & 0.127 & 2.8 & 2.5\\
\hline
BP3(Fig.3b) & 229 & 245 & 220  & 248 & 249   & 246 & 2000  & 245 & 233   & 0.109 & 8.1 & 3.4\\
\hline
BP4(Fig.4a) & 63 & 682 & 61 & 695 & 695 & 374 & 374  &355 & 366   &0.137  &1.9  &0.3 \\
\hline
BP5(Fig.4b)  & 357 & 478 & 350 & 491 & 491 & 420 & 420 &354 & 413 & 0.098  & 0.7 & 1.3\\
\hline
BP6(Fig.6b)  & 193 & 281 & 187 & 287 & 286 & 263  &263 & 197 & 251. & 0.125  & 1.6 & 3.6\\
\hline
BP7(Fig.8)  &179 & 194 & 171 & 196 & 196 & 397  &397 & 371 & 390  & 0.127  &16.2  &3.1 \\
\hline
BP8(Fig.7)  & 190 & 206 & 183 & 208 & 208 & 408  &203 & 194 & 401     &0.108	&13.1	&2.8 \\
\hline
BP9(Fig.9)  & 89 & 700 & 86 & 709 & 709 & 122  &1000 & 109 & 95 &0.111   &0.1	&1.4 \\
\hline
BP10(Fig.10a)  & 124 & 800 & 120&799 &799& 163  &163 & 129& 145  & 0.121  & 0.5 	&2.0 \\
\hline

       \end{tabular}
       \end{center}
          \caption{ {\it  The sparticle spectra corresponding to different benchmark points (BPs)
chosen from  Fig.1 to Fig.10.}}
\label{tab1}
}
          \end{table}
\begin{table}[!htb]
{\small
\begin{center}
\hspace{-1 cm}
\begin{tabular}{|c|c|c|c|c|c|c|c|c|c|c|c|c|c|c|}
\hline
\hline
Decay 	Modes				&BP1 	&BP2	&BP3	&BP4	&BP5	 &BP6	&BP7	&BP8	 &BP9 	&BP10	\\
\hline
\hline

$\gl    \ra \lspone  q \bar q  $	&9.3	&9.5	&9.3	&50.6	 &17.1	&10.4	&8.0	&8.2	&76.1 	&75.2	\\
$\quad  \ra  \lsptwo q q\prime$		&22.5	&22.4	&22.3	&16.7	 &27.7	&23.5 	&18.3	&18.9	& 8.5	&8.8	\\
$\quad  \ra  \chonepm q \bar q  $	&45.0	&44.8	&44.9	&32.6	 &55.2	&46.8	&37.3	&38.6	&15.2 	&15.9	\\
$\quad  \ra  \chtwopm q \bar q  $	&12.4	&12.7	&12.4	&-	 &-	&10.4	&18.6	&17.6	& -	&-	\\

\hline
\hline

$\chonepm \ra \lspone q q\prime $	&-	&-	&-	&-	 &-	&-	&65.8	&-	&- 	&-	\\
$\quad    \ra \lspone \ell \nu_{\ell}$&-	&-	&	-&	 -&-	-&-	&34.2	&-	&- 	&-	\\
$\quad    \ra\snutau \tau	$	&27.8	&22.1	&33.2	&17.0	&16.3	&14.5	&-	&-	&17.3 	&17.0	\\
$\quad    \ra\stau_{1} \nutau  $	&6.4	&11.8	&1.2	&9.0	&24.5	&44.5	&-	&100	&16.5	&8.9	\\
$\quad    \ra \stau_{2} \nutau        $	&-	&-	&-	&7.4	 &-	&-	&-	&-	&- 	&7.4	\\
$\quad    \ra\snu_l l	$		&53.8	&43.2	&63.6	&34.0	&32.2	&28.2	&-	&-	&33.8	&34.0	\\
$\quad    \ra \wt{l}_L \nu_{l}$		&12.0	&22.8	&2.0	&32.6 	&26.4	&12.2	&-	&-	&32.2	 &32.4\\

\hline\hline
$\lsptwo  \ra  \lspone \gamma 	    $	&-	&-	&-	&-	 &-	&-	&15.0	&-	&- 	&-	\\
$\quad  \ra  \slepl^\pm l^\mp       $	&16.5	&26.4	&4.6	&32.9	 &27.3	&13.2	&-	&-	&33.1 	&33.4	\\
$\quad  \ra\snu_l \bar{\nu_{l}}      $	&49.5	&39.8	&61.8	&33.6	 &31.3	&26.2	&-	&-	&32.9 	&32.9	\\
$\quad  \ra  \slepr^\pm l^\mp       $	&-	&-	&-	&-	 &-	&-	&-	&13.0	& -	&-	\\
$\quad  \ra  \stauonepm \tau^\mp    $	&9.1	&13.9	&2.8	&9.1	 &25.5	&47.4	&-	&87.0	&17.2 	&9.2	\\
$\quad \ra {{\wt\tau_2}^\pm}\tau^\mp$	&-	&-	&-	&7.4	 &-	&-	&-	&-	&- 	&7.6	\\
$\quad  \ra  \snutau \bar\nutau	$	&24.8	&19.9	&30.8	&16.8	 &15.7	&13.1	&-	&-	&16.6 	&16.5	\\
$\quad  \ra  \lspone q \bar q       $	&-	&-	&-	&-	 &-	&-	&85.0	&-	& -	&-	\\
\hline
       \end{tabular}
       \end{center}
          \caption{ {\it The BRs ($\%$) of the dominant decay modes of $\gl$ (for $\mgl$ = 1.2 TeV),
 $\chonepm$ and $\lsptwo$  for the benchmark points. Here $l$ stands for e and $\mu$, but $\ell$ denotes all three generations of 
leptons. All leptonic sparticles arising from the decays of the $\chonepm$ and the $\lsptwo$ decay into their SM partner and the LSP with 100 \% BR.} }
\label{tab2}

}
        \end{table}

The BPs correspond to different DM producing mechanisms and mass hierarchies
among the EW sparticles. The mass hierarchies determine the relevant BRs as well as the 
efficiencies of the kinematical cuts for isolating the desired signals from the backgrounds.  
Below we summarize the main features of the above BPs.

\begin{itemize}
\item For BP1 - BP3 and BP9 $\snu$ is the NLSP and $\snu -\lspone$ coannihilation is the main DM producing mechanism. 
\item LSP pair annihilation via the Higgs resonance is one of the DM relic density producing mechanism for BP 4. 
However, this point is consistent with the $(g-2)_{\mu}$ constraint at the level of 3$\sigma$ only.
\item For BPs 4 - 6, 8 and 10 ~$\stauone$ is the NLSP and is responsible for DM production via coannihilation with the LSP.
\item For BP7 $\chonepm$ is the NLSP and along with $\lsptwo$ it efficiently coannihilates with the LSP.
\end{itemize}

For BP1 - BP10 (except BP7) both $\chonepm$ and $\lsptwo$ decay into two body channels 
involving all the three lepton generations (see Table~\ref{tab2}). As a result, final states enriched 
with leptons -both charged and neutral, are obtained from gluino decays. However, the abundance of
$e$ and $\mu$ in the final state varies from case to case. In the first three scenarios, $\lsptwo$ decays 
dominantly into the invisible mode $\wt{\nu}_{l} 
\nu_{l}$ or $\wt{\nu}_{\tau} \nu_{\tau}$. The extreme example is provided by BP3 where the combined BR of the invisible decays is 
92.6\%. This weakens the trilepton signature.
On the other hand for BP6 and BP8, $\stauone$ is the NLSP leading to gluino signatures 
with $\tau$-rich final states. Since $\tau$ decays primarily into 
hadrons, the final states with $e$ and/or $\mu$ will be suppressed. In BP8 with heavy L-sleptons  the final 
state is entirely $\tau$-dominated. BP7 represents a scenario where both L and R-type sleptons are heavy and 
$\chonepm$ as well as $\lsptwo$ both decay dominantly via three body modes into hadronic channels leading 
to weaker mass limits from gluino searches requiring $e$ and $\mu$ in the final states. 
            
We now summarise the ATLAS SUSY search results in the $n$ = 0, 1 and 2 (same sign dilepton(SSD)) channels.  
The ATLAS group has updated their result for SUSY search in the jets + $\met$ channel  (n = 0) 
for $\lum$ = 20.3  $\ifb$ at 8 TeV \cite{atlas0l}. 
Corresponding  to jet multiplicities from two to six, they introduced  five inclusive analyses 
channels labelled as A to E (for the details of the cuts see Table~1 of \cite{atlas0l}).  
Each channel is further divided as `Tight',`Medium' and `Loose' depending on the final 
cuts on the observables $\met$ / $m_{eff}$ and $m_{eff}$(incl.)\footnote{$m_{eff}$ is defined 
as the scalar sum of the transverse momenta of the leading N jets which defines the signal region 
 and $\met$. $m_{eff}$(incl.) is defined as the scalar sum of the transverse momenta of the jets having 
$P_T$ greater than 40 GeV and $\met$. }. The constraints are presented 
in terms of an upper limit on the effective cross-section $\sigma_{BSM}$/fb 
or the number of events from BSM physics ($N_{BSM}$) for each of the 10 signal regions. 
We use these model independent limits to derive new limits on $\mgl$ in this section.    
The observed upper limits on $N_{BSM}$ at 95 $\%$ Confidence Level (CL) for 
signal regions SRA-Light, SRA-Medium, SRB-Medium, SRB-Tight, SRC-Medium, SRC-Tight, 
SRD, SRC-Loose, SRE-Tight, SRE-Medium, SRE-Loose are 
1341, 51.3, 14.9, 6.7,  81.2,  2.4, 15.5, 92.4, 28.6, 8.3 respectively \cite{atlas0l}.

For single lepton ($n = 1$) analysis we use the ``hard single-lepton channel"  introduced 
in \cite{atlas1l}. Selection criteria for the signal regions are listed in Table~4 of Ref.\cite{atlas1l}. 
For each jet multiplicity ATLAS collaboration defined two sets of requirements - 
an inclusive signal region and a binned one. In the absence of signal events they put 
upper limits on $N_{BSM}$ at 95 $\%$ CL with $\lum$ = 20.3 $\ifb$ for 6 signal regions (see Table~17 of Ref.\cite{atlas1l}). 
Furthermore, in this analysis the electron and the muon channels are treated independently. 
For the binned hard single-lepton channels 3-jet (electron), 3-jet (muon), 5-jet (electron), 5-jet (muon), 
6-jet (electron), 5-jet (muon) the upper limits on number of events are 19.8, 11.6, 12.7, 7.7, 6.6, 7.1 respectively. 
For inclusive hard single-lepton channels 3-jet (electron), 3-jet (muon), 5-jet (electron), 5-jet (muon), 
6-jet (electron), 5-jet (muon) the upper limits on the number of events are 6.0, 7.7, 6.0, 4.6, 4.6, 3.0 respectively.

For the $n = 2$ (SSD) analysis, ATLAS group defined three signal regions 
(SR0b, SR1b, SR3b) depending on the number of tagged b jets \cite{atlas2l}.  
Since we consider all three generations of squarks, including $\lstop$ to be heavier than the gluino 
and mass degenerate, here the signal events are mainly sensitive to the 0b tagged data. 
Details of the selection cuts are discussed in Table~1 of \cite{atlas2l}. 
Analysing 20.7 $\ifb$ data recorded during LHC 8 TeV run, ATLAS collaboration 
obtained the upper limits on the number of signal events in SR0b, SR1b and 
SR3b are 6.7, 11.0 and 7.0 respectively at 95 $\%$ CL. 

We adopt the different selection criteria for varying signal regions discussed above.  
For b-tagging we use the $P_T$ dependent b-tagging efficiency obtained by ATLAS 
collaboration \cite{btagging}. 
We check that our efficiencies for different cuts used in various signal regions 
match 
with what ATLAS obtained for some benchmark points in Refs.\cite{atlas0l,atlas1l,atlas2l}.

\begin{table}[!htb]
\begin{center}\
\begin{tabular}{||c||c||c||c||}
\hline
Points		& \multicolumn{3}{c|}{Limit on $\mgl$ (GeV)} 		\\
\cline{2-4}
& $jets+ 0l + \met$\cite{atlas0l} 	&$jets+ 1l + \met$ \cite{atlas1l}	& $jets+ 2l + \met$ \cite{atlas2l}	\\
\hline
BP1	 	& 	950 		&	1125 		& 885	\\
\hline
BP2	 	&	860 		& 	1140		&950	\\
\hline
BP3	 	& 	1015 		&	1110		&810	\\
\hline
BP4	 	&	 1150		& 	1175		&-	\\
\hline
BP5	 	&	750 		&  	1155		&945	\\
\hline
BP6	 	&	1015		& 	1140		&875	\\
\hline
BP7		&	1105		& 	1080		&-	\\
\hline
BP8		&	1110		& 	1025		&-	\\
\hline
BP9		&	1250		& 	1010		&-	\\
\hline
BP10		&	1240		& 	1010		&-	\\
\hline
       \end{tabular}\
       \end{center}
           \caption{ {\it Limits on $\mgl$ using the ATLAS jets + $0l$ + $\met$ data\cite {atlas0l}, 
jets + $1l$ + $\met$ data\cite{atlas1l} and the $jets+ 2l + \met$ (SSD) data\cite {atlas2l}. }}
\label{tab3}
          \end{table}


Next we compute the number of events in different channels from gluino pair production 
for a given $\mgl$ for different benchmark points in Table~\ref{tab1}. 
For signal event generation we use PYTHIA (v6.428) \cite{pythia} and the NLO cross-section 
for the $\gl \gl$ pair production is computed by PROSPINO 2.1 \cite {prospino} with 
CTEQ6.6M PDF \cite {cteq6.6}. 
Comparing the computed number 
with the corresponding upper limits on $N_{BSM}$ in the relevant SRs and adjusting $\mgl$ accordingly, we derive 
the new limits on $\mgl$ in $0l, 1l, 2l$ (SSD) channels. The results are presented in Table~\ref{tab3}. 

It may be noted that in most cases the strongest limit on $\mgl$ 
comes from the hard single lepton (1$l$) data \cite{atlas1l}. 
This limit varies from 1010 to 1175 GeV. The results are in the same ball park as the 
limits obtained by ATLAS for heavy squarks  in mSUGRA and in many simplified 
simplified models \cite{atlas1l}.  

As has already been discussed, most of the scenarios considered by us lead to leptonically 
enriched final states. However, leptons are soft in many cases due to small energy release in 
 the concerned decay processes. As a result although the $1l$ signal is strong , the dilepton signal 
is rather weak in such cases. Moreover, the presence of at least one hard lepton in most events tends 
to weaken the bound from the  $n = 0$ channel.  
In fact a comparison of the $\mgl$ limits in the $n = 0$ and $n = 2$ (SSD) channels in different scenarios suggests a 
suitable strategy for discriminating among them as we will see below. 

In BP9 and BP10 $\chonepm$ is much heavier than the LSP. As a result, for relatively light gluinos, the  
gluino decays dominantly into the $\bar{q} q \lspone$ channel. This suppresses the 1$l$ events and practically depletes
the dilepton signal in spite of the fact that $\chonepm$ and $\lsptwo$ decay copiously into $e$ and $\mu$ . 
The strongest limits come from the $n$ = 0 channel for BP9 and BP10. The same effect is seen for BP4 albeit to a lesser
extent. Here the limits from the $n$ = 0 and $n$ = 1 channels are comparable.

Depletion of the SSD channel is also seen for BP7 and BP8. In the former case BR of chargino decay to 
$q \bar q \lspone$ is 66$\%$. In the latter case with a heavy L-slepton,  the suppression is due to the fact that $\chonepm$ and 
$\lsptwo$ decay mainly into $\tau$ rich final states with almost 100$\%$ BR.
In contrast to  BP2 and BP5 the mass difference between $\chonepm$ and the L-slepton as well as that between L-slepton and the LSP is
relatively large. Thus the $n = 2$ (SSD) channel yields stronger limits than that for $n = 0$.

\begin{table}[!htb]
\begin{center}\
\begin{tabular}{||c||c||c||c||}
\hline

Points		& $r_1$ = $\frac{S(0l+j+ \etslash)} {S(1l+j+ \etslash)}$& $r_2$ = $\frac{ S(0l+j+ \etslash)} {S(2l+j+ \etslash)}$ & $r_3$ = $\frac{S(1l+j+ \etslash)} {S(2l+j+ \etslash)}$\\
\hline
BP1	 	& 1.85			&	13.16 		&	7.12 		\\
\hline
BP2	 	& 1.35			&	6.30 		&	 4.67		\\
\hline 		
BP3	 	& 2.42			&	24.10 		&	 9.94		\\
\hline
BP4	 	& 1.45			&	 8.31		&	5.75		\\
\hline
BP5	 	& 1.17			&	 4.48		&	 3.84		\\
\hline
BP6	 	& 1.91			&	19.04 		&	9.98 		\\
\hline
BP7	 	& 4.16			&	 215.91		&	51.96 		\\
\hline
BP8	 	& 4.88			&	 287.36		&	58.91 		\\
\hline
BP9	 	& 1.70			&	 11.29		&	 6.64		\\
\hline
BP10	 	& 3.06			&	 34.02		&	 11.10		\\
\hline
\hline

       \end{tabular}\
       \end{center}
          \caption{Here $r_1$ ( $r_2$ ) represents the ratio of number of events from SRD 0l signal region \cite{atlas0l} 
with  ``Inclusive-5j1$\mu$"  single lepton signal  \cite{atlas1l}(``SR0b" SSD signal \cite{atlas2l}) region  
for $\mgl$ = 1.25 TeV.}
\label{tab4}
          \end{table}

It is worth noting from Table~\ref{tab3} that irrespective of the EW sector considered in Sec.~\ref{actualanalysis}, 
$\mgl$ limit is unlikely to be way below 1.1 TeV when limits from all channels are taken into account.

We next consider the three ratios $r_1, r_2$ and $r_3$ with relatively small theoretical errors 
introduced at the beginning of this section and defined in Table~\ref{tab4}. This table is 
computed for $\mgl=$ 1.25 TeV which is just beyond the reach of recently concluded 
LHC experiments (see Table~\ref{tab3}). Of course all three ratios are not independent. 
But their associated errors though expected to be small, may be different in each case. We quote the 
results for all three with the hope that the two having the least errors may settle the issue once 
sufficient data is accumulated. 
It follows from Table~\ref{tab4} that if one of the ratios for two benchmark points appears 
to be similar, the others will discriminate between the two. The correlation between the size 
of the ratios and the corresponding gluino mass limits may easily be noted.

\section{Conclusion} 
\label{Section:Conclusion}
The LHC searches during the 7/8 TeV runs in the $m$-jets + $n$-leptons + $\met$ channels, 
where m $\geq$ 2, have obtained important limits on the masses of the strongly interacting sparticles - 
the squarks and the gluinos (see Refs.\cite{atlas0l, atlas1l, atlas2l, atlas-susy,cms-susy}). 
These limits, however,  provide little information on the EW sparticles unless
very specific SUSY breaking mechanisms like mSUGRA \cite{msugramodel} are
invoked to relate masses of the strong and EW sparticles.
 
The purpose of this paper is to investigate the EW sector of  
pMSSM \cite{pmssm}. In order to achieve our goal we focus on the  bounds from  ATLAS and CMS searches for the 
direct production of $\chonepm$$\lsptwo$ \cite{atlas3lew,cmsew} and slepton pairs \cite{atlas2lew,cmsew} 
via the hadronically quiet channels with large $\met$.
We also include in our analysis the WMAP/PLANCK constraints \cite{wmap,planck} on the observed DM 
relic density and 
require  $a_\mu^{\rm SUSY}$ to agree with $\Delta a_\mu$ at 
the level of 2$\sigma$ (Sec.~\ref{Section:DetailsOfConstraints})\cite{g-2exp}. 
The observables 
under consideration while sensitive to the EW sectors of SUSY models, are by and large  
independent of the strongly interacting sparticles. 
Moreover, the measurement of $m_h$ enables us to study 
LSP pair annihilation into the h-resonance more precisely.    

The main conclusion of this paper is that for a fairly large number 
of pMSSM models \cite{pmssm} without specific assumptions for soft SUSY breaking, the EW sectors are 
constrained by the above data (see Figs.\ref{LL_0.5_0.5} - \ref{LLR_lsp_slep}). In many cases the constraints are quite severe while 
they are a little relaxed in the other cases. However, in all cases the 
allowed parameter space (APS) is a bounded region 
indicating both upper and lower bounds on EW sparticle masses. 

Using the model independent limits on $N_{BSM}$ (defined in Sec.~\ref{section3.1}) as obtained by ATLAS and CMS, 
we constrain the EW sectors of several pMSSM models closely related to the simplified models considered 
by  the LHC collaborations. The models are characterized by different mass hierarchies among the EW sparticles.
The simplified models showcase the basic features of  dedicated LHC searches but it is important to relate
the search results with indirect observables like the DM relic density and $\gmin2$. They also 
involve unrealistic assumptions like $M_{{\tilde l}_L}=M_{\tilde {\nu}}$ (see Sec.~\ref{section3.1}) and 
consequently miss some phenomenologically interesting possibilities like
the invisible decays of $\lsptwo$ with $\approx$ 100 \% BR (see Sec.~\ref{section3.1.1}).
We have used 
the ATLAS and CMS data to derive new 
constraints in several models which are interesting in their own 
right but not included in Refs.\cite{atlas3lew,atlas2lew,cmsew}.

We  focus on models with bino dominated LSP, wino dominated $\chonepm$ and $\lsptwo$ along with light sleptons.  
All strongly interacting sparticles and the heavier Higgs bosons are assumed to be decoupled. 
These models are highly sensitive to the trilepton signal from $\chonepm$$\lsptwo$  pair production. 
 In this analysis we have also taken into account the limits from direct slepton searches  
(Sec.\ref{section3.5} and Sec.\ref{section3.6})
which sometimes cover parameter spaces insensitive to the trilepton data.

We now summarize the results for the models with relatively light
$\chonepm$ and $\lsptwo$ and sleptons (L-type or R-type or both) lighter than the above gauginos 
(Figs.\ref{LL_0.5_0.5} - \ref{30_R_slep_gt_m2}). 
The tilted LGLS-$\lspone$ model (Sec.~\ref{section3.1.1}, Fig.\ref{LL_0.75_0.25_A}), for low values of 
$\tan\beta$,  is disfavoured by the combined constraints.  
The LGLRS-$\lspone$ model (Sec.~\ref{section3.2.1} Figs.\ref{LLR_0.75_0.25_A}, \ref{LLR_0.75_0.25_B}) 
for both low and high $\tan\beta$ is also not viable. 
The last two constraints follow from both 
chargino-neutralino and  direct slepton searches and illustrate the interplay between different    
search channels. All the other models in this category have APS consistent with combined constraints.

Within pMSSM a few DM producing mechanisms are possible which are not viable in specific  models like 
mSUGRA \cite{arg_jhep1}. LSP-sneutrino coannihilation is a case in point. However, the combined constraints used in our analysis 
put severe restrictions on some of the pMSSM allowed mechanisms. Bulk annihilation, for example, is disfavoured as the dominant relic density producing mechanism in all models except for one (see Fig.\ref{LLR_lsp_slep_A})). Only in the LLRS model with small $\tan\beta$ the tip of the near vertical red dotted region representing bulk annihilation is consistent with all constraints. The LSP pair 
annihilation into a light Higgs resonance can produce the required DM relic density for low $\tan\beta$ only. But the LHC constraints rule this out for low $\mchonepm$. As a result the SUSY contribution to $\gmin2$ is suppressed leading to a tension with the measured value. Only if the $\gmin2$ constraint
is relaxed to the level of 3$\sigma$, this option is viable in a few 
cases (see Figs.\ref{LLR_0.5_0.5_A}, \ref{LLR_0.75_0.25_A}, \ref{LLR_lsp_slep_A})\footnote{It may be recalled that in the LGHS model
(see Sec. 3.4) the Higgs resonance mechanism can not be excluded beyond
doubt since the spoiler mode may weaken the trilepton signal.}.
For similar reasons LSP annihilation into the Z-resonance is also 
not viable. Thus, in contrast to the LSP pair annihilation, 
various coannihilation processes survive as 
the main DM producing mechanisms favoured in most scenarios over large 
regions of parameter space. 

It is well known that the coannihilation mechanisms operate on narrow strips in each parameter space. The combination of theoretical constraints/ LEP limits, the LHC exclusion contours  and the $\gmin2$ constraint at the level of 
2$\sigma$ restrict the lower and
the upper edges of this strip. Thus in  each of the APS  under consideration  the EW sparticles have their masses bounded from both above and below.

We have also analysed models with heavy sleptons and lighter 
$\chonepm$, $\lsptwo$ (Sec.~\ref{section3.4} and Fig.\ref{30_slep_gt_m2}).  
In this case the LHC constraints are relatively weak. Nevertheless the 
strip allowed by WMAP/PLANCK data arising from LSP - 
$\chonepm$/ $\lsptwo$ coannihilation is bounded
by the $\gmin2$ constraint at the level of 2$\sigma$.

Models with light sleptons and heavy as well as decoupled $\chonepm$, $\lsptwo$ 
have also been considered in this analysis. We have analysed the 
 LLS (Sec.~\ref{section3.5}, Fig.\ref{30_LL_lsp_slep}) and the LLRS (Sec.~\ref{section3.6} and Fig.\ref{LLR_lsp_slep}) 
models with high and low $\tan\beta$. 
In all cases $\mchonepm$ is assumed to be beyond the direct 
LHC search limit. We find a bounded APS in each case. For the LLS model LSP-sneutrino coannihilation is 
responsible for the right amount of relic density. 
In the LLRS model $\mu$ has to be large to ensure a wino dominated
 chargino. As a result for both choices of $\tan\beta$ we find $\stauone$ to be the NLSP and LSP
 undergoes coannihilation with it to produce the required amount of DM relic density. For low $\tan\beta$, 
LSP pair annihilation into the h-resonance is also viable for slepton masses beyond the LHC reach. 
This possibility, however, is in conflict with the $\gmin2$ constraint at the $2 \sigma$ level. 
We note in passing that the light right slepton (LRS) model is inconsistent with the $\gmin2$ limit.

We have also studied the impact of the direct and indirect searches of DM on the APS of different 
models after filtering them through the above three constraints. We would however like to remind 
the readers of the inherent theoretical, experimental and astrophysical uncertainties and ambiguities 
involved in the analysis as reviewed in details in the text (see Sec.\ref{section1} and Sec.\ref{section2.3}). 

After including the DM direct detection limits, it follows from Fig.\ref{dd_LL} that there is a 
tension between two models and the XENON100 \cite{xenon100}/ LUX \cite{lux} data. These are 
the LGLS model (Fig.\ref{LL_0.5_0.5_A}) and the tilted LGLS-$\chonepm$ model (Fig.\ref{LL_0.25_0.75_A}) at low 
$\tan\beta$.
Modulo the aforesaid uncertainties the LGLRS 
(Fig.\ref{LLR_0.5_0.5_A}) and the tilted LGLRS-$\chonepm$ (Fig.\ref{LLR_0.25_0.75_A}) scenarios  at low $\tan\beta$ are also in conflict with
the direct detection data (Fig.\ref{dd_LLR}). The XENON1T experiment \cite{xenon1t} is  
expected to scrutinize all the remaining models closely.

It follows from Fig.\ref{dd_fig7_to_10} that the other cases namely the LGRS, LGHS, LLS and LLRS models (see Fig.\ref{30_R_slep_gt_m2}
to Fig.\ref{LLR_lsp_slep}) are fairly insensitive to XENON100\cite{xenon100} and LUX \cite{lux} data. XENON1T \cite{xenon1t} 
can spell the final verdict on the  LGHS and LGRS models. The remaining models 
will be probed by the XENON1T \cite{xenon1t}  
if the theoretical and astrophysical uncertainties are brought under control.
    
Next we consider the possible impacts of the above scenarios on the next 
round of experiments at LHC. 
However, it will be hard to 
establish the underlying model and the DM producing mechanism in the early stages of the experiment even if SUSY is discovered. 
Therefore we explore the possibility of identifying the 
observables which are sensitive to different DM producing 
mechanisms. This may be possible if at least one of the strongly 
interacting sparticles are relatively 
light.  The feasibility of this approach has already been demonstrated 
by considering the light stop, 
the light stop-gluino and the light gluino scenarios and 
observables based on the 
$n$-leptons + $m$-jets + $\met$ signal for different values of 
n\cite{arg_jhep1, arg_jhep2}.

In this paper we focus on  the light gluino scenario (see Sec.~\ref{Section:gluinomasslimit}). 
We choose characteristic benchmark points 
from Figs.\ref{LL_0.5_0.5} to \ref{LLR_lsp_slep} (excluding Figs.\ref{LL_0.75_0.25_A}, \ref{LLR_0.75_0.25_A}
 and \ref{LLR_0.75_0.25_B}) which are allowed by the combined constraints and 
correspond to different relic density producing mechanisms (see Tables 1 and 2). Using the latest 
ATLAS data in search channels with $n$ = 0 \cite{atlas0l}, $n$ = 1 \cite{atlas1l} and $n$ = 2 (same sign dilepton)
\cite{atlas2l}. we reanalyse the gluino mass limits in all cases (see Table~\ref{tab3}). In our generator level simulation 
we have adopted the selection criteria of Refs.\cite{atlas0l,atlas1l,atlas2l}. 

It is worth noting that the $\mgl$ limit varies considerably with the 
search channel for each BP.  For different scenarios the strongest  
limit comes from channels corresponding to different $n$. 
For all scenarios with a L-slepton lighter than the $\chonepm$ 
(BP 1-6), these limits come from the $n$ = 1 channel. In the 
remaining cases (BP 7 - 10) the $n$ = 0 channel yields the best  
limits. However, the above limits for all scenarios lie in a 
reasonably narrow range: 1105 - 1250 GeV. Thus the limit 
on $\mgl$ is only moderately sensitive to the EW sector if it is 
derived from a multichannel analysis. 

Taking cue from the above discussion the observables which may 
potentially discriminate among various scenarios can be introduced.
We define three ratios $r_1$, $r_2$ and $r_3$ 
(Table~\ref{tab4}) that are associated with relatively small theoretical errors (see Sec.~\ref{Section:gluinomasslimit}). 
They are derived using the event rates for $n$ = 0, 1 and 2 for a gluino 
mass of 1.25 TeV which is just beyond the reach of 
the recently concluded LHC experiments
(see Table~\ref{tab3}). The values of these ratios indeed illustrate that 
sufficiently accurate measurements may discriminate among the 
underlying scenarios.     

{ \bf Acknowledgments : } AD acknowledges the award of a 
Senior Scientist position by the Indian National Science Academy.
MC would like to thank Council of Scientific and Industrial Research, 
Government of India for financial support. 


\end{document}